\documentclass[aps,prd,onecolumn,eqsecnum,amsmath,nofootinbib,preprintnumbers]{revtex4}%

\usepackage[dvips]{color,graphicx}
\usepackage{amsfonts,amssymb,theorem,mathrsfs,times}
{\theorembodyfont{\upshape}
}
{\theorembodyfont{\upshape}
}
{\theorembodyfont{\upshape}
}
{\theorembodyfont{\upshape}
}
{\theorembodyfont{\upshape}
}
{\theorembodyfont{\upshape}
}

\newcommand{\dalm}{\kern1pt\vbox{\hrule height 0.9pt\hbox{\vrule width
0.9pt\hskip 2.5pt\vbox{\vskip 5.5pt}\hskip 3pt\vrule width 0.3pt}\hrule height
0.3pt}\kern1pt}

\begin{document}
\preprint{\hfill {\small {ICTS-USTC-13-13}}}
\title{Generalized Formalism in Gauge-Invariant Gravitational Perturbations
}

\author{Rong-Gen Cai$^b$}

\email{cairg@itp.ac.cn}

\author{ Li-Ming Cao$^{a,b}$}

\email{caolm@ustc.edu.cn}


\affiliation{$^a$
Interdisciplinary Center for Theoretical Study\\
University of Science and Technology of China, Hefei, Anhui 230026, China}

\affiliation{$^b$ State Key Laboratory of Theoretical Physics,
Institute of Theoretical Physics, Chinese Academy of Sciences,
P.O. Box 2735, Beijing 100190, China}

\date{\today}

\begin{abstract}
By use of the gauge-invariant variables proposed by Kodama and Ishibashi, we obtain the most general perturbation equations in the
$(m+n)$-dimensional spacetime with a warped product metric. These equations do not depend on the spectral expansions of the Laplace-type operators on the $n$-dimensional Einstein manifold. These equations enable us to have a complete gauge-invariant perturbation theory and a well-defined spectral expansion for all modes and the gauge invariance is kept for each mode.  By studying perturbations of some projections of Weyl tensor in the case of $m=2$, we define three Teukolsky-like gauge-invariant variables and obtain the perturbation equations of these variables  by considering perturbations of the Penrose wave equations in the $(2+n)$-dimensional Einstein spectime. In particular, we find the relations between the  Teukolsky-like gauge-invariant  variables and the Kodama-Ishibashi  gauge-invariant variables. These relations imply that the Kodama-Ishibashi gauge-invariant variables all come from the perturbations of Weyl tensor of the spacetime.
\end{abstract}


\maketitle


\section{Introduction}

Any gravitational perturbation theory is gauge dependent. To define the perturbation variables, we have to chose an identification mapping between the manifold of background spacetime and the manifold of perturbed spacetime~\cite{Stewart:1974uz}, therefore choosing a mapping just defines a gauge.
A perturbation variable under two different gauges is related by gauge transformation. On the level of linear perturbation, this gauge transformation is simply the Lie derivative to the corresponding unperturbed tensor field along the generator of some differmorphism of the background spacetime. We have to face the problem of gauge when one studies the perturbation of a spacetime. One way is to choose some physically preferred gauges, and another one is to establish a so called gauge-invariant perturbation theory. In the gauge-invariant theory, the perturbation variables remain unchanged under gauge transformation.  The complete set of all the gauge-invariant variables in the gauge-invariant perturbation theory determines a gauge in some sense.

The linear perturbation of a four dimensional Schwarzschild black hole was first studied by Regge and Wheeler in 1957~\cite{Regge:1957td}. By selcting a gauge for the waves with odd and even parities (Regge-Wheeler gauge), they showed the final radial wave equations are second order ordinary differential equations. Further, in the odd parity case,  the radial equations can be put into a form of a single second order Schrodinger-type equation (Regge-Wheeler equation), and the four dimensional Schwarzschild black hole is showed stable under the linear perturbation. In 1970, Zerilli showed that with the Regge-Wheeler gauge, in the even parity case, the perturbation equations can also be transformed into a Schordinger-type wave equation (Zerrili equation)~\cite{Zerilli:1970se}. In the middle of 1970's, based on the Hamiltonian formalism for spherically symmetric gravitational fields, Moncrief proved that the perturbed constraints commute to each other (under Possion brackets) and are the generators of the gauge transformation. Based on this result, some gauge-invariant canonical perturbation variables were defined for a Reissner-Nordstrom black hole, and the Regge-Wheeler like equations were obtained by  reduced Hamilton equations~\cite{Moncrief:1975sb}. This work showed that although in the discussions by Regge, Wheeler, and Zerilli, the so-called  Regge-Wheeler gauge is used, the final perturbation equations turn out to be gauge-invariant~\cite{Frolov}.

In those works, all the perturbation variables are expanded by the  harmonic tensors (scalars, vectors and symmetric rank-2 tensors) on the two dimensional transverse sphere of the four dimensional spacetime. Further, the radial equations of perturbations are second order differential equations on the two  dimensional spacetime spanning by time and radial coordinates. This structure suggests an idea on the split of the four dimensional spacetime into a product of a two dimensional sphere and a corresponding two dimensional orbit space. In 1979, Gerlach and Senguputa proposed a gauge-invariant perturbation theory for a general four dimensional spherically symmetric spacetime~\cite{Gerlach:1979rw}. The metric of the spacetime is put into a form with $``2+2"$ warped product, and gauge-invariant perturbation variables are defined by the combination of the usual perturbation variables of the metric and energy-momentum tensor. The linear perturbation gravitational equations can be nicely transformed into the equations of these gauge-invariant variables. Odd-parity and even-parity master equations are both obtained as wave equations on the two dimensional orbit space.

In recent years, with the development of supergravity, string theory,  brane world theory and AdS/CFT correspondence etc.,  gravity in higher dimensional spacetimes has attracted a lot of attention. At the beginning of 2000, Kodama, Ishibashi, and Seto have generalized the method by Gerlach and Senguputa to a general $(m+n)$-dimensional spacetime with a warped product of  an $m$-dimensional Lorentian manifold and  an $n$-dimensional maximal symmetric Riamannian space~\cite{Kodama:2000fa} (see also~\cite{Mukohyama:2000ui} for a similar discussion in $(2+n)$-dimensional maximally symmetric spacetimes).  They have not used the classifications of tensor according to the tensor's parity following Regge and Wheeler, instead the decomposition theorems of tensor on the $n$ dimensional submanifold have been used~\cite{Kodama:1985bj, Ishibashi:2004wx}. In this way, the idea to define gauge-invariant variables proposed by Gerlach and Senguputa can be generalized to the higher dimensional case.  The Kodama-Ishibashi formalism of perturbation theory has been used to study the gravitational perturbation in brane world cosmology, and the stability of higher dimensional static black holes and  higher dimensional rotating black holes with some special rotating parameters~\cite{Kodama:2003jz, Kodama:2003kk, Ishibashi:2003ap}.

Note that some discussions in most of the perturbation theories mentioned above are heavily dependent upon the spectral expansion on some Einstein manifolds which are submanifolds of the spacetimes under consideration, and the perturbation equations can be studied mode by mode only. In other words, these gauge-invariant theories have strong dependency on the spectral expansion of some Einstein manifold. However, the spectral expansion method will cause some inconvenience if one wants to compare two different gauge-invariant perturbation theories. Further, it also will lead some trouble in discussing the gauge-invariance of some special modes:  Some perturbation quantities have to be chosen to be vanishing in the spectral expansion. The absence of these perturbation quantities blocks our way to define the gauge-invariant variables for these modes. In this sense, the gauge-invariant theories by Gerlach and Sengupta~\cite{Gerlach:1979rw} or Kodama-Ishibashi~\cite{Kodama:2000fa} are incomplete. It is therefore deserved to define gauge-invariant variables and find the most general perturbation equations without using the spectral expansion. These kinds of gauge-invariant variables have been proposed in~\cite{Ishibashi:2011ws} by Ishibashi and Kodama in 2011. However,  to get the perturbation equations, the spectral expansion is still used there, and the gauge dependence of the special modes exists yet~\cite{Ishibashi:2011ws}. So it is quite necessary to obtain the general perturbation equations for these gauge-invariant variables without using the spectral expansion.

On the other hand, based on the Newmann-Penrose formalism~\cite{Newman:1961qr}, in 1970's, Teukolsky~\cite{Teukolsky:1972my,Teukolsky:1973ha} found that perturbations of gravitational fields and some matter fields in type D spacetime (for example, Kerr black hole spactime) can be cast into a unified equation, namely, Teukolsky equation (see~\cite{Misao} for new developments). This equation can describe not only gravitational radiation but also the dynamics of matter fields with some spins.  Furthermore, the corresponding equations for different spins are decoupled. Because the perturbations of some components of Weyl tensor in the Newmann-Penrose frame are gauge-invariant, the Teukolsky equation for the gravitational perturbation is gauge-invariant.  Note that in deriving  the Teukolsky equation, one does not use the harmonic expansion of the waves.  This is very different from the theories mentioned in the previous paragraphs. Unfortunately, the procedure of Teukolsky cannot be generalized to the case in higher dimensional spacetime, partially due to the absence of the Newmann-Penrose formalism beyond four dimensions, and maybe more important due to the difference of optical properties between in higher dimensional spacetime and four dimensional spacetime~\cite{Durkee:2010qu}.

In this paper, using the Kodama-Ishibashi gauge-invariant variables~\cite{Ishibashi:2011ws}, we obtain the most general perturbation equations for the linear perturbations in $(m+n)$-dimensional spacetimes with  warped product metrics.  These perturbation equations are related through the perturbations of Bianchi identity.  These  equations are independent of the spectral expansion (or the harmonic expansions according to the Laplace-Beltrami and Lichnerowcz operators) on the $n$-dimensional Einstein manifolds. Thus the incompleteness problem is remedied, and a complete gauge-invariant perturbation theory is accomplished. This improved formalism provides a useful toolkit to study other gauge-invariant perturbation theories. If we further make spectral expansions for these perturbation equations, it is found that those vector-type and scalar-type harmonic tensors introduced in~\cite{Kodama:2000fa, Ishibashi:2011ws} are no longer necessary in order to derive the mode equations for gravitational perturbations. In our method, those problems for some special modes appearing in the Kodama-Ishibashi formalism can be avoided.

To see the physical meanings of the Kodama-Ishibashi gauge-invariant variables, we study the perturbations of some projection of Weyl tensor in the $(2+n)$-dimensional Einstein spacetime with warped product metric. By defining three Teukolsky-like  gauge-invariant variables, we study the perturbation of Penrose wave equations and obtain the perturbation equations of the three Teukolsky-like variables.  These perturbation equations form a closed system and they couple into each other, and decouple only in some special cases. In particular, we find that the three Teukolsky-like gauge-invariant variables can be explicitly expressed in terms of the Kodama-Ishibashi variables. This sets up the relations between the Teukolsky-like variables and Kodama-Ishibashi variables and gives the geometric origin of the Kodama-Ishibashi gauge-invariant variables.

This paper is organized as follows. In Sec.\ref{sec:background}, according to the warped spacetime, the decompositions of the Einstein equations and linearly perturbed  Einstein equations are given. In Sec.\ref{sec:GivMaster} we first give a brief review on the gauge-invariant variables proposed by Kodama and Ishibashi, and then present the general perturbation equations of the Kodama-Ishibashi gauge-invariant variables  without using spectral expansion.  The spectral analysis of these perturbation equations is discussed in Sec.\ref{sec:Spectral}, by paying some attention on some special modes. In Sec.\ref{sec:cdim2}, we show that a single master equation can be obtained for the vector perturbation when $m=2$, and that  for an Einstein spacetime, we can obtain a single master equation for the scalar perturbation. We discuss the perturbation of the Penrose wave equation in Sec.\ref{sec:Teukolsky}, and by defining three Teukolsky-like gauge-invariant variables, we obtain corresponding perturbation equations and find the relations between the   Teukolsky-like gauge-invariant variables and the Kodama-Ishibashi gauge-invariant variables.  Sec.\ref{sec:summary} is devoted to the summary and discussions.

While in the finale stage of writing the manuscript, we noticed that the authors of a recent work~\cite{Chaverra:2012bh} obtained the perturbation equations of gauge-invariant variables proposed by Gerlach and Senguputa without using spectral expansion in four dimensional spherically symmetric spacetimes.

\section{The Decomposition of Einstein Equations and Perturbation Equations}
\label{sec:background}

\subsection{Background Spacetime and the Decomposition of Einstein Equations}
\label{subsec:background}
Let us consider a $D=m+n$ dimensional spacetime $(\mathscr{D}^D, g_{MN})$ which has a local direct product manifold $\mathscr{D}^D=\mathscr{M}^m\times \mathscr{N}^n$ and a metric with warped product
\begin{equation}
\label{metric1}
g_{MN}dx^Mdx^N= g_{ab}(y)dy^ady^b + r^2(y)\gamma_{ij}(z)dz^idz^j\, ,
\end{equation}
where coordinates $x^M=\{y^1\, ,\cdots\, ,y^m\, ;\, z^1\, ,\cdots\, ,z^n\}$. The two-element set denoted by $(\mathscr{M}^m,g_{ab})$ has a Lorentian signature, while  $(\mathscr{N}^n,\gamma_{ij})$ is Riemannian. The metric compatible covariant derivatives associated with $g_{MN}$, $g_{ab}$, and $\gamma_{ij}$ are denoted by $\nabla_M$, $D_a$, and $\hat{D}_i$, respectively. Further, the Reimannian manifold $(\mathscr{N}^n,\gamma_{ij})$ is assumed to be Einstein, i.e.,
\begin{equation}
\hat{R}_{ij}=(n-1)K \gamma_{ij}\, ,
\end{equation}
where $\hat{R}_{ij}$ is the Ricci tensor of $(\mathscr{N}^n,\gamma_{ij})$, and $K$ is a constant. Based on these definitions and assumption, the nontrivial components of Riemann tensor $R_{MNL}{}^{K}$ is given by
\begin{eqnarray}
&&R_{abc}{}^{d}={}^m\!R_{abc}{}^{d}\, ,\quad R_{aib}{}^{j}=-\frac{D_aD_b r}{r}\delta_{i}{}^{j}\, ,\nonumber\\
&&R_{ijk}{}^{l}=\hat{R}_{ijk}{}^{l}-(Dr)^2\left(\delta_{j}{}^{l}\gamma_{ki}-\delta_{i}{}^{l}\gamma_{kj}\right)\, .
\end{eqnarray}
Here ${^m}\!R_{abc}{}^{d}$ and $\hat{R}_{ijk}{}^{l}$ are the Riemann tensors of  $(\mathscr{M}^m,g_{ab})$ and $(\mathscr{N}^n,\gamma_{ij})$, respectively, and $(Dr)^2=D_arD^ar=g^{ab}D_arD_br$.
By these, it is easy to find the expressions for the components of the Ricci tensor of the spacetime,
\begin{eqnarray}
\label{Riccicomponets}
R_{ab}={^m}\!R_{ab}-n\frac{D_aD_br}{r}\, ,\qquad R_{ai}=0\, ,\nonumber\\
R_{ij}= \Big[-\frac{{}^{m}\!\Box r}{r}+(n-1)\frac{K-(Dr)^2}{r^2}\Big]r^2\gamma_{ij}\, ,
\end{eqnarray}
and then the scalar curvature of the spacetime
\begin{equation}
R={^m}\!R -2 n\frac{{}^{m}\!\Box r}{r} + n(n-1)\frac{K-(Dr)^2}{r^2}\, .
\end{equation}
Thus the Einstein equations can be decomposed into two parts:
\begin{equation}
{}^{m}\!G_{ab}-n\frac{D_aD_b r}{r}-\left[\frac{1}{2}n(n-1)\frac{K-(Dr)^2}{r^2}-n\frac{{}^{m}\!\Box r}{r}\right]g_{ab}+\Lambda g_{ab}=\kappa^2 T_{ab}\, ,
\end{equation}
\begin{equation}
-\frac{1}{2} {^m}\!R -\frac{1}{2}(n-1)(n-2)\frac{K-(Dr)^2}{r^2} + (n-1)\frac{{}^{m}\!\Box r}{r}+\Lambda=\kappa^2 P\, .
\end{equation}
In the above equations, ${^m}\!R$ and ${}^{m}\!G_{ab}$ are scalar curvature and Einstein tensor of $(\mathscr{M}^m, g_{ab})$, and ${}^{m}\!\Box=g^{ab}D_aD_b$ is the usual D'Alembertian in $(\mathscr{M}^m, g_{ab})$. According to the metric (\ref{metric1}), any energy-momentum tensor $T_{MN}$ has a decomposition $T_{MN}={\rm diag} \{T_{ab}(y), r^2P(y)\gamma_{ij}\}$, where  $T_{ab}$ and $P$ both depend only on the coordinates $\{y^a\}$.  And $\kappa^2=8\pi G$.

By using the decomposition of the Einstein equations, we find that in the case $m=2$, the nontrivial components of the Wyel tensor (denoted by $W_{M_1M_2M_3M_4}$) of the spacetime are given by
\begin{eqnarray}
\label{weylwarped}
&&W_{abcd}=2c_1 w g_{a[c}g_{d]b}
-\frac{2}{n}\kappa^2(g_{a[c}\psi_{b]d}-g_{b[c}\psi_{d]a})\, ,\nonumber\\
&&W_{iajb}=-c_2 w r^2 g_{ab}\gamma_{ij}\, ,\nonumber\\
&&W_{ijkl}=2 c_3 w r^4\gamma_{i[k}\gamma_{l]j}+r^2\hat{W}_{ijkl}\, ,
\end{eqnarray}
where
\begin{equation}
\label{eq2.9}
c_1=\frac{n-1}{2(n+1)}\, ,\qquad c_2= \frac{n-1}{2n(n+1)}\, ,\qquad c_3= \frac{1}{n(n+1)}\, ,
\end{equation}
and $\psi_{ab}$ is the traceless part of $T_{ab}$ according to the metric of $(\mathscr{M}^2, g_{ab})$ , i.e., $\psi_{ab}=T_{ab}-(1/2)g_{ab} T^c_{~c}$.
Since the indices of $$\frac{2}{n}\kappa^2(g_{a[c}\psi_{b]d}-g_{b[c}\psi_{d]a})$$ have the symmetry of the Riemann tensor in the two dimensional space $(\mathscr{M}^2, g_{ab})$, one can easily find that term  is identically vanishing~\cite{Maeda:2007uu}.
The scalar $w$ is defined by
\begin{equation}
\label{eq210}
w={}^2\!R+2\frac{{}^{2}\!\Box r}{r} + 2\frac{K-(Dr)^2}{r^2}\, .
\end{equation}
The tensor $\hat{W}_{ijkl}$ has a form
\begin{equation}
\hat{W}_{ijkl}=\hat{R}_{ijkl}-2K\gamma_{i[k}\gamma_{l]j}\, ,
\end{equation}
 which is just the Weyl tensor of the Einstein manifold $(\mathscr{N}^n,\gamma_{ij})$ when $n>3$.  Note that two or three dimensional Einstein manifolds are maximally symmetric, therefore $\hat{W}_{ijkl}$ always vanishes when $n=2$ or $n=3$.

\subsection{Decomposition of Linear Perturbation Equations}
\label{subsec:linearPT}
To extract the tensor, vector, and scalar parts of a given perturbation variable according to the tensor decomposition theorem of the Einstein manifold $(\mathscr{N}^n,\gamma_{ij})$,  we have to decompose the perturbation variable in the same way like the background geometric quantities.

Considering a metric perturbation $g_{MN}\rightarrow g_{MN}+h_{MN}$,  the linear perturbation equations of Einstein gravity are given by
\begin{equation}
\label{linearEinsteinE}
\delta G_{MN}+\Lambda h_{MN}=\kappa^2 \delta T_{MN}
\end{equation}
or
\begin{eqnarray}
&&-\Box h_{MN}+R_{ML}h_{N}{}^{L}+R_{NL}h_{M}{}^{L}-2R_{MLNK}h^{LK}+(\Box h) g_{MN}\nonumber\\
&&+\nabla_M\nabla_Lh_N{}^{L}+\nabla_N\nabla_Lh_M{}^{L}-\nabla_M\nabla_N h -\nabla^L\nabla^Kh_{LK}g_{MN}\nonumber\\
&&+ R^{LK}h_{LK}g_{MN}-Rh_{MN}+2\Lambda h_{MN}\nonumber\\
&&=2\kappa^2 \delta T_{MN}\, ,
\end{eqnarray}
where $h=g^{MN}h_{MN}$. The decomposition of the equations are complicated. Considering
$$
2\delta G_{MN}=2\delta R_{MN}-R h_{MN} - \delta R g_{MN}\, ,
$$
 and $$\delta R= g^{MN}\delta R_{MN}-R^{MN}h_{MN}\, ,$$  one can see that the decomposition is completed once the decomposition of $\delta R_{MN}$ is done (The expression of $\delta R_{MN}$ can be found in Appendix\ref{sec:pweyl}). After some calculations, we have
\begin{eqnarray}
&&2\delta R_{ab}=-{}^{m}\!\Box h_{ab}  + {}^m\!R_{a}{}^{c}h_{cb}+  {}^m\!R_{b}{}^{c}h_{ac}-2({}^m\!R_{acbd}h^{cd})+ D_aD^ch_{cb}+D_bD^ch_{ac}\nonumber\\
&&+n\frac{D^cr}{r}\Big(-D_ch_{ab}+D_ah_{cb}+D_bh_{ac}\Big) -\frac{1}{r^2}\hat{\Delta}h_{ab}+\frac{1}{r^2}\Big(D_a\hat{D}^ih_{bi}+D_b\hat{D}^ih_{ai}\Big)\nonumber\\
&&-\frac{1}{r^3}\Big[D_arD_b(h_{ij}\gamma^{ij})+D_brD_a(h_{ij}\gamma^{ij})\Big]+ 4\frac{D_arD_br}{r^4}(h_{ij}\gamma^{ij})-D_aD_bh\, ,
\end{eqnarray}
and
\begin{eqnarray}
&&2\delta R_{ai}=\hat{D}_iD^bh_{ab}+(n-2)\frac{D^br}{r}\hat{D}_ih_{ab}-r~{}^{m}\!\Box \Big(\frac{h_{ai}}{r}\Big)-n\frac{D^br}{r}D_bh_{ai}\nonumber\\
&&-D_arD^b\Big(\frac{h_{bi}}{r}\Big)+(n+1)\frac{D^br}{r}D_ah_{bi} + rD_aD^b\Big(\frac{h_{bi}}{r}\Big)+\frac{D_arD^br}{r^2}h_{bi}\nonumber\\
&&+(n+1)rD_a\Big(\frac{D^br}{r^2}\Big)h_{bi}-rD_a\Big(\frac{\hat{D}_ih}{r}\Big)-(n+2)\frac{D_aD^br}{r}h_{bi}+ {}^m\!R_{a}^{~b}h_{bi}\nonumber\\
&&+\Big[(n+1)\frac{(Dr)^2}{r^2}+(n-1)\frac{K-(Dr)^2}{r^2}-\frac{{}^m\!\Box r}{r}\Big]h_{ai}  -\frac{1}{r^2}\hat{\Delta} h_{ai}\nonumber\\
 &&+ \frac{1}{r^2}\hat{D}_i\hat{D}^jh_{aj} + r D_a\Big(\frac{\hat{D}^jh_{ij}}{r^3}\Big) + \frac{D_ar}{r^3}\hat{D}^jh_{ij}- \frac{D_ar}{r^3}\hat{D}_i(h_{jk}\gamma^{jk})
\, ,
\end{eqnarray}
where $\hat{\Delta}=\gamma^{ij}\hat{D}_i\hat{D}_j$ corresponds to the Laplace-Beltrami operator of $(\mathscr{N}^n,\gamma_{ij})$. These two equations actually are the same as the results in the Appendix of~\cite{Kodama:2000fa}. The difference appears in the perturbation of $R_{ij}$, which is given by
\begin{eqnarray}
&&2\delta R_{ij}=2\Big[r D^arD^bh_{ab} + (n-1)D^arD^br h_{ab}+ rD^aD^brh_{ab}\Big]\gamma_{ij}\nonumber\\
&&+ r\hat{D}_iD^a\Big(\frac{h_{aj}}{r}\Big)+r\hat{D}_jD^a\Big(\frac{h_{ai}}{r}\Big)+(n-1)\frac{D^ar}{r}\Big(\hat{D}_ih_{aj}+\hat{D}_jh_{ai}\Big)\nonumber\\
&&+2\frac{D^ar}{r}\hat{D}^kh_{ak}\gamma_{ij}-r^2~{}^m\!\Box\Big(\frac{h_{ij}}{r^2}\Big)-n\frac{D^ar}{r}D_ah_{ij}+\frac{1}{r^2}
\Big(\hat{D}_i\hat{D}^{k}h_{jk}+\hat{D}_j\hat{D}^{k}h_{ik}\Big)\nonumber\\
&&-\frac{1}{r^2}\hat{\Delta}h_{ij} + 2\Big[(n-1)\frac{K}{r^2}+ \frac{(Dr)^2}{r^2}-\frac{{}^m\!\Box r}{r}\Big]h_{ij}
 -\frac{2}{r^2}\hat{R}_{i}{}^k{}_j{}^lh_{kl}\nonumber\\
 &&-\hat{D}_i\hat{D}_jh-rD^arD_ah\gamma_{ij}\, .
\end{eqnarray}
The Riemann tensor  $\hat{R}_{ijkl}$ appears in the fourth line of this equation because $(\mathscr{N}^n,\gamma_{ij})$ is assumed to be a general Einstein manifold but without the assumption of maximal symmetry. When  $(\mathscr{N}^n,\gamma_{ij})$ is maximally symmetric, this equation reduces to the one in~\cite{Kodama:2000fa}. Substituting above results into Eq.(\ref{linearEinsteinE}), we can obtain the decompositions of the linearly perturbed Einstein equations.

\section{Gauge-invariant Perturbation variables and perturbation Equations}
\label{sec:GivMaster}
\subsection{Gauge invariant  variables of perturbations}
\label{subsec:gipv}
In this section, we give a brief review on the gauge-invariant variables introduced by Ishibashi and Kodama in 2011~\cite{Ishibashi:2011ws}.
The existence of the Einstein manifold $(\mathscr{N}^n,\gamma_{ij})$ allows us to classify the metric perturbation $h_{MN}$ into tensor, vector and scalar parts. The variable $h_{ab}$  clearly belongs to the scalar part according to the differmorphism of $(\mathscr{N}^n,\gamma_{ij})$, while $h_{ai}$ and $h_{ij}$ are vector and rank-2 symmetric tensor, respectively. They are decomposed as
\begin{equation}
\label{hai}
h_{ai}=\hat{D}_ih_a+h_{ai}^{(1)}\, ,
\end{equation}
\begin{equation}
\label{hij}
h_{ij}=h^{(2)}_{Tij}+\hat{D}_ih_{Tj}^{(1)}+\hat{D}_jh_{Ti}^{(1)} + h_L\gamma_{ij} + \hat{L}_{ij}h_T\, ,
\end{equation}
where $$\hat{L}_{ij}=\hat{D}_i\hat{D}_j- \frac{1}{n} \gamma_{ij}\hat{\Delta}\, .$$ The tensor $h^{(2)}_{Tij}$ is transverse trace free, and vectors $h_{ai}^{(1)}$ and $h_{Ti}^{(1)}$ are both divergence free. $h_a$, $h_T$, and $h_L$ are scalars on $(\mathscr{N}^n,\gamma_{ij})$. So the scalar part of $h_{MN}$ is given by $(h_{ab}, h_{a}, h_L, h_T)$, and the vector part includes $(h_{ai}^{(1)}, h_{Ti}^{(1)})$, while tensor components are simply $h^{(2)}_{Tij}$. Similarly, for the perturbation of the energy-momentum tensor, i.e., $\delta T_{MN}$, we have
\begin{equation}
\delta T_{ai}= \hat{D}_i\delta T_a + \delta T^{(1)}_{ai}\, ,
\end{equation}
and
\begin{equation}
\delta T_{ij}=\delta T^{(2)}_{Tij} + \hat{D}_i\delta T_{Tj}^{(1)} + \hat{D}_j\delta T_{Ti}^{(1)}+ \delta T_L \gamma_{ij} + \hat{L}_{ij}\delta T_T\, .
\end{equation}
So the perturbations of the energy-momentum tensor can also be classified into scalar, vector, and tensor parts, i.e., $(\delta T_{ab}, \delta T_a, \delta T_L, \delta T_T)$, $(\delta T^{(1)}_{ai}, \delta T^{(1)}_{Ti})$, and $\delta T^{(2)}_{Tij}$. These components are not gauge-invariant except for the tensor components. However, by studying their gauge transformations, one can consider some combination of them and construct some gauge-invariant variables. For the metric perturbation $h_{MN}$, one finds the gauge-invariant quantities
\begin{equation}
\left\{h_{Tij}^{(2)}\, ;\, \, F_{ai}^{(1)};\, \, F^{(0)}_{ab}\, ,\, \, F^{(0)}\right \}\, ,
\end{equation}
while for the perturbation of energy-momentum tensor $\delta T_{MN}$, the gauge-invariant variables are
\begin{equation}
\left\{\delta T_{Tij}^{(2)}\, ;\,  \tau_{ai}^{(1)}\, , \,\tau_{ij}^{(1)}\, ;\, \Sigma^{(0)}_{ab}\, ,\, \Sigma^{(0)}_{ai}\, , \, \Sigma^{(0)}\, , \, \Pi^{(0)}_{ij}\right \}\, .
\end{equation}
These gauge-invariant variables are defined as follows~\cite{Ishibashi:2011ws}:
\begin{eqnarray}
\label{giv}
&&F^{(1)}_{ai}=h_{ai}^{(1)}-r^2D_a\Big(\frac{h_{Ti}^{(1)}}{r^2}\Big)\, ,\nonumber\\
&&\tau_{ai}^{(1)}=\delta T_{ai}^{(1)}-Ph_{ai}^{(1)}\, ,\nonumber\\
&&\tau_{ij}^{(1)}=2\hat{D}_{(i}\delta T_{Tj)}^{(1)}-2 P\hat{D}_{(i}\delta h_{Tj)}^{(1)}\, ,\nonumber\\
&&F^{(0)}_{ab}=h_{ab}+2D_{(a}X_{b)}\, ,\nonumber\\
&&F^{(0)}=h_L+ 2 rD^a r X_a + \frac{2}{n}\hat{\Delta}X_L\, ,\nonumber\\
&&\Sigma^{(0)}_{ab}=\delta T_{ab} + X^cD_cT_{ab} + T_a^{~c}D_bX_c + T_b^{~c}D_a X_c\, ,\nonumber\\
&&\Sigma^{(0)}_{ai}=\hat{D}_i\left[\delta T_a + T_{ab}X^b + r^2 P D_a\left(\frac{X_L}{r^2}\right) \right]\, ,\nonumber\\
&&\Sigma^{(0)}=\delta T_L-P h_L + r^2 X^a D_aP\, ,\nonumber\\
&&\Pi^{(0)}_{ij}=\hat{L}_{ij}\left(\delta T_T + 2 P X_L\right)\, ,
\end{eqnarray}
where $X_a$ and $X_L$ are respectively given by
\begin{equation}
\label{XaXL}
X_a=-h_a + \frac{1}{2}r^2 D_a\left(\frac{h_T}{r^2}\right)\, ,\quad \mathrm{and}\quad X_L = -\frac{1}{2}h_T\, .
\end{equation}
 These gauge-invariant variables are independent of
the spectral expansions on the Einstein manifold $(\mathscr{N}^n, \gamma_{ij})$.  These are different from those in~\cite{Kodama:2000fa} where the gauge-invariant variables are defined mode by mode.

The perturbation equations of these gauge-invariant variables will be given in the next subsection. Here we give some remarks on these gauge-invariant variables.
First we notice that the geometric meanings of some variables are absent. For example, it is not so easy to understand the geometric meaning of the variable $F^{(0)}_{ab}$. We will give an appropriate explanation of these variables in the case of $m=2$ in subsection~\ref{subsec:decompos}.
Second, in the tensor decomposition theorems on $(\mathscr{N}^n, \gamma_{ij})$, there are some ambiguities~\cite{Ishibashi:2004wx}. These ambiguities will
cause that these gauge-invariant variables are defined up to some uncertainty.  These uncertainties of gauge-invariant variables are just the tensors belonging to the kernel of some elliptic operators (such as $\hat{\Delta}$ and $\hat{\Delta}(\hat{\Delta}+nK)$). For example,  $h_{a}(y,z)$ in Eq.(\ref{hai}) can be changed to $h_a(y,z)+f_a(y)$, where $f_a$  depends on the coordinates $\{y^a\}$ only.  However, it is clear that these ambiguities can be removed by redefining the variables in (\ref{giv})~\cite{Ishibashi:2004wx}.

\subsection{ Equations of the Gauge-invariant Variables}
\label{subsec:gme}
\subsubsection{Tensor perturbation}

Let us first discuss the tensor perturbation  by setting
\begin{eqnarray}
&&h_{ab}=0\, ,\quad h_{ai}=0\, ,\quad h_{ij}=h^{(2)}_{Tij}\, ,\nonumber\\
&&\delta T_{ab}=0\, ,\quad \delta T_{ai}=0\, ,\quad \delta T_{ij}=\delta T^{(2)}_{Tij}\, .
\end{eqnarray}
From the linear perturbation $\delta G_{MN}+ \Lambda h_{MN}=\kappa^2\delta T_{MN}$, we find that the nontrivial component of this equation is $\delta G_{ij}+\Lambda h_{ij}=\kappa^2 \delta T_{ij}$, and it can be expressed as
\begin{equation}
\label{T}
-{}^{m}\!\Box \Big(\frac{h^{(2)}_{Tij}}{r^2}\Big)-n\frac{D^ar}{r}D_a\Big(\frac{h^{(2)}_{Tij}}{r^2}\Big)
+\frac{\hat{\Delta}_L-2(n-1)K}{r^2}\Big(\frac{h^{(2)}_{Tij}}{r^2}\Big)=2\kappa^2 \Big[\Big(\frac{\delta T_{Tij}^{(2)}}{r^2}\Big)- P\Big(\frac{h^{(2)}_{Tij}}{r^2}\Big)\Big]\, ,
\end{equation}
where $\hat{\Delta}_L$ is Lichnerowicz operator acting on the symmetric rank-2 tensor on $(\mathscr{N}^n,\gamma_{ij})$. The relation between this operator and usual Laplace-Beltrami operator is given by Weitzenb$\mathrm{\ddot{o}}$ck formula:
\begin{equation}
\label{Lichnerowicz}
\hat{\Delta}_{L}s_{ij}=-\hat{\Delta}s_{ij}+\hat{R}_{i}{}^{k}s_{kj}+\hat{R}_{j}{}^{k}s_{ik}-2\hat{R}_{ikjl}s^{kl}\, ,
\end{equation}
where $s_{ij}$ is an arbitrary symmetric tensor field on $(\mathscr{N}^n,\gamma_{ij})$.  Since we  are considering $(\mathscr{N}^n,\gamma_{ij})$ as an Einstein manifold, we can replace the Ricci tensor $\hat{R}_{ij}$ by $(n-1)K\gamma_{ij}$, and obtain
\begin{equation}
\label{Lichnerowicz1}
\hat{\Delta}_{L}s_{ij}=-(\hat{\Delta}-2nK)s_{ij}-2\hat{W}_{ikjl}s^{kl}\, ,
\end{equation}
where $\hat{W}_{ikjl}$ is the Weyl tensor of the Einstein manifold $(\mathscr{N}^n,\gamma_{ij})$. Furthermore, in the maximally symmetric space case, one has $\hat{\Delta}_{L}s_{ij}=(-\hat{\Delta}+2nK)s_{ij}$. It should be stressed here that $h^{(2)}_{Tij}$ does not exist when $n=2$, because it is transverse trace free, $h^{(2)}_{Tij}$ must have vanishing degrees of freedom.  This in fact reflects the fact that in four dimensions, there does not exist radial gravitational radiation.

\subsubsection{Vector perturbation}
To get the vector part of the perturbation equations, we consider the perturbations with
\begin{eqnarray}
\label{vector1}
&&h_{ab}=0\, ,\quad h_{ai}=h_{ai}^{(1)}\, ,\quad h_{ij}=\hat{D}_i h^{(1)}_{Tj}+ \hat{D}_j h^{(1)}_{Ti}\, ,\nonumber\\
&&\delta T_{ab}=0\, ,\quad \delta T_{ai}=\delta T_{ai}^{(1)}\, ,\quad \delta T_{ij}=\hat{D}_i \delta T^{(1)}_{Tj}+ \hat{D}_j \delta T^{(1)}_{Ti}\, .
\end{eqnarray}
  In this case, the nontrivial component of the linear perturbation equations is $\delta G_{ai}+\Lambda h_{ai}=\kappa^2 \delta T_{ai}$. After some calculations, this equation can be expressed as
\begin{equation}
\label{V1}
-\frac{1}{r^{n}}D^b\Bigg{\{}r^{n+2}\Big[D_b\Big(\frac{F_{ai}^{(1)}}{r^2}\Big)-D_a\Big(\frac{F_{bi}^{(1)}}{r^2}\Big)\Big]\Bigg{\}}
-\big[\hat{\Delta}+(n-1)K\big]\Big(\frac{F_{ai}^{(1)}}{r^2}\Big)=2\kappa^2 \tau^{(1)}_{ai}\, .
\end{equation}
Another nontrivial component is $\delta G_{ij}+\Lambda h_{ij}=\kappa^2 \delta T_{ij}$. This equation can be written as
\begin{equation}
\label{V2}
\frac{1}{r^{n-2}}D^a\Big{\{}r^{n-2}\Big{[}\hat{D}_i F_{aj}^{(1)}+ \hat{D}_j F_{ai}^{(1)}\Big{]}\Big{\}}=2\kappa^2 \tau_{ij}^{(1)}\, .
\end{equation}
Eqs. (\ref{V1}) and (\ref{V2}) are just the perturbation equations for the vector-type gauge-invariant variables.



\subsubsection{Scalar perturbation}

 Let us consider the scalar perturbation as
\begin{eqnarray}
\label{scalar}
&&h_{ab}=h_{ab}\, ,\quad h_{ai}=\hat{D}_ih_{a}\, ,\quad h_{ij}=h_L\gamma_{ij}+ \hat{L}_{ij}h_T\, ,\nonumber\\
&&\delta T_{ab}=\delta T_{ab}\, ,\quad \delta T_{ai}=\hat{D}_i\delta T_{a}\, ,\quad \delta T_{ij}=\delta T_L\gamma_{ij}+ \hat{L}_{ij}\delta T_T\, .
\end{eqnarray}
After long and tedious calculations, we find that the perturbation equation $\delta G_{ab}+\Lambda h_{ab}=\kappa^2 \delta T_{ab}$ can be expressed as
\begin{eqnarray}
\label{S1}
&&-{}^{m}\!\Box F^{(0)}_{ab}  + {}^m\!R_{a}{}^{c}F^{(0)}_{cb}+  {}^m\!R_{b}{}^{c}F^{(0)}_{ac}-2~{}^m\!R_{acbd}F^{(0)cd}+ D_aD^cF^{(0)}_{cb}+D_bD^cF^{(0)}_{ac}\nonumber\\
&&+n\frac{D^cr}{r}\Big{[}-D_cF^{(0)}_{ab}+D_aF^{(0)}_{cb}+D_bF^{(0)}_{ac}\Big{]} -\frac{1}{r^2}\hat{\Delta}F^{(0)}_{ab}+2\Lambda F^{(0)}_{ab}\nonumber\\
&&-D_aD_bF^{(0)c}_{~~c}-\Big{[}{}^m\!R-2n\frac{{}^m\!\Box r}{r}+n(n-1)\frac{K-(Dr)^2}{r^2}\Big{]}F^{(0)}_{ab}\nonumber\\
&&-n\Big{[}D_aD_b\Big{(}\frac{F^{(0)}}{r^2}\Big{)}+\frac{D_ar}{r}D_b\Big{(}\frac{F^{(0)}}{r^2}\Big{)}+\frac{D_br}{r}D_a\Big{(}\frac{F^{(0)}}{r^2}\Big{)}\Big{]}\nonumber\\
&&-\Big{\{}D_cD_dF^{(0)cd} + 2n \frac{D^cr}{r}D^d F^{(0)}_{cd}-\Big{[}{}^m\!R^{cd}-2n\frac{D^cD^dr}{r}-n(n-1)\frac{D^crD^dr}{r^2}\Big{]}F^{(0)}_{cd}\nonumber\\
&&-{}^{m}\!\Box F^{(0)c}_{~~c}-n\frac{D^cr}{r}D_cF^{(0)d}_{~~d}-n~{}^{m}\!\Box \Big{(}\frac{F^{(0)}}{r^2}\Big{)}-n(n+1)\frac{D^cr}{r}D_c \Big{(}\frac{F^{(0)}}{r^2}\Big{)}\nonumber\\
&&-(n-1)\frac{\hat{\Delta}+nK}{r^2}\Big{(}\frac{F^{(0)}}{r^2}\Big{)}-\frac{1}{r^2}\hat{\Delta}F^{(0)c}_{~~c}\Big{\}}g_{ab}\nonumber\\
&&=2\kappa^2 \Sigma^{(0)}_{ab}\, .
\end{eqnarray}
The perturbation equation $\delta G_{ai}+\Lambda h_{ai}=\kappa^2 \delta T_{ai}$  gives
\begin{eqnarray}
\label{S2}
\hat{D}_i\Bigg{\{}\frac{1}{r^{n-2}}D^b\Big{(}r^{n-2}F^{(0)}_{ab}\Big{)}-r D_a \Big{(}\frac{F^{(0)c}_{~~c}}{r}\Big{)}-(n-1) D_a \Big{(}\frac{F^{(0)}}{r^2}\Big{)}\Bigg{\}}=2\kappa^2 \Sigma_{ai}^{(0)}\, .
\end{eqnarray}
The trace part of the equation $\delta G_{ij} + \Lambda h_{ij}=\kappa^2 \delta T_{ij}$ can be written as
\begin{eqnarray}
\label{S3}
&&-D_aD_b F^{(0)ab} -2(n-1)\frac{D^ar}{r}D^b F^{(0)}_{ab}+{}^{m}\!\Box F^{(0)c}_{~~c}+(n-1)\frac{D^ar}{r}D_aF^{(0)c}_{~~c}\nonumber\\
 &&+ \Big{[}{}^m\!R^{ab}-2(n-1)\frac{D^aD^br}{r}-(n-1)(n-2)\frac{D^arD^br}{r^2}\Big{]}F^{(0)}_{ab}\nonumber\\
 &&+ (n-1)~{}^m\!\Box \Big{(}\frac{F^{(0)}}{r^2}\Big{)}+ n(n-1)\frac{D^ar}{r}D_a\Big{(}\frac{F^{(0)}}{r^2}\Big{)} +\frac{n-1}{n}\frac{1}{r^2}\hat{\Delta}F^{(0)c}_{~~c}\nonumber\\
 &&+ \frac{(n-1)(n-2)}{n }\frac{\hat{\Delta}+nK}{r^2}\Big{(}\frac{F^{(0)}}{r^2}\Big{)}\nonumber\\
 &&=2\kappa^2\frac{\Sigma^{(0)}}{r^2}\, ,
\end{eqnarray}
while the trace free part of the equation $\delta G_{ij} + \Lambda h_{ij}=\kappa^2 \delta T_{ij}$ gives
\begin{equation}
\label{S4}
-\hat{L}_{ij}\Big{[}F^{(0)c}_{~~c} + (n-2)\Big{(}\frac{F^{(0)}}{r^2}\Big{)} \Big{]}=2\kappa^2 \Pi^{(0)}_{ij}\, .
\end{equation}
Eqs. (\ref{S1}), (\ref{S2}), (\ref{S3}) and (\ref{S4}) are the equations of the scalar perturbations. By now, we have obtained all  the
equations for the gauge-invariant variables. These equations are related to each other by the perturbation of Bianchi identity. This will be shown in next subsection.

\subsection{The perturbation of Bianchi identity}
\label{subsec:bianchipertur}

Consider the conservation equation of energy-momentum tensor, i.e.,
$\nabla_{M}T^{M}{}_{N}=0$, which is equivalent to the  Bianchi identity in Einstein gravity theory,
we have
\begin{eqnarray}
\label{eq1}
&&\delta (\nabla_MT^{M}{}_{N})
=-h^{LM}\nabla_{M}T_{LN}+\nabla^M\delta T_{MN}\nonumber\\
&&-g^{ML}\delta C_{ML}{}^{K}T_{KN} -\delta C_{MN}{}^{K}T^{M}{}_{K}=0\, ,
\end{eqnarray}
where
\begin{equation}
\label{tensorCabc}
\delta C_{MN}{}^{L}= \frac{1}{2}g^{LK}\left(\nabla_Mh_{KN}+\nabla_Nh_{MK}-\nabla_Kh_{MN}\right)\, .
\end{equation}
Substituting (\ref{tensorCabc}) into (\ref{eq1}) yields
\begin{eqnarray}
\label{insert}
\delta (\nabla_MT^{M}{}_{N})=\nabla^{M}\delta T_{MN}-\nabla^{M}(T^{L}{}_{N}h_{LM})-\frac{1}{2}T^{ML}\nabla_N h_{ML}+\frac{1}{2} T^{M}{}_{N}(\nabla_M h)\, .
\end{eqnarray}
This equation can also be decomposed according to the warped product of the spacetime. After some calculations, we find
\begin{eqnarray}
\label{eq2}
&&\delta(\nabla_{M}T^{M}{}_{i})= \frac{1}{r^2}\hat{D}^{k}\Pi^{(0)}_{ki}+\frac{1}{r^2}\hat{D}_i\Sigma^{(0)}+ \frac{1}{r^n}D^a\left(r^n\Sigma^{(0)}_{ai}\right)\nonumber\\
&&-\frac{1}{2}T^{ab}\hat{D}_i F^{(0)}_{ab} +\frac{1}{2} P \hat{D}_i F^{(0)c}_{~~c}+\frac{1}{r^2}\hat{D}^{k}\tau^{(1)}_{ki}+\frac{1}{r^n}D^a\left(r^n\tau^{(1)}_{ai}\right)
\, .
\end{eqnarray}
 Note that from Eqs. (\ref{V1}), (\ref{V2}), (\ref{S2}), (\ref{S3}) and (\ref{S4}),  we  have the following two relations
\begin{equation}
\label{Bianchi1}
\frac{1}{r^2}\hat{D}^{k}\tau^{(1)}_{ki}+\frac{1}{r^n}D^a\left(r^n\tau^{(1)}_{ai}\right)=0\, ,
\end{equation}
and
\begin{equation}
\label{Bianchi2}
\frac{1}{r^2}\hat{D}^{k}\Pi^{(0)}_{ki}+\frac{1}{r^2}\hat{D}_i\Sigma^{(0)}+ \frac{1}{r^n}D^a\left(r^n\Sigma^{(0)}_{ai}\right)
-\frac{1}{2}T^{ab}\hat{D}_i F^{(0)}_{ab}+\frac{1}{2} P \hat{D}_i F^{(0)c}_{~~c}=0\, .
\end{equation}
This suggests that the equation (\ref{eq2}) can be split into two parts: one is the scalar part with superscripts $(0)$  and the other vector part with superscripts $(1)$, and $\delta (\nabla_MT^M{}_{i})=0$ is automatically satisfied once those perturbation equations (\ref{V1}), (\ref{V2}), (\ref{S2}), (\ref{S3}) and (\ref{S4}) are satisfied. In addition, we have from (\ref{insert}) that
\begin{eqnarray}
\label{Bianchi3}
&&0=\delta(\nabla_{M}T^{M}{}_{a})= \frac{1}{r^n}D^b\left[r^n\left(\Sigma_{ab}^{(0)}-T_a{}^{c}F^{(0)}_{bc}\right)\right]
-n\frac{D_ar}{r}\Big(\frac{\Sigma^{(0)}}{r^2}\Big)+\frac{1}{r^2}\hat{D}^{i}\Sigma_{ai}^{(0)}\nonumber\\
&&+\frac{1}{2}\left[T_a{}^{b}D_bF^{(0)c}_{~~c}-T^{bc}D_a F^{(0)}_{bc}\right]+\frac{n}{2}\Big[T_a{}^{b}D_b\Big(\frac{F^{(0)}}{r^2}\Big)-PD_a\Big(\frac{F^{(0)}}{r^2}\Big)\Big]\, .
\end{eqnarray}
This scalar equation is also automatically satisfied if equations (\ref{S1}), (\ref{S2}), and (\ref{S3}) hold.


Thus, according to the decomposition theorem of tensor on the Einstein manifold, we have obtained perturbation equations of gauge-invariant variables for Einstein gravity in the $(m+n)$-dimensional spacetime with a warped product metric. It was shown that these equations are related to each other through the perturbation equation of Bianchi identity.  With the gauge-invariant variables in(\ref{giv}) and our perturbation equations presented above, now a complete gauge-invariant perturbation theory can be accomplished.


\section{Spectral Expansion}
\label{sec:Spectral}
One can use the eigen-tensors defined in~\cite{Ishibashi:2011ws} to expand all the perturbation equations above, and get the same results as in~\cite{Ishibashi:2011ws} (and references therein). However, we will here adopt a little bit different expansion. In this spectral expansion, the physical meanings of some special modes are clear, and the gauge-invariant properties of these modes are naturally preserved.

\subsection{Tensor perturbation}

In this case, our analysis is the same as the one in~\cite{Ishibashi:2011ws}. Here we present the corresponding result for completeness.  Consider the eigenvalue problem of the Licherowicz $\hat{\Delta}_L$ on the Einstein manifold $(\mathscr{N}^n,\gamma_{ij})$:
\begin{equation}
(\hat{\Delta}_L -\lambda_L)\mathbb{T}_{ij}=0\, ,
\end{equation}
where the symmetric tensor $\mathbb{T}_{ij}$ satisfies $\gamma^{ij}\mathbb{T}_{ij}=\hat{D}^{i}\mathbb{T}_{ij}=0$. By assuming
\begin{equation}
\Big(\frac{h_{Tij}^{(2)}}{r^2}\Big)=2 H_{T} \mathbb{T}_{ij}\, ,\qquad \Big(\frac{\delta T_{Tij}^{(2)}}{r^2}\Big)- P\Big(\frac{h^{(2)}_{Tij}}{r^2}\Big)=\tau_T \mathbb{T}_{ij}
\end{equation}
with $H_T=H_T(y)$ and $\tau_T=\tau_T(y)$,  one can get the master equation on the expansion coefficients $H_T$ and $\tau_T$:
\begin{equation}
{}^{m}\!\Box H_T + n\frac{D^ar}{r}D_a H_T -\frac{1}{r^2}\left[\lambda_L - 2(n-1)K\right]H_T=-\kappa^2 \tau_T\, .
\end{equation}
 Note that in general the eigenvalue of $\hat{\Delta}_{L}$ is not easy to find out. From Eq.(\ref{Lichnerowicz1}), we have
\begin{equation}
\hat{\Delta}_{L}\mathbb{T}_{ij}=-(\hat{\Delta}-2nK)\mathbb{T}_{ij}-2\hat{W}_{i}{}^k{}_j{}^l\mathbb{T}_{kl}\, ,
\end{equation}
so it might be possible to choose the coordinates $z^i$ such that $\mathbb{T}_{ij}$ is an algebraic eigenvector of the matrix $\hat{W}_{i}{}^k{}_j{}^l$ (which is viewed as a mapping acting on symmetric transverse traceless tensors) to estimate the eigenvalue $\lambda_L$~\cite{Gibbons:2002pq, Gibbons:2002th}.

\subsection{Vector perturbation}

Since only the Laplace-Beltrami operator $\hat{\Delta}$ appears in the equations (\ref{V1}) and (\ref{V2}),  this suggests that the harmonic expansions are enough. Considering the harmonic vector field (one-form) on $(\mathscr{N}^n,\gamma_{ij})$
\begin{equation}
\label{vectorharmonic}
(\hat{\Delta}+k^2)\mathbb{V}_i=0\, ,\qquad \hat{D}^i\mathbb{V}_i=0\, ,
\end{equation}
and the expressions in (\ref{vector1}), we can expand the gauge-invariant variables as
\begin{equation}
\label{vectorexpansion}
F_{ai}^{(1)}= r F_a \mathbb{V}_i\, ,\qquad \tau^{(1)}_{ai}=r\tau_a \mathbb{V}_i\, ,\qquad \tau^{(1)}_{ij}=-2 r^2 \tau_T \hat{D}_{(i}\mathbb{V}_{j)}\, .
\end{equation}
Note that here  we have not used  the vector-type tensor
$\mathbb{V}_{ij}=-(\hat{D}_i\mathbb{V}_j+\hat{D}_j\mathbb{V}_i)/(2 k)$ defined in~\cite{Kodama:2000fa, Ishibashi:2003ap, Kodama:2003jz, Kodama:2003kk, Ishibashi:2011ws}. The expansion of $\tau_{ij}^{(1)}$ can be easily understood from the expression of the $\tau_{ij}^{(1)}$ in (\ref{vector1}) in terms of the perturbation variables $h_{Ti}^{(1)}$ and $\delta T_{Ti}^{(1)}$:
\begin{equation}
h_{Ti}^{(1)}=H_T\mathbb{V}_{i}\, ,\qquad \delta T_{Ti}^{(1)}=\delta T_T \mathbb{V}_i\, .
\end{equation}
Note that from (\ref{giv}), we can have
\begin{equation}
\label{tautottht}
r^2 \tau_T=\delta T_T-PH_T\, .
\end{equation}
Substituting the expansions (\ref{vectorexpansion}) into the vector perturbation equation (\ref{V1}), we have
\begin{equation}
\label{vectormodes1}
\frac{1}{r^{n+1}} D^{b}\Bigg{\{}r^{n+2}\Big[D_b\Big(\frac{F_a}{r}\Big)-D_a\Big(\frac{F_b}{r}\Big)\Big]\Bigg{\}}-\frac{1}{r^2}\big[k^2-(n-1)K\big]F_a =-2 \kappa^2 \tau_a\, .
\end{equation}
Similarly,  Eq. (\ref{V2}) and Eq.(\ref{Bianchi1}) can be changed into
\begin{equation}
\label{vectormodes2}
\frac{1}{r^n} D^a\big(r^{n-1}F_a\big) \hat{D}_{(i}\mathbb{V}_{j)}=-\kappa^2 \tau_T \hat{D}_{(i}\mathbb{V}_{j)}\, ,
\end{equation}
and
\begin{equation}
\label{vectormodes3}
[k^2-(n-1)K]\tau_T + \frac{1}{r^{n}}D^a(r^{n+1}\tau_a)=0\, .
\end{equation}
It can be seen clearly that the above three equations are valid for any modes, even for the $k=0$ mode. However, for the modes with $\hat{D}_{(i}\mathbb{V}_{j)}= 0$ , Eq.(\ref{vectormodes2}) is trivially satisfied. These modes are called exceptional modes, and the remains are generic modes~\cite{Ishibashi:2011ws}. They will be discussed separately.

\subsubsection{generic modes}
 let us first  consider the generic modes, i.e. $\hat{D}_{(i}\mathbb{V}_{j)}\ne 0$, such that  $\hat{D}_{(i}\mathbb{V}_{j)}$ can be eliminated from Eq.(\ref{vectormodes2}). This means $$\hat{D}^i(\hat{D}_i\mathbb{V}_j+\hat{D}_j\mathbb{V}_i)=[-k^2+(n-1)K]\mathbb{V}_j\ne 0\, .$$
Therefore when $k^2-(n-1)K\ne 0$, we have three  equations: Eq. (\ref{vectormodes1}), Eq.(\ref{vectormodes3}),  and
\begin{equation}
\label{vectormodes4}
\frac{1}{r^n} D^a\big(r^{n-1}F_a\big) = -\kappa^2 \tau_T\, .
\end{equation}
These three equations are not independent. Actually, by considering the differential of Eq.(\ref{vectormodes1}) and Eq.(\ref{vectormodes3}), we can obtain  Eq.(\ref{vectormodes4}).  Eq.(\ref{vectormodes3}) just means that $\tau_T$ is totally determined by $\tau_a$.
Thus, for each such kind of modes, we have $m$ equations for $m$ independent variables coming from $F_a$ if $\tau_a$ is regarded as a known source term.

\subsubsection{exceptional modes}
For the special modes with $\hat{D}_{(i}\mathbb{V}_{j)}= 0$, Eq.(\ref{vectormodes2}) is trivially satisfied and does not give any constraint between $F_a$ and $\tau_T$.
From the perturbation equation of the Bianchi identity, i.e., Eq.(\ref{vectormodes3}), we have
\begin{equation}
\label{vectormodes5}
D^a(r^{n+1} \tau_a)=0\, ,
\end{equation}
and Eq.(\ref{vectormodes1}) becomes
\begin{equation}
\label{vectormodes6}
 D^{b}\Bigg{\{}r^{n+2}\Big[D_b\Big(\frac{F_a}{r}\Big)-D_a\Big(\frac{F_b}{r}\Big)\Big]\Bigg{\}} =-2 \kappa^2 r^{n+1} \tau_a\, .
\end{equation}
These two  equations are also not independent. By considering the divergence of Eq.(\ref{vectormodes6}), one can obtain Eq.(\ref{vectormodes5}). Then, we still have $m$ equations for $m$ variables (from $F_a$).  In this case, the source term has to satisfy Eq.(\ref{vectormodes5}) and $\tau_a$ has nothing to do with $\tau_T$.

Note that in the Kodama-Ishibashi formalism, for these exceptional modes,  $h_{Ti}^{(1)}$, $\delta T_{Ti}^{(1)}$ and $\tau_T$ are not well defined~\cite{Ishibashi:2011ws}. Instead, in our expansion, these perturbation variables are well defined and are of obviously  physically meaningful:  $h_{Ti}^{(1)}$ and $\delta T_{Ti}^{(1)}$ correspond to Killing vectors of $(\mathscr{N}^n,\gamma_{ij})$ for the exceptional modes.  Further, $\tau_T$ is no longer arbitrary and instead  it is determined by the combination of the coefficients of $h_{Ti}^{(1)}$ and $\delta T_{Ti}^{(1)}$ in Eq.(\ref{tautottht}).  In addition, let us mention that $\tau_T$  does not enter into the vector perturbation equations, and this is quite different from the case with generic modes.

\subsection{Scalar perturbation}

In this case, as the case of vector perturbation, only the Laplace-Beltrami operator appears in the scalar perturbation equations (\ref{S1}), (\ref{S2}), (\ref{S3}) and (\ref{S4}), we therefore need only to consider the harmonic function on $(\mathscr{N}^n, \gamma_{ij})$:
\begin{equation}
(\hat{\Delta}+k^2)\mathbb{S}=0\, .
\end{equation}
From the forms in (\ref{scalar}), it is natural to expand the scalar-type gauge-invariant  variables in (\ref{giv}) as
\begin{eqnarray}
&&F^{(0)}_{ab}=F_{ab}\mathbb{S}\, ,~\qquad F^{(0)}=2 r^2 F \mathbb{S}\, ,\nonumber\\
&&\Sigma^{(0)}_{ab}=\Sigma_{ab}\mathbb{S}\, ,\qquad ~\Sigma^{(0)}=r^2\Sigma\mathbb{S}\, ,\nonumber\\
&&\Sigma^{(0)}_{ai}=r \Sigma_a \hat{D}_{i}\mathbb{S}\, ,\quad \Pi^{(0)}_{ij}= r^2 \tau_T \hat{L}_{ij}\mathbb{S}\, .
\end{eqnarray}
Note that these  expansions have some differences from those in \cite{Kodama:2000fa, Ishibashi:2003ap, Kodama:2003jz, Kodama:2003kk, Ishibashi:2011ws}, here we have not used the scalar-type tensors as in~\cite{Kodama:2000fa, Ishibashi:2003ap, Kodama:2003jz, Kodama:2003kk, Ishibashi:2011ws}.
Substituting these expansions into the scalar equation (\ref{S1}), we have
\begin{eqnarray}
\label{scalarmodes1}
&&-{}^{m}\!\Box F_{ab}  + {}^m\!R_{a}{}^{c}F_{cb}+  {}^m\!R_{b}{}^{c}F_{ac}-2~{}^m\!R_{acbd}F^{cd}+ D_aD^cF_{cb}+D_bD^cF_{ac}\nonumber\\
&&+n\frac{D^cr}{r}\Big{[}-D_cF_{ab}+D_aF_{cb}+D_bF_{ac}\Big{]} +\frac{k^2}{r^2}F_{ab}+2\Lambda F_{ab}\nonumber\\
&&-\Big{[}{}^m\!R-2n\frac{{}^m\!\Box r}{r}+n(n-1)\frac{K-(Dr)^2}{r^2}\Big{]}F_{ab}-D_aD_bF^{~c}_{c}\nonumber\\
&&-2n\Big{[}D_aD_bF+\frac{D_ar}{r}D_bF+\frac{D_br}{r}D_aF\Big{]}\nonumber\\
&&-\Big{\{}D_cD_dF^{cd} + 2n \frac{D^cr}{r}D^d F_{cd}-\Big{[}{}^m\!R^{cd}-2n\frac{D^cD^dr}{r}-n(n-1)\frac{D^crD^dr}{r^2}\Big{]}F_{cd}\nonumber\\
&&-{}^{m}\!\Box F^{~c}_{c}-n\frac{D^cr}{r}D_cF^{~d}_{d}-2n~{}^{m}\!\Box F-2n(n+1)\frac{D^cr}{r}D_c F\nonumber\\
&&+2(n-1)\frac{k^2-nK}{r^2}F+\frac{k^2}{r^2}F^{~c}_{c}\Big{\}}g_{ab}\nonumber\\
&&=2\kappa^2 \Sigma_{ab}\, .
\end{eqnarray}
This is exactly the same as in~\cite{Kodama:2000fa}. The second equation (\ref{S2}) of scalar perturbation is transformed to
\begin{eqnarray}
\label{scalarmodes2}
\Bigg{\{}\frac{1}{r^{n-2}}D^b\Big{(}r^{n-2}F_{ab}\Big{)}-r D_a \Big{(}\frac{F^{~c}_{c}}{r}\Big{)}-2(n-1) D_aF \Bigg{\}}(\hat{D}_i\mathbb{S})=2\kappa^2 r\Sigma_{a}(\hat{D}_i\mathbb{S})\, .
\end{eqnarray}
For the zero mode, i.e., $k=0$, the harmonic function $\mathbb{S}$ must be a constant because the Laplace-Beltrami is nonnegative. In this case, the above equation is trivially satisfied.
The third equation (\ref{S3}) becomes
\begin{eqnarray}
\label{scalarmodes3}
&&-D_aD_b F^{ab} -2(n-1)\frac{D^ar}{r}D^b F_{ab}+{}^{m}\!\Box F^{~c}_{c}+(n-1)\frac{D^ar}{r}D_aF^{~c}_{c}\nonumber\\
 &&+ \Big{[}{}^m\!R^{ab}-2(n-1)\frac{D^aD^br}{r}-(n-1)(n-2)\frac{D^arD^br}{r^2}\Big{]}F_{ab}\nonumber\\
 &&+ 2(n-1)~{}^m\!\Box F+ 2n(n-1)\frac{D^ar}{r}D_a F-\frac{n-1}{n}\frac{k^2}{r^2}F^{~c}_{c}\nonumber\\
 &&-2 \frac{(n-1)(n-2)}{n}\frac{(k^2-nK)}{r^2}F\nonumber\\
 &&=2\kappa^2\Sigma\, .
\end{eqnarray}
This expression is the same as the one in~\cite{Kodama:2000fa}. Finally, the last  equation (\ref{S4}) for the scalar perturbation can be reduced to
\begin{equation}
\label{scalarmodes4}
-\Big{[}F^{~c}_{c} + 2(n-2)F \Big{]}(\hat{L}_{ij}\mathbb{S})=2\kappa^2 \tau_T (\hat{L}_{ij}\mathbb{S}) \, .
\end{equation}
Once again, for the zero mode, this equation is trivially satisfied. Further, a simple investigation
\begin{equation}
\hat{D}^j\hat{D}^i\hat{L}_{ij}\mathbb{S}=\frac{n-1}{n}k^2(k^2-nK)\mathbb{S}
\end{equation}
suggests that this equation is also trivially satisfied when $k^2=nK$. To see this, we notice the following identity
\begin{equation}
\label{LijSLijS}
(\hat{L}^{ij}\mathbb{S})(\hat{L}_{ij}\mathbb{S})=
\hat{D}_{i}(\hat{D}_j\mathbb{S}\hat{D}^j\hat{D}^i\mathbb{S})
-\hat{D}_i(\mathbb{S}\hat{\Delta}\hat{D}^i\mathbb{S})-\frac{1}{n}\hat{D}_i(\hat{D}^i\mathbb{S}\hat{\Delta}\mathbb{S})
+\mathbb{S}\hat{D}^j\hat{D}^i\hat{L}_{ij}\mathbb{S}\, .
\end{equation}
This implies that $\hat{L}_{ij}\mathbb{S}$ will be vanishing once $\hat{D}^j\hat{D}^i\hat{L}_{ij}\mathbb{S}=0$. This can be obtained by integrating the both sides of above identity on $\mathscr{N}^n$, with the assumption that $\mathscr{N}^n$ is closed or the function $\mathbb{S}$ is of required asymptotic behaviors.

On the other hand, for the perturbation equations of the energy-momentum conservation equations (\ref{Bianchi2}) and (\ref{Bianchi3}), we have
\begin{equation}
\label{Bianchimodes2}
\Bigg{[}\frac{1}{r^n}D^a\left(r^{n+1}\Sigma_{a}\right)-\frac{n-1}{n}(k^2-nK)\tau_T+\Sigma
-\frac{1}{2}T^{ab}F_{ab}+\frac{1}{2} P  F^{~c}_{c}\Bigg{]}(\hat{D}_i\mathbb{S})=0\, ,
\end{equation}
and
\begin{eqnarray}
\label{Bianchimodes3}
&&\frac{1}{r^n}D^b\left[r^n\left(\Sigma_{ab}-T_a^{~c}F_{bc}\right)\right]
-n\frac{D_ar}{r}\Sigma-\frac{k^2}{r}\Sigma_{a}+\frac{1}{2}\left(T_a^{~b}D_bF^{~c}_{c}-T^{bc}D_a F_{bc}\right)\nonumber\\
&&+n\left(T_a^{~b}D_bF-PD_aF\right)=0\, .
\end{eqnarray}
 These two  equations give the relations among the coefficients of the source terms, i.e., $\Sigma_{ab}$, $\Sigma_a$, $\Sigma$ and $\tau_T$, and at the same time reveal the relations among the mode equations for the scalar perturbations.  Here, we have corrected a typo in~\cite{Kodama:2000fa}, the factor $``n"$ in the last two terms of the
  equation (\ref{Bianchimodes3}) has been missed in ~\cite{Kodama:2000fa}.

\subsubsection{generic modes}

In the case $k^2(k^2-nK)\ne 0$, $\hat{D}_i \mathbb{S}$ and $\hat{L}_{ij}\mathbb{S}$ are both non-vanishing, and they can be removed from both sides of Eqs.(\ref{scalarmodes2}) and (\ref{scalarmodes4}).  Thus for generic modes, we have four equations, i.e., Eqs.(\ref{scalarmodes1}), (\ref{scalarmodes3}) and  following two:
\begin{eqnarray}
\label{genericscalar1}
\frac{1}{r^{n-2}}D^b\Big{(}r^{n-2}F_{ab}\Big{)}-r D_a \Big{(}\frac{F^{~c}_{c}}{r}\Big{)}-2(n-1) D_aF =2\kappa^2 r\Sigma_{a}\, ,
\end{eqnarray}
\begin{equation}
\label{genericscalar2}
F^{~c}_{c} + 2(n-2)F =-2\kappa^2r^2 \tau_T\, .
\end{equation}
In addition, we have from  (\ref{Bianchimodes2})
\begin{equation}
-\frac{n-1}{n}(k^2-nK)\tau_T+\Sigma+ \frac{1}{r^n}D^a\left(r^{n+1}\Sigma_{a}\right)
-\frac{1}{2}T^{ab}F_{ab}+\frac{1}{2} P  F^{~c}_{c}=0.
\end{equation}
Eq.(\ref{Bianchimodes3}) together with this equation provides $m+1$ constraints on the coefficients of sources $\Sigma_{ab}$, $\Sigma_a$, $\Sigma$ and $\tau_T$.
Thus one has $m(m+1)/2 +1$ gauge-invariant variables of scalar perturbations and  the independent equations of perturbations with the same numbers of
gauge-invariant variables . Therefore, in principle,  this system can
be solved once the source terms are given.

\subsubsection{zero modes}

In the case of $k=0$,  $\hat{D}_i \mathbb{S}$ and $\hat{L}_{ij}\mathbb{S}$ are both vanishing. In this case,  Eqs.(\ref{scalarmodes2}) and (\ref{scalarmodes4})
are trivially satisfied.  Thus we  have only two equations of scalar perturbations, i.e., Eqs.(\ref{scalarmodes1}) and (\ref{scalarmodes3}). For the perturbations of the Bianchi identity,  Eq.~(\ref{Bianchimodes2}) becomes trivial, while  Eq.~(\ref{Bianchimodes3}) remains nontrivial.

 Note that in this case $\tau_T$ and $\Sigma_a$ do not appear in Eqs.(\ref{Bianchimodes3}), (\ref{scalarmodes1}) and (\ref{scalarmodes3}). Hence they are totally free and have no contribution to the dynamics of zero modes.


\subsubsection{exceptional modes}

When $k^2=nK$, $\hat{L}_{ij}\mathbb{S}$ is vanishing,  Eq.(\ref{scalarmodes4}) therefore is trivially satisfied. We have three equations of perturbations, namely, Eqs.(\ref{scalarmodes1}), (\ref{scalarmodes2}) and (\ref{scalarmodes3}). The perturbation equations of the Bianchi identity (\ref{Bianchimodes2}) and (\ref{Bianchimodes3}) are both nontrivial.  In this case, it is easy to see that $\tau_T$ is totally free and has no contribution to the dynamics of these modes.

 Note that as in the vector perturbation case,  for some exceptional modes and zero mode of scalar perturbations, some variables are not well defined in the  Kodama-Ishibashi formalism, instead these cases do not appear here in our expansion.

\section{The case of $m=2$}
\label{sec:cdim2}

When $m=2$,  we find that the perturbation equations of gauge-invariant variables can be simplified. For the vector perturbations, we can write down
the master equation of the perturbations, while for the scalar perturbations, we can also obtain the master equation of perturbation in the case
without source. In this section we show these results.

\subsection{Vector Perturbation}

 From the vector perturbation equation (\ref{V1}), we can obtain
\begin{equation}
D^a\Big{\{}r^{n+2}\Big[D_b\Big(\frac{\epsilon^{bc}F_{ci}^{(1)}}{r^2}\Big)\Big]\Big{\}}
+\big[\hat{\Delta}+(n-1)K\big]r^n\Big(\frac{\epsilon^{ab}F_{bi}^{(1)}}{r^2}\Big)=-2\kappa^2 r^n\epsilon^{ab}\tau^{(1)}_{bi}\, ,
\end{equation}
where $\epsilon_{ab}$ is the Levi-Civita tensor on $(\mathscr{M}^2, g_{ab})$.
The first term in the left side of above equation is an exact 1-form in $(\mathscr{M}^2, g_{ab})$. Thus we can introduce a function
$\Omega_i$ so that
\begin{equation}
\big[\hat{\Delta}+(n-1)K\big]r^n\Big(\frac{\epsilon^{ab}F_{bi}^{(1)}}{r^2}\Big)+2\kappa^2 r^n\epsilon^{ab}\tau^{(1)}_{bi}=D^a\Omega_i\, ,
\end{equation}
where $\Omega_i$ is a scalar on the two dimensional space $(\mathscr{M}^2, g_{ab})$, but it is a vector field on the Einstein manifold $(\mathscr{N}^n, \gamma_{ij})$, and satisfies $\hat{D}^i\Omega_{i}=0$. Then, we have
\begin{equation}
D^a\Big{\{}r^{n+2}\Big[D_b\Big(\frac{\epsilon^{bc}F_{ci}^{(1)}}{r^2}\Big)\Big]\Big{\}}
+D^a\Omega_i=0\, .
\end{equation}
By acting operator $\big[\hat{\Delta}+(n-1)K\big]$ on both sides of the above equation, we get
\begin{equation}
D^a\Big{\{}r^{n+2}\Big[D_b\Big(\frac{D^b\Omega_i}{r^n}-2\kappa^2 \epsilon^{bc}\tau^{(1)}_{ci}\Big)\Big]
+\big[\hat{\Delta}+(n-1)K\big]\Omega_i\Big{\}}=0\, ,
\end{equation}
or
\begin{equation}
r^{n+2}\Big[D_b\Big(\frac{D^b\Omega_i}{r^n}-2\kappa^2 \epsilon^{bc}\tau^{(1)}_{ci}\Big)\Big]
+\big[\hat{\Delta}+(n-1)K\big]\Omega_i=C\, .
\end{equation}
By redefining $\Omega_i$, we can always set $C=0$. In this way,  we can obtain the master equation of vector perturbation
\begin{equation}
\label{insert1}
{}^2\!\Box \Omega_i -n\frac{D^ar}{r}D_a\Omega_i
+\frac{1}{r^2}\big[\hat{\Delta}+(n-1)K\big]\Omega_i=\kappa^2 r^n\epsilon^{bc}\big(D_b\tau^{(1)}_{ci}-D_c\tau^{(1)}_{bi}\big)\, .
\end{equation}
We can expand $\Omega_i$ and $\tau^{(1)}_{ai}$ as $\Omega_i=\Omega\mathbb{V}_i$ and $\tau^{(1)}_{ai}=r\tau_a \mathbb{V}_i$ by using the harmonic vectors in Eq.(\ref{vectorharmonic}), and can obtain the equations of $\Omega$ for each mode.   This equation (\ref{insert1}) is our main result of this subsection.



\subsection{Scalar Perturbation}
\label{subsec:scalarmaster}

In this case, for simplicity, we consider the case without matter fields.
To simplify the expressions of equations, we first define
\begin{equation}
X=\frac{F^{(0)}}{r^2}\, ,\qquad F^{(0)c}_{~~c}=W\, ,\qquad J= W + (n-2)X\, ,\qquad Z_{ab}= F^{(0)}_{ab}-\frac{1}{2}F^{(0)c}_{~~c}g_{ab}\, .
\end{equation}
The traceless part of the perturbation equation (\ref{S1}) then becomes
\begin{eqnarray}
\label{S11}
&&-{}^{2}\!\Box Z_{ab} + D_aD^cZ_{cb}+D_bD^cZ_{ac}-D_cD_dZ^{cd}g_{ab}-\frac{1}{r^2}\hat{\Delta}Z_{ab}+2~{}^2\!R Z_{ab}\nonumber\\
&&+n\frac{D^cr}{r}\Big{(}-D_cZ_{ab}+D_aZ_{cb}+D_bZ_{ac}-D^d Z_{cd}g_{ab}\Big{)}+(2\Lambda - R) Z_{ab} \nonumber\\
&&+\frac{n}{2}\Big{(}\frac{D_b r}{r}D_aJ+\frac{D_ar}{r}D_bJ-\frac{D^c r}{r}D_cJg_{ab}\Big{)}-n\Big(D_{a}D_bX-\frac{1}{2}{}^{2}\!\Box Xg_{ab}\Big)\nonumber\\
&&-\frac{1}{2}n^2\Big{(}\frac{D_ar}{r}D_bX
+\frac{D_br}{r}D_aX-\frac{D^cr}{r}D_c X g_{ab}\Big{)}\nonumber\\
&&=2\kappa^2\Big[ \Sigma^{(0)}_{ab}-\frac{1}{2}\Sigma^{(0)c}_{~~c}g_{ab}\Big]\, ,
\end{eqnarray}
while the trace part of (\ref{S1}) is given by
\begin{eqnarray}
\label{traceS1}
&&+n({}^{2}\!\Box X) + 2n^2 \frac{D^cr}{r}D_cX + \frac{n}{r^2}\big[\hat{\Delta}+n(n-1)K\big]X-2(n-2)\Lambda X\nonumber\\
&&-2n \frac{D^cr}{r}D^dZ_{cd} - 4n\frac{D^cD^dr}{r}Z_{cd} - 2n(n-1)\frac{D^crD^dr}{r^2}Z_{cd}\nonumber\\
&&+\frac{1}{r^2}\big[\hat{\Delta}-n(n-1)K\big]J + 2\Lambda J\nonumber\\
&&=2\kappa^2\Sigma^{(0)c}_{~~c}\, .
\end{eqnarray}
It is found that  the difference between Eq. (\ref{traceS1}) and Eq.(\ref{S3}) is given by
\begin{eqnarray}
&&n(n-1)\frac{D^cr}{r}D_cX + \frac{(n-1)}{r^2}\big[\hat{\Delta}+\frac{1}{2}n^2 K \big]X+\frac{1}{2}(n-2)({}^{2}\!\Box X) \nonumber\\
&&+D^cD^dZ_{cd} - 2(n-1)\frac{D^cD^dr}{r}Z_{cd}-n(n-1)\frac{D^crD^dr}{r^2}Z_{cd}\nonumber\\
&&-(n-2)\Big(\frac{n-2}{n}\Lambda +\kappa^2 P \Big)X -\frac{1}{2}({}^{2}\!\Box J)\nonumber\\
&&+\Big[\frac{n-2}{n}\Lambda +\kappa^2 P -\frac{1}{2}n(n-1)\frac{K}{r^2}\Big]J\nonumber\\
&&=2\kappa^2\Big[\frac{n-1}{n} \Sigma^{(0)c}_{~~c}-\frac{\Sigma^{(0)}}{r^2}\Big]\, .
\end{eqnarray}
In this case,  Eq.(\ref{S2}) becomes
\begin{eqnarray}
\label{S21}
\hat{D}_i\Bigg{\{}\frac{1}{r^{n-2}}D^b\Big{[}r^{n-2}\Big(Z_{ab}-\frac{1}{2}Xg_{ab}\Big)\Big{]}-\frac{1}{2}D_aJ+\frac{n}{2}\frac{D_ar}{r}J\Bigg{\}}=2\kappa^2 \Sigma_{ai}^{(0)}\, .
\end{eqnarray}
And Eq.(\ref{S4}) is changed to
 $$\hat{L}_{ij}J=-2\kappa^2\Pi^{(0)}_{ij}\, .$$
 Note that when the matter fields are absent, we have
\begin{eqnarray}
&&~~R=2\Big(1+\frac{2}{n}\Big)\Lambda \, ,\qquad R_{ab}=\frac{2}{n}\Lambda g_{ab}\, ,\nonumber\\
&&{}^{2}\!R=\frac{4}{n}\Lambda + n\frac{{}^{2}\!\Box r}{r}
\, ,\qquad D_aD_br=\frac{1}{2}{}^{2}\!\Box rg_{ab}\, .
\end{eqnarray}
The last one in the above equations implies that $D^aD^brZ_{ab}=0$ and  we can have a Killing vector field $$\xi^a=\epsilon^{ab}D_br$$  in
$(\mathscr{M}^2, g_{ab})$.  This can be easily checked by the following equation:
\begin{equation}
D_a\xi_b+D_b\xi_a=\epsilon_{bc}D_aD^cr+\epsilon_{ac}D_bD^cr=0\, .
\end{equation}
In fact, $\xi^a$ is just the so-called Kodama vector proposed in~\cite{Kodama:1979vn}. When the source terms are absent, we have  $\hat{L}_{ij}J=0$, and then  the equation (\ref{S21}) can be written as
\begin{equation}
\label{s211}
D^bZ_{ab}+(n-2)\frac{D^br}{r}Z_{ab}=\frac{1}{2}n D_aX +\frac{1}{2}n(n-2) \frac{D_ar}{r} X\, .
\end{equation}
Note that, here and after, the action of $\hat{L}_{ij}$ on both sides of some equations is assumed so that the terms including $``J"$ can be
dropped out.  Using (\ref{s211}) and considering $$\hat{L}_{ij}\hat{\Delta}J=-\hat{\Delta}_L\hat{L}_{ij}J\, ,$$  we see that
Eq.(\ref{traceS1}) can be changed to
\begin{eqnarray}
\label{procedure1}
&&  2\frac{D^crD^dr}{r^2}Z_{cd}={}^{2}\!\Box X + n \frac{D^cr}{r}D_cX + \frac{1}{r^2}\big(\hat{\Delta}+nK\big)X\nonumber\\
&&-\Big[2\Big(1-\frac{2}{n}\Big)\Lambda -n(n-2)\frac{K-(Dr)^2}{r^2}\Big]X\, .
\end{eqnarray}
By defining $\sigma=r^n X$, this equation becomes
\begin{equation}
\label{Xiequation}
E[\sigma]+U_{\sigma}\cdot\sigma=-2 r^{n}W_{ab}\Big(\frac{D^aD^br}{r}-\frac{D^arD^br}{r^2}\Big)\, .
\end{equation}
where $W_{ab}=Z_{ab}-(1/2)nXg_{ab}$ and the operator $E[\phi]$ is defined by
\begin{equation}
E[\phi]={}^{2}\!\Box \phi - n \frac{D^cr}{r}D_c\phi + \frac{\hat{\Delta}+nK}{r^2}\phi\, ,
\end{equation}
and  the potential $U_{\sigma}$ is given by
\begin{equation}
U_\sigma=-\Big[2\Big(1-\frac{2}{n}\Big)\Lambda -n(n-2)\frac{K-(Dr)^2}{r^2}\Big]\, .
\end{equation}
By use of $W_{ab}$, Eq.(\ref{S11}) can be simplified to
\begin{eqnarray}
\label{simplifiedS1}
&&D^c\Big[r^n\big(-D_c W_{ab} + D_aW_{cb}+D_bW_{ac}-D^dW_{cd}g_{ab}\big)\Big]-\frac{1}{r^2}\hat{\Delta}(r^n W_{ab})\nonumber\\
&&+(2\Lambda - R) (r^n W_{ab})-n (r^nW_{cd})\Big(\frac{D^cD^dr}{r}+\frac{D^crD^dr}{r^2}\Big)g_{ab}=0\, .
\end{eqnarray}
Since $\xi^a$ is a Killing vector field on $(\mathscr{M}^2,g_{ab})$, we have
\begin{equation}
D^a(r^{n-2}W_{ab}\xi^b)=r^{n-2}W_{ab}D^a\xi^b=0\, .
\end{equation}
This implies  that one can define a scalar $\tau$ on $(\mathscr{M}^2,g_{ab})$ as
\begin{equation}
\label{tau}
r^{n-2}\epsilon_{a}^{~c}W_{cb}\xi^b=-D_a\tau\, .
\end{equation}
Considering $\epsilon_{ab}\epsilon_{cd}=g_{ad}g_{bc}-g_{ac}g_{bd}$, we have
\begin{equation}
\label{taudefinition}
r^{n-2}(W_{ab}D^br-W^c_{c}D_ar)=D_a\tau\, .
\end{equation}
By these relations, it is easy to find
\begin{eqnarray}
r^{n-2}W_{ab}\xi^a\xi^b=D^arD_a\tau\, ,
\end{eqnarray}
\begin{eqnarray}
\label{eq524}
r^{n-2}W_{ab}D^ar\xi^b=\xi^aD_a\tau\, ,
\end{eqnarray}
\begin{eqnarray}
r^{n-2}W_{ab}D^arD^br=D^arD_a\tau+r^{n-2}W^c_c (Dr)^2\, .
\end{eqnarray}
Now, we can  express  $W_{ab}$  in terms of  $\tau$. For instance, the right hand of Eq.(\ref{Xiequation}) can be rewritten as
\begin{equation}
V[\sigma,\tau]=n\Big[\frac{{}^{2}\!\Box r}{r}-2 \frac{(Dr)^2}{r^2}\Big]\sigma+2D^crD_c\tau\, .
\end{equation}
And Eq.(\ref{simplifiedS1}) can be expressed as
\begin{equation}
\label{xitau}
E[\Psi]+U_{\tau}\cdot \Psi=V\left[\Phi,~\frac{\Psi}{r}\right]\, ,
\end{equation}
where $\Psi=r\xi^cD_c\tau$, $\Phi=\xi^cD_c\sigma$ and
\begin{equation}
U_{\tau}=-\Big[2\Lambda+(n+1)\frac{{}^{2}\!\Box r}{r} -n(n-2)\frac{K-(Dr)^2}{r^2}\Big]\, .
\end{equation}
Furthermore, taking use of the Killing vector $\xi$, Eq.~(\ref{Xiequation}) can be
changed to
\begin{equation}
\label{xisigma}
E[\Phi]+U_{\sigma}\cdot \Phi=V\left[\Phi,~\frac{\Psi}{r}\right]\, .
\end{equation}
Thus, we have obtained two equations of scalar perturbation (\ref{xitau}) and (\ref{xisigma}). The difference between Eq.(\ref{xitau}) and Eq.(\ref{xisigma}) is given by
\begin{equation}
\label{eq530}
E[\Phi-\Psi]+U_{\sigma}\cdot\Phi-U_{\tau}\cdot\Psi=0\, .
\end{equation}
Note that in general, $U_{\sigma}\neq U_{\tau}$. Thus we can not  naively rewrite the above equation as an equation of a single variable $(\Phi-\Psi)$.  However, after some
manipulations, we can arrive at our goal. To see this, let us notice that from  (\ref{s211}) and (\ref{eq524}), we can have
\begin{equation}
{}^{2}\!\Box \Psi-2\frac{D^ar}{r}D_a\Psi -\frac{{}^{2}\!\Box r}{r}\Psi +2\frac{(Dr)^2}{r^2}\Psi
=n\Big[\frac{1}{2}\frac{{}^{2}\!\Box r}{r}-2 \frac{(Dr)^2}{r^2}\Big]\Phi+n\frac{D^ar}{r}D_a\Phi\, .
\end{equation}
The difference between this equation and Eq.(\ref{xitau}) gives
\begin{equation}
\label{eq532}
\Big(-\frac{\hat{\Delta} + nK}{r^2} + w \Big)\Phi = \Big(-\frac{\hat{\Delta} + nK}{r^2}
+ \omega -\frac{n}{2}\frac{{}^{2}\!\Box r}{r}\Big)(\Phi-\Psi)+n\frac{D^ar}{r}D_a(\Phi-\Psi)\, ,
\end{equation}
where
\begin{equation}
\omega=\frac{n}{2(n-1)}\Big[\frac{4\Lambda}{n} + (n+1)\frac{{}^{2}\!\Box r}{r}\Big]=\frac{n}{2(n-1)}(U_{\sigma}-U_{\tau})\, .
\end{equation}
Substituting (\ref{eq532}) into (\ref{eq530}), we arrive at
\begin{eqnarray}
\label{Xiequation1}
&&\Big(-\frac{\hat{\Delta} + nK}{r^2} + \omega \Big)E[\Xi] + 2(n-1)(-(\hat{\Delta} + nK) + r^2\omega)\frac{D^ar}{r}D_a\Xi+\Big(-\frac{\hat{\Delta} + nK}{r^2} + \omega \Big)U_{\tau}\cdot\Xi\nonumber\\
&&+\frac{2(n-1)}{n}\omega\Big(-\frac{\hat{\Delta} + nK}{r^2}
+ \omega -\frac{n}{2}\frac{{}^{2}\!\Box r}{r}\Big)\Xi=0\, ,
\end{eqnarray}
where $\Xi=\Phi-\Psi$. This is a second order PDE of $\Xi$ in the orbit space $(\mathscr{M}^2, g_{ab})$.  To further simplify this equation, let us define
 a scalar $\Omega$ as
\begin{equation}
\Xi=\Big[-(\hat{\Delta} + nK) + r^2\omega  \Big]\Omega \equiv H\Omega\, .
\end{equation}
Thus Eq.~(\ref{Xiequation1}) can be rewritten as
\begin{eqnarray}
&&HE[H\Omega] + 2(n-1)rHD^crD_c(H\Omega)+U_{\tau}H^2\Omega\nonumber\\
&&+\frac{2(n-1)}{n}\omega\Big[H -\frac{n}{2}r({}^{2}\!\Box r)\Big]H\Omega=0\, .
\end{eqnarray}
This equation can be further changed to
\begin{eqnarray}
H^2E[\Omega] +  U_{\sigma}H^2\Omega+2(n-1)\Big[n\omega(Dr)^2H-\omega r({}^{2}\!\Box r)H-(n-1)\omega^2r^2(Dr)^2\Big]\Omega =0\, .
\end{eqnarray}
Finally we obtain the master equation of scalar perturbation
\begin{equation}
\label{scalarsingle}
E[\Omega] + U_S\Omega=0\, ,
\end{equation}
where we have assumed that the operator $H$ is invertible, and the effective potential is given by
\begin{equation}
U_S= U_{\sigma}+2(n-1)\Big{\{}\omega r^2\Big[n\frac{(Dr)^2}{r^2}-\frac{{}^{2}\!\Box r}{r} \Big]H-(n-1)\omega^2r^4\frac{(Dr)^2}{r^2}\Big{\}}H^{-2}\, .
\end{equation}
Some remarks on the master equation are in order.

(1). Equation (\ref{scalarsingle}) is a wave equation on $(\mathscr{M}^2, g_{ab})$ for the master variable $\Omega$. We can expand $\Omega$ by using the harmonic functions on $(\mathscr{N}^n, \gamma_{ij})$, and get the mode equations.  Similar mode equations have been obtained in~\cite{Kodama:2003jz,Ishibashi:2003ap}, where the Fourier transformation with respect to the Killing time coordinate has been used. Here we have obtained the master equation for the scalar $\Omega$ by using the properties of the Kodama (Killing) vector.

(2). We notice that the inverse operator $H^{-2}$ appears in the effective potential of scalar perturbation and similar situation also appears in Refs.~\cite{Kodama:2003jz,Ishibashi:2003ap}. But such inverse operators do not occur in the case of tensor and vector perturbations.

(3). When matter fields are present, the situation becomes complicated. We believe similar master equation can also be obtained if the Killing vector $\xi$ exists in the warped spcetime.

(4). It should be stressed here that the equation (\ref{scalarsingle}) is valid in the sense
$$\hat{L}_{ij}\Big\{E[\Omega] + U_S\Omega\Big\}=0\, .$$
With the same assumptions below Eq.(\ref{LijSLijS}), this is equivalent to
\begin{equation}
\label{Lij}
\frac{n-1}{n}\hat{\Delta}(\hat{\Delta}+nK)\Big\{E[\Omega] + U_S\Omega\Big\}=0\, .
\end{equation}
This means that if the eigenvalue of the operator $\hat{\Delta}(\hat{\Delta}+nK)$ vanishes, the above equation is trivially satisfied. In this case, one is not able to study these special modes of perturbations, and other methods are required.  This is because we have applied the operator $\hat{L}_{ij}$ on the general equation for the scalar perturbation in order to get the master equation (\ref{scalarsingle}). Similar situation also appears in \cite{Kodama:2003jz,Ishibashi:2003ap}.

(5). The limitation on Eq.(\ref{scalarsingle}) discussed in item (4) can be removed as follows: We can always redefine the gauge-invariant variables by the ambiguities  mentioned at the end of subsec.\ref{subsec:gipv} such that $J$ is vanishing (but not merely $\hat{L}_{ij}J=0$). Similar consideration can be applied to get Eq.(\ref{s211}). By these redefinitions and the same procedure starting from Eq.(\ref{procedure1}), one can find that Eq.(\ref{scalarsingle}) exactly holds without the limitation.

\section{Wave Equation of Weyl Tensor for Einstein spacetime}
\label{sec:Teukolsky}

Based on the Newmann-Penrose formalism~\cite{Newman:1961qr}, Teukolsky found the celebrated perturbation equation--Teukolsky equation for general Type D spacetimes in four dimensions~\cite{Teukolsky:1972my, Teukolsky:1973ha}.
 Actually, one can also get the gravitational perturbation equation from the perturbed Penrose wave
equation, for example, see references~\cite{Ryan:1974nt} and~\cite{Bini:2002jx} , where the Newmann-Penrose frame  also  plays a crucial role. Unfortunately, in general, one
cannot define the Newmann-Penrose frame in higher dimensions.  To classify the Weyl tensor in higher dimensions, one has to introduce some generalized frame~\cite{Coley:2004jv, Coley:2007tp}. For example, the so called Geroch-Held-Penrose (GHP) frame in four dimensions~\cite{Geroch:1973am} has been generalized to the case in higher dimensions~\cite{Durkee:2010xq}.


In this section, we will introduce some Teukolsky-like gauge-invariant variables without using GHP frame or its higher dimensional generalization and obtain perturbation equations of these variables from the perturbed Penrose wave equations.

\subsection{Penrose wave equation}
From now on, we focus on an Einstein spacetime. In this case, the Ricci tensor of the spacetime satisfies
\begin{equation}
\label{EinsteinSpacetime}
R_{MN}=\frac{2\Lambda}{n} g_{MN}\, .
\end{equation}
 By considering the covariant derivative of the Bianchi identity for Riemann tensor, the Weyl tensor of the spacetime satisfies (see Appendix \ref{sec:curvature} for details)
\begin{equation}
\label{penrosewave}
\Box W_{M_1M_2M_3M_4}+W^{MN}{}_{M_1M_2}W_{MNM_3M_4}+4 W_{M_1}{}^{MN}{}_{[M_3}W_{M_4]NMM_2}-\frac{4\Lambda}{n} W_{M_1M_2M_3M_4}=0\, ,
\end{equation}
This is the Penrose wave equation with a cosmological constant in $(n+2)$ dimensions.
For the warped spacetime (\ref{metric1}), this equation can be decomposed into three parts:
\begin{eqnarray}
\label{eq63}
&&\Big[ {}^{2}\!\Box w  -\frac{4\Lambda}{n} w + \frac{n-1}{n} w^2 + n\frac{D^cr}{r}D_c w\Big](g_{ac}g_{bd}-g_{ad}g_{bc})
+2(n+1)\Big(\frac{D_arD_dr}{r^2}g_{cb}\nonumber\\
&&-\frac{D_arD_cr}{r^2}g_{db}+\frac{D_brD_cr}{r^2}g_{ad}-\frac{D_brD_dr}{r^2}g_{ac}\Big)w=0\, ,
\end{eqnarray}
\begin{equation}
\label{eq64}
\Big[ {}^{2}\!\Box w  - \frac{4\Lambda}{n} w  + \frac{n-1}{n} w^2+ n\frac{D^cr}{r}D_c w -2(n+1)\frac{(Dr)^2}{r^2}w\Big]r^2g_{ab}\gamma_{ij}=0\, ,
\end{equation}
and
\begin{eqnarray}
\label{eq65}
&&\frac{1}{n(n+1)}\Big[ {}^{2}\!\Box w    - \frac{4\Lambda}{n}  w  + \frac{n-1}{n}w^2+ n\frac{D^cr}{r}D_c w -2(n+1)\frac{(Dr)^2}{r^2}w\Big]r^4(\gamma_{ik}\gamma_{jl}-\gamma_{il}\gamma_{jk})\nonumber\\
&&+\hat{\Delta}\hat{W}_{ijkl}+\hat{W}^{mn}{}_{ij}\hat{W}_{mnkl}+4 \hat{W}_{i}{}^{mn}{}_{[k}\hat{W}_{l]nmj}- 2 \Bigg( \frac{2\Lambda}{n}+\frac{{}^{2}\!\Box r}{r} +(n-1) \frac{(Dr)^2}{r^2} \Bigg)r^2\hat{W}_{ijkl}
  =0\, ,
\end{eqnarray}
where $w$ is defined in Eq.(\ref{eq210}). Note that in the $n$-dimensional Einstein manifold $(\mathscr {N}^n,\gamma_{ij})$, we have
$$\hat{\Delta}\hat{W}_{ijkl}+\hat{W}^{mn}{}_{ij}\hat{W}_{mnkl}+4 \hat{W}_{i}{}^{mn}{}_{[k}\hat{W}_{l]nmj}=2(n-1)K\hat{W}_{ijkl}\, ,$$
one can show that (\ref{eq64}) and (\ref{eq65}) lead to a single equation
\begin{equation}
\label{reducedpenrose}
{}^{2}\!\Box w  - \frac{4\Lambda}{n} w  + \frac{n-1}{n} w^2+ n\frac{D^cr}{r}D_c w -2(n+1)\frac{(Dr)^2}{r^2}w=0\, ,
\end{equation}
by using Eqs.(\ref{Riccicomponets}) and (\ref{EinsteinSpacetime}).  Substituting the above equation into (\ref{eq63}), it becomes
\begin{eqnarray}
\frac{D_arD_dr}{r^2}g_{cb}
-\frac{D_arD_cr}{r^2}g_{db}+\frac{D_brD_cr}{r^2}g_{ad}-\frac{D_brD_dr}{r^2}g_{ac}=-\frac{(Dr)^2}{r^2}(g_{ac}g_{bd}-g_{ad}g_{bc})\, .
\end{eqnarray}
This equation is trivially satisfied by considering the symmetry property of a tensor in the two dimensional space~$(\mathscr{M}^2, g_{ab})$.
As a result, we have the nontrivial equation (\ref{reducedpenrose}) only. In addition,  we have from the Bianchi identity for the Einstein spacetime that
\begin{equation}
\label{BianchiEinstein}
\nabla^MW_{MNLP}=0\, .
\end{equation}
We can obtain from the above equation that
\begin{equation}
\label{reducedBianchi}
D_aw+(n+1)\frac{D^cr}{r}D_cw=0\, .
\end{equation}
This equation will be used to simplify the perturbation of the Penrose wave equation in following subsections.

\subsection{Perturbation of Penrose Wave Equation and Gauge-invariant Variables}

Consider the linear perturbation of the Penrose wave equation (\ref{penrosewave}), we can have
\begin{eqnarray}
\label{perturPenrose}
&&\Big(\Box- \frac{4\Lambda}{n}\Big)\Omega_{M_1M_2M_3M_4}+W^{MN}{}_{M_1M_2}\Omega_{MNM_3M_4}+\Omega^{MN}{}_{M_1M_2}W_{MNM_3M_4}
+4\Omega_{M_1}{}^{MN}{}_{[M_3}W_{M_4]NMM_2}\nonumber\\
&&+4W_{M_1}{}^{MN}{}_{[M_3}\Omega_{M_4]NMM_2}-h^{MN}\nabla_M\nabla_N W_{M_1M_2M_3M_4}-\delta C_{M}{}^{MN}\nabla_{N}W_{M_1M_2M_3M_4}\nonumber\\
&&-2\sum_{i=1}^4 \delta C_{MM_i}{}^{N_i}\nabla^{M}W_{M_1\cdots N_i \cdots M_4}-\sum_{i=1}^4 \nabla^{M}\delta C_{MM_i}{}^{N_i}W_{M_1\cdots N_i\cdots M_4}
-2h^M{}_LW^{LN}{}_{M_1M_2}W_{MNM_3M_4}\nonumber\\
&&-4h^M{}_LW_{M_1}{}^{LN}{}_{[M_3}W_{M_4]NMM_2}-4h^N{}_LW_{M_1}{}^{ML}{}_{[M_3}W_{M_4]NMM_2}=0\, ,
\end{eqnarray}
where $\Omega_{M_1M_2M_3M_4}=\delta W_{M_1M_2M_3M_4}$, and $\delta C_{M N}{}^{L}$ can be found in Eq.(\ref{tensorCabc}). Obviously, $\Omega_{M_1M_2M_3M_4}$ has
the symmetry of Riemann tensor inheriting from the symmetry of the Weyl tensor. In this subsection, the indices $(M, N, \cdots)$ should be understood as abstract indices~\cite{Wald1984}.

In a general case,  $\Omega_{M_{1}M_{2}M_{3}M_{4}}$ is not gauge-invariant.
 To construct gauge-invariant  variables of perturbations, let us introduce two null vectors $\ell^M$ and $n^M$ so that the metric of the spacetime can be written as
\begin{equation}
g_{MN}=-\ell_{M}n_{N}-n_M\ell_N+q_{MN}\, ,
\end{equation}
where $$\ell_M\ell^M=n_Mn^M=\ell^{M}q_{MN}=n^Mq_{MN}=0\, ,\qquad \ell_Mn^M=-1\, ,$$
and $q_{MN}$ is the metric of $n$-dimensional submanifold of the spacetime.  For the warped product spacetime we are considering, concretely we have
\begin{equation}
-\ell_Mn_N - n_M \ell_N=g_{ab}(dy^a)_M(dy^b)_N\, ,
\end{equation}
\begin{equation}
q_{MN}=g_{ij}(dz^i)_M(dz^j)_N=r(y)^2\gamma_{ij}(z)(dz^i)_M(dz^j)_N
\end{equation}
in the coordinate system $\{y^{a},z^i\}$. Note that there is a freedom to re-scale the null vectors $\{\ell^M, n^M\}$ as
\begin{equation}
\ell^M\rightarrow \alpha \ell^M\, ,\qquad n^M\rightarrow \alpha^{-1}n^M\, ,
\end{equation}
where $\alpha$ is a scalar function of the spacetime.  However, it is easy to see that such rescaling will not change our results.

In this frame,  it is easy to find that the nontrivial projections of the Weyl tensor are
\begin{eqnarray}
\label{nonvanishingproject}
&&\Psi_0=W_{MNLP}\ell^{M}n^{N}\ell^{L}n^{P}\, ,\nonumber\\
&&\Phi_{MN}= q_{M}{}^{S}q_{N}{}^{T}W_{SLTP}\ell^{L}n^{P}\, ,\nonumber\\
&&\mathcal{W}_{MNLP}=q_{M}{}^{S}q_{N}{}^{T}q_{L}{}^{U}q_{P}{}^{V}W_{STUV}\, ,
\end{eqnarray}
where $q_{M}{}^{N}=q_{M}{}^{L}q_{L}{}^{N}$ is the projection operator onto the $n$-dimensional Riemannian submanifold (The Einstein manifold $(\mathscr{N}^n,\gamma_{ij})$ with ``radius" $r$).   For the warped product spacetime (\ref{metric1}), we have
\begin{equation}
\Psi_0=-c_1 w\, ,\qquad \Phi_{MN}=c_2 w q_{MN}\, .
\end{equation}
 $\Psi_0$ is a scalar and $\Phi_{MN}$ is a symmetric tensor. Further, we can expand $W_{MNLP}$ as
\begin{eqnarray}
\label{expandWeyl}
W_{MNLP}&=&-c_1 w \epsilon_{MN}\epsilon_{LP}- 2c_2 w q_{M[L}f_{P]N} - 2c_2 w q_{N[P}f_{L]M} \nonumber\\
&&+2c_3 w q_{M[L}q_{P]N}+ r^2\hat{W}_{MNLP}\, ,
\end{eqnarray}
where the coefficients $c_1$, $c_2$ and $c_3$ are given in (\ref{eq2.9}), and
\begin{equation}
f_{MN}=-\ell_{M}n_N-n_M\ell_N\, ,\qquad \epsilon_{MN}=\ell_{M}n_N-n_M\ell_N\, .
\end{equation}
Thus the tensor $\mathcal{W}_{MNLP}$ can be  decomposed into traceless part and trace part, i.e.,
\begin{equation}
\mathcal{W}_{MNLP}=2c_3 w q_{M[L}q_{P]N}+ r^2\hat{W}_{MNLP}\, .
\end{equation}
 To define gauge-invariant perturbation variables,  let us focus on the projections
\begin{equation}
\label{Theta}
\Theta_{MN}=q_{M}{}^{S}q_{N}{}^{T}W_{SLTP}\ell^{L}\ell^{P}\, ,\qquad \bar{\Theta}_{MN}=q_{M}{}^{S}q_{N}{}^{T}W_{SLTP}n^{L}n^{P}\, .
\end{equation}
These two quantities both vanish for the background warped product spacetime.
This implies that the linear perturbations of $\Theta_{MN}$  and $\bar{\Theta}_{MN}$ are gauge invariant~\cite{Stewart:1974uz}.
 By the definition of $\Omega_{MNLP}$, we have
\begin{eqnarray}
&&\Omega_{MN}=q_M{}^{S}q_{N}{}^{T}\Omega_{SLTP}\ell^L\ell^P
+ q_M{}^{S}q_{N}{}^{T} W_{SLTP}\delta\ell^L\ell^P +  q_M{}^{S}q_{N}{}^{T} W_{SLTP}\ell^L\delta\ell^P
\nonumber\\
&&+ \delta q_M{}^{S}q_{N}{}^{T} W_{SLTP}\ell^L\ell^P +  q_M{}^{S}\delta q_{N}{}^{T} W_{SLTP}\ell^L\ell^P\, ,
\end{eqnarray}
where we have defined $\Omega_{MN}=\delta\Theta_{MN}$.
Substituting $W_{SLTP}$ in (\ref{expandWeyl}) into the above equation yields
\begin{eqnarray}
&&\Omega_{MN}=q_M{}^{S}q_{N}{}^{T}\Omega_{SLTP}\ell^L\ell^P
-2c_2 q_{MN}\ell_P\delta\ell^P
\, .
\end{eqnarray}
This indicates that $\Omega_{MN}$ is totally tangent to the $n$-dimensional submanifold, i.e.,
\begin{equation}
q_{M}{}^{L}q_{N}{}^{P}\Omega_{LP}=\Omega_{MN}\, .
\end{equation}
Considering
\begin{equation}
\label{hMNln}
-h^{MN}=-\delta\ell^{M}n^{N}-\ell^{M}\delta n^{N}-\delta n^{M}\ell^N-n^M\delta\ell^N +\delta q^{MN}\, .
\end{equation}
we have
\begin{equation}
-h^{MN}\ell_N\ell_M=2\ell_M\delta\ell^{M}\, .
\end{equation}
Substituting this into  $\Omega_{MN}$, we obtain
\begin{eqnarray}
\Omega_{MN}=q_M{}^{S}q_{N}{}^{T}\Omega_{SLTP}\ell^L\ell^P
+ c_2 q_{MN}(h_{LP}\ell^L\ell^P)\, .
\end{eqnarray}
Similarly, we can define a gauge-invariant quantity $\bar{\Omega}_{MN}$ by using $\bar{\Theta}_{MN}$.

Now let us write down the components of $\Omega_{MN}$ in the coordinate system $\{y^a, z^i\}$. For the warped product spacetime,  the null vectors $\ell_{M}$ and $n_M$ can be expressed as
\begin{equation}
\ell_{M}=\ell_a(dy^a)_M\, ,\qquad n_{M}=n_a(dy^a)_M\, .
\end{equation}
where $\ell_a$ and $n_a$ are the nonvanishing coordinate components of $\ell_M$ and $n_M$. Clearly, $\ell_a=\ell_a(y)$ and $n_a=n_a(y)$, they are independent of the coordinates
$z^i$.   The projection operator $q_{M}{}^{N}$ can be expressed as
\begin{equation}
q_{M}{}^{N}=(dz^i)_M\Big(\frac{\partial}{\partial z^i}\Big)^N\, .
\end{equation}
By these expressions, we can obtain all the components of the tensors we are studying. For example, $\Phi_{MN}$ can be expressed as $$\Phi_{MN}=\ell^a n^bW_{aibj}(dz^i)_M(dz^j)_N=c_2w g_{ij}(dz^i)_M(dz^j)_N\, ,$$
and $\Omega_{MN}$ has a form
\begin{eqnarray}
&&\Omega_{MN}=q_M{}^{S}q_{N}{}^{T}\Omega_{MLNP}\ell^L\ell^P+ c_1q_{MN}(h_{LP}\ell^L\ell^P)\nonumber\\
&&=\Big[\ell^a\ell^b\Omega_{aibj}+c_2 w g_{ij} (\ell^a\ell^b h_{ab})\Big](dz^i)_M(dz^j)_N\nonumber\\
&&=\Omega_{ij}(dz^i)_M(dz^j)_N\, .
\end{eqnarray}
This indicates that components of $\Omega_{MN}$ in this set of coordinates are given by
\begin{equation}
\label{Omegaijweyl}
\Omega_{ij}=\ell^a\ell^b\Omega_{aibj}+c_2 w g_{ij} (\ell^a\ell^b h_{ab})\, .
\end{equation}
Similarly, we can have the gauge-invariant tensor
\begin{equation}
\bar{\Omega}_{MN}=\bar{\Omega}_{ij}(dz^i)_M(dz^j)_N\, ,
\end{equation}
where
\begin{equation}
\bar{\Omega}_{ij}=n^an^b\Omega_{aibj}+c_2 w g_{ij} (n^an^b h_{ab})\, .
\end{equation}
Note that $\Omega_{MN}$ and $\bar{\Omega}_{MN}$ are both traceless tensors. To see this, let us notice that  Weyl tensor obeys,
$W_{L M P N}g^{MN} =0$. Its perturbation equation is given by
\begin{equation}
\label{traclessweyl}
\Omega_{L M P N}g^{M N} = W_{L M P N}h^{MN}\, .
\end{equation}
Multiplying  $\ell^L$ and $\ell^P$ on both sides of the above equation , we have
\begin{eqnarray}
\label{eq6.34}
&&\ell^L\ell^P\Omega_{L M P N}g^{M N} - \ell^L \ell^P W_{L M P N}h^{MN}
=\ell^L\ell^P\Omega_{L M P N}q^{M N} - \ell^L\ell^P W_{L M P N}h^{MN}\nonumber\\
&&=\ell^L\ell^P\Omega_{L M P N}q^{M N}+ q^{MN}\Phi_{MN}(\ell_S\ell_Th^{ST})\nonumber\\
&&=q^{MN}\Omega_{MN}=0.
\end{eqnarray}
This completes our proof. Here we have used (\ref{expandWeyl}) and the symmetry of $\Omega_{MNLP}$.
Using  Eq.(\ref{Omegaijweyl}), we have from (\ref{eq6.34}) that
\begin{equation}
g^{ij}\Omega_{ij}=g^{ij}\ell^a\ell^b\Omega_{aibj} + n c_2 w \ell^a\ell^b h_{ab}=0\, .
\end{equation}

Now, let us consider another  gauge-invariant variable. Taking use of $\Phi_{MN}$ , we can construct a tensor as
\begin{equation}
\mathcal{B}_{MN}:=\Phi_{MN}+\Phi_{NM}-\frac{2}{n}q^{LP}\Phi_{LP}q_{MN}\, ,
\end{equation}
which is identically vanishing.
This suggests that
\begin{equation}
C_{MN}:=\delta\mathcal{B}_{MN}
\end{equation}
is an gauge-invariant variable. From the definition of $\Phi_{MN}$ in Eqs.(\ref{nonvanishingproject}), one can have
\begin{eqnarray}
&&\delta\Phi_{MN}=\delta\big(q_{M}{}^{S}q_{N}{}^{T}W_{SLTP}\ell^{L}n^{P}\big)=\delta q_{M}{}^{S}q_{N}{}^{T}W_{SLTP}\ell^{L}n^{P}+ q_{M}{}^{S}\delta q_{N}{}^{T}W_{SLTP}\ell^{L}n^{P}\nonumber\\
&&+q_{M}{}^{S}q_{N}{}^{T}\Omega_{SLTP}\ell^{L}n^{P}+q_{M}^{~~S}q_{N}{}^{T}W_{SLTP}\delta\ell^{L}n^{P}
+q_{M}{}^{S}q_{N}{}^{T}W_{SLTP}\ell^{L}\delta n^{P}\nonumber\\
&&=q_{M}{}^{S}q_{N}{}^{T}\Omega_{SLTP}\ell^{L}n^{P}+c_2 w\Big[\delta q_{M}{}^{S}q_{NS}+ q_{MS}\delta q_{N}{}^{S}- q_{MN}\big(n_{L}\delta\ell^{L}
+\ell_{L}\delta n^{L}\big)\Big]\, .
\end{eqnarray}
Thus we obtain
\begin{eqnarray}
\label{calculationCMN}
&&C_{MN}=\delta\Phi_{MN}+\delta\Phi_{NM}-\frac{2}{n}\delta (q^{LP}\Phi_{LP}q_{MN})=\delta\Phi_{MN}+\delta\Phi_{NM}-\frac{2}{n} q^{LP}\delta\Phi_{LP}q_{MN}\nonumber\\
&&-\frac{2}{n}\delta q^{LP}\Phi_{LP}q_{MN}-\frac{2}{n} q^{LP}\Phi_{LP}\delta q_{MN}=\delta\Phi_{MN}+\delta\Phi_{NM}-\frac{2}{n} q^{LP}\delta\Phi_{LP}q_{MN}\nonumber\\
&&-\frac{2}{n}c_2wq_{LP}\delta q^{LP}q_{MN}-2c_2w \delta q_{MN}=\Big(q_{M}{}^{S}q_{N}{}^{T}+q_{N}{}^{S}q_{M}{}^{T}-\frac{2}{n}q_{MN}q^{ST}\Big)\Omega_{SLTP}\ell^{L}n^{P}\nonumber\\
&&+2c_2 w\Big[q_{NS}\delta q_{M}{}^{S}+ q_{MS}\delta q_{N}{}^{S}-\frac{1}{n}q_{LP}\delta q^{LP}q_{MN}- \delta q_{MN}\Big]\, .
\end{eqnarray}
From $q_{M}{}^{L}q_{N}{}^{P}q_{LP}=q_{MN}$, we have
\begin{equation}
q_{NS}\delta q_{M}{}^{S}+ q_{MS}\delta q_{N}{}^{S}=\delta q_{MN}-q_{M}{}^{L}q_{N}{}^{P}\delta q_{LP}\, .
\end{equation}
From Eq.(\ref{hMNln}), we get
\begin{equation}
q_{M}{}^{L}q_{N}{}^{P}\delta q_{LP}=q_{M}{}^{L}q_{N}{}^{P}h_{LP}\, ,\qquad q^{LP}\delta q_{LP}=q^{LP}h_{LP}\, ,\qquad q_{LP}\delta q^{LP}=-q^{LP}h_{LP}\, .
\end{equation}
Substituting  these  into Eq.(\ref{calculationCMN}), we find that $C_{MN}$ can be written as
\begin{eqnarray}
&&C_{MN}=\Big(q_{M}{}^{S}q_{N}{}^{T}+q_{N}{}^{S}q_{M}{}^{T}-\frac{2}{n}q_{MN}q^{ST}\Big)\Omega_{SLTP}\ell^{L}n^{P}\nonumber\\
&&-2c_2 w\Big(q_{M}{}^{L}q_{N}{}^{P}h_{LP}-\frac{1}{n}q^{LP}h_{LP}q_{MN}\Big)\, .
\end{eqnarray}
In the coordinate system $\{y^a, z^i\}$, this gauge-invariant variable can be expressed as
\begin{equation}
C_{MN}=C_{ij}(dz^i)_M(dz^j)_N\, ,
\end{equation}
where
\begin{equation}
\label{Cij}
C_{ij}= \ell^an^b\Big(\Omega_{aibj}+\Omega_{ajbi}-\frac{2}{n}g^{kl}\Omega_{akbl}g_{ij}\Big)-2c_2 w\Big(h_{ij}-\frac{1}{n}g^{kl}h_{kl}g_{ij}\Big)\, .
\end{equation}
In principle we can construct more gauge-invariant variables by considering other vanishing projections of Weyl tensor. But we limit us to the above three variables, $\Omega_{MN}$, $\bar{\Omega}_{MN}$ and $C_{MN}$
in this paper. In addition, let us mention here that in fact, these three variable can be expressed in terms of the gauge-invariant variables proposed
by Kodama and Ishibashi. This has been shown in Eqs.(\ref{Aomegaij}), (\ref{Aomegabarij}) and (\ref{Acij}) in Appendix \ref{sec:pweyl}.

\subsection{Perturbation Equations of Gauge-invariant Variables}

Substituting Eqs.(\ref{weylwarped}) into Eq.(\ref{perturPenrose}), after a long and tedious calculation, we get the equation of component $``aibj"$. The projection of this equation along $\ell^a\ell^b$ is given by
\begin{eqnarray}
&&\ell^a\ell^b\Box\Omega_{aibj}+ \ell^a\ell^b\Box (c_2 w g_{ij}h_{ab}) -w\ell^a\ell^b(\Omega_{aibj}+c_2 w g_{ij}h_{ab})-\frac{4\Lambda}{n}\ell^a\ell^b(\Omega_{aibj}+c_2 w g_{ij}h_{ab})\nonumber\\
&&+\frac{2}{r^2}\gamma^{km}\gamma^{ln}\ell^a\ell^b(\Omega_{ambn}+c_2 w g_{mn}h_{ab}) \hat{W}_{ikjl}-(c_2g_{ij}\ell^a\ell^bh_{ab})\Big(D^cD_c w +\frac{n-1}{n}w^2-\frac{4\Lambda}{n} w\nonumber\\
&&+ n\frac{D^cr}{r}D_cw\Big)
+2\frac{\ell^aD_{a}r}{r^3}\gamma^{km}\gamma^{lm}\hat{W}_{imjn}\ell^bD_{b}h_{kl}-6\frac{\ell^a\ell^bD_{a}rD_{b}r}{r^4}\gamma^{km}\gamma^{lm}\hat{W}_{imjn}h_{kl}\nonumber\\
&&-\frac{\ell^a D_{a}r}{r}w\ell^bD_{b}h_{ij}
+\frac{n+2}{n}\frac{\ell^a\ell^bD_{a}rD_br}{r^2}w h_{ij}+\frac{n-1}{n}\frac{\ell^aD_{a}r}{r}w\ell^b\big(\hat{D}_{i}h_{bj}+\hat{D}_{j}h_{bi}\big)
\nonumber\\
&&+2\frac{n-1}{n}w\ell^a\ell^b D_{a}rD^cr h_{bc}\gamma_{ij}+\frac{1}{n}\frac{\ell^aD_{a}r}{r}w\gamma_{ij}\gamma^{kl}\ell^bD_{b}h_{kl}-\frac{3}{n}\frac{\ell^a\ell^bD_{a}rD_{b}r}{r^2}w\gamma_{ij}\gamma^{kl}h_{kl}
\nonumber\\
&&=0\, .
\end{eqnarray}
By using  Eq. (\ref{reducedpenrose}) and Eq.(\ref{Omegaijweyl}) the above equation can be further simplified to
\begin{eqnarray}
\label{omegaaiaj}
&&\ell^a\ell^b\Box\Omega_{aibj}+ \ell^a\ell^b\Box (c_2 w g_{ij}h_{ab}) -w\Omega_{ij}-\frac{4\Lambda}{n}\Omega_{ij}+\frac{2}{r^2}\gamma^{km}\gamma^{ln}\Omega_{ij} \hat{W}_{ikjl}\nonumber\\
&&-2(n+1)\frac{(D r)^2}{r^2}\ell^a\ell^b(c_2 w g_{ij}h_{ab})
+2\frac{\ell^aD_{a}r}{r^3}\gamma^{km}\gamma^{lm}\hat{W}_{imjn}\ell^bD_{b}h_{kl}\nonumber\\
&&-6\frac{\ell^a\ell^bD_{a}rD_{b}r}{r^4}\gamma^{km}\gamma^{lm}\hat{W}_{imjn}h_{kl}-\frac{\ell^a D_{a}r}{r}w\ell^bD_{b}h_{ij}
+\frac{n+2}{n}\frac{\ell^a\ell^bD_{a}rD_br}{r^2}w h_{ij}\nonumber\\
&&+\frac{n-1}{n}\frac{\ell^aD_{a}r}{r}w\ell^b\big(\hat{D}_{i}h_{bj}+\hat{D}_{j}h_{bi}\big)
+2\frac{n-1}{n}w\ell^a\ell^bD_{a}rD^cr  h_{bc}\gamma_{ij}\nonumber\\
&&+\frac{1}{n}\frac{\ell^aD_{a}r}{r}w\gamma_{ij}\gamma^{kl}\ell^bD_{b}h_{kl}-\frac{3}{n}\frac{\ell^a\ell^bD_{a}rD_{b}r}{r^2}w\gamma_{ij}\gamma^{kl}h_{kl}
\nonumber\\
&&=0\, .
\end{eqnarray}
In the spacetime with warp product, one has
\begin{equation}
\label{eq6.47}
\Box \Omega_{aibj} + \Box (c_2 w g_{ij}h_{ab})=\nabla^c\nabla_c\Omega_{aibj}+\nabla^c\nabla_c (c_2 w g_{ij}h_{ab})+g^{kl}\nabla_k\nabla_l\Omega_{aibj}+g^{kl}\nabla_k\nabla_l (c_2 w g_{ij}h_{ab})\, .
\end{equation}
For the former two terms in the right hand side of the above equation, we have
\begin{eqnarray}
\label{nanb}
&&\ell^a\ell^b\nabla^c\nabla_c\Omega_{aibj}+ \ell^a\ell^b\nabla^c\nabla_c(c_2 w g_{ij}h_{ab})=\ell^a\ell^b({}^{2}\!\Box\Omega_{aibj}) -4\ell^a\ell^b\frac{D^cr}{r}D_c\Omega_{aibj}\nonumber\\
&&-\Big[2\frac{{}^{2}\!\Box r}{r}-6\frac{(Dr)^2}{r^2}\Big](\ell^a\ell^b\Omega_{aibj})+\ell^a\ell^b\big[{}^{2}\!\Box (c_2 w g_{ij}h_{ab})\big] -4\ell^a\ell^b\frac{D^cr}{r}D_c(c_2 w g_{ij}h_{ab})\nonumber\\
&&-\Big[2\frac{{}^{2}\!\Box r}{r}-6\frac{(Dr)^2}{r^2}\Big]\ell^a\ell^b(c_2 w g_{ij}h_{ab})\, .
\end{eqnarray}
On the other hand, for the latter two terms in the right hand side of the equation (\ref{eq6.47}), we have
\begin{eqnarray}
\label{gijninj}
&&\ell^a\ell^bg^{kl}\nabla_{k}\nabla_{l}\Omega_{aibj}+\ell^a\ell^bg^{kl}\nabla_{k}\nabla_{l}(c_2 w g_{ij}h_{ab})=\ell^a\ell^b\frac{1}{r^2}\hat{\Delta}\Omega_{aibj} +2\frac{\ell^a\ell^bD_{a}rD_{b}r}{r^4}\gamma^{kl}\Omega_{ikjl}\nonumber\\
&&+n\ell^a\ell^b\frac{D^cr}{r}D_c\Omega_{aibj}-2\frac{\ell^aD_{a}r}{r^3}\ell^b\hat{D}^{k}(\Omega_{kibj}+\Omega_{bikj})-(n-2)\frac{\ell^aD_a rD^cr}{r^2}\ell^b(\Omega_{cibj}+\Omega_{bicj})\nonumber\\
&&+2\frac{D^{c}r}{r}\ell^a\ell^b(\hat{D}_i\Omega_{acbj}
+\hat{D}_j\Omega_{aibc})+(D^crD^dr)\ell^a\ell^b(\Omega_{adbc}+\Omega_{acbd})\gamma_{ij}+\ell^a\ell^b\frac{1}{r^2}\hat{\Delta}(c_2 w h_{ab} g_{ij})\nonumber\\
&&-2(n+1)\frac{(Dr)^2}{r^2}(\ell^a\ell^b\Omega_{aibj})+nc_2(r D^crD_cw)\gamma_{ij}(\ell^a\ell^bh_{ab})+c_2 w  \gamma_{ij}\Big[-4\frac{\ell^aD_{a}r}{r}\ell^b\hat{D}^{k}h_{bk}\nonumber\\
&&+nrD^cr\ell^a\ell^b(D_ch_{ab})-2n(\ell^aD_{a}rD^{c}r)\ell^bh_{bc}+2\frac{\ell^a\ell^bD_{a}rD_{b}r}{r^2}\gamma^{kl}h_{kl}\Big]\, .
\end{eqnarray}
To simplify this expression, let us consider the perturbations of the Bianchi identity (\ref{BianchiEinstein}) and
\begin{equation}
\label{eq6.50}
\nabla_{S}W_{MNLP}+\nabla_{L}W_{MNPS}+\nabla_{P}W_{MNSL}=0\, .
\end{equation}
Eq. (\ref{BianchiEinstein}) leads to
\begin{eqnarray}
\label{Bianchiweyl}
&&-h^{UV}\nabla_{U}W_{VNLP}+\nabla^{M}\Omega_{MNLP}-\delta C_{M}^{~~~MU}W_{UNLP} \nonumber\\
&&-\delta C^{M}_{~~~N}{}^UW_{MULP}-\delta C^{M}_{~~~L}{}^UW_{MNUP}-\delta C^{M}_{~~~P}{}^UW_{MULU}\nonumber\\
&&=0\, ,
\end{eqnarray}
while Eq. (\ref{eq6.50}) gives
\begin{eqnarray}
\label{Bianchiweyl2}
&&\nabla_{S}\Omega_{MNLP}-\delta C_{SM}^{~~~~~U}W_{UNLP}-\delta C_{SN}^{~~~~~U}W_{MULP}-\delta C_{SL}^{~~~~~U}W_{MNUP}-\delta C_{SP}^{~~~~~U}W_{MNLU}\nonumber\\
&&+\nabla_{L}\Omega_{MNPS}-\delta C_{LM}^{~~~~~U}W_{UNPS}-\delta C_{LN}^{~~~~~U}W_{MUPS}-\delta C_{LP}^{~~~~~U}W_{MNUS}-\delta C_{LS}^{~~~~~U}W_{MNPU}\nonumber\\
&&+\nabla_{P}\Omega_{MNSL}-\delta C_{PM}^{~~~~~U}W_{UNSL}-\delta C_{PN}^{~~~~~U}W_{MUSL}-\delta C_{PS}^{~~~~~U}W_{MNUL}-\delta C_{PL}^{~~~~~U}W_{MNSU}\nonumber\\
&&=0\, .
\end{eqnarray}
From Eq.(\ref{Bianchiweyl}), we have
\begin{eqnarray}
&&-\frac{1}{r^2}\ell^b(\hat{D}^{k}\Omega_{kibj}+\hat{D}^{k}\Omega_{kjbi})-\ell^bD^{c}\Omega_{cibj}-\ell^bD^{c}\Omega_{cjbi}
=2c_2 r^2\gamma_{ij}\ell^bh_{b}{}^{a}D_a w + c_3 w \ell^bD_{b}h_{ij}\nonumber\\
&&+c_2 w\ell^b (3D_{b}h_{ij}-\hat{D}_{i}h_{bj}-\hat{D}_{j}h_{bi})+2(n-1)c_2 wrD^ar\ell^b h_{ab}\gamma_{ij}-(4c_2+2c_3) w \frac{\ell^bD_br}{r}h_{ij}
\nonumber\\
&&-(c_3+c_2) w \gamma_{ij}\gamma^{kl}\ell^bD_{b}h_{kl}+2c_2 w r^2\gamma_{ij}\ell^bD^{c}h_{bc}+2c_3 w \gamma_{ij}\gamma^{kl}\frac{\ell^bD_br}{r}h_{kl}+ 2c_2 w \gamma_{ij}\ell^b\hat{D}^kh_{bk}\nonumber\\
&&+4\frac{\ell^bD_{b}r}{r^3}\gamma^{km}\gamma^{ln}h_{mn}\hat{W}_{kilj}-\frac{1}{r^2}\gamma^{km}\gamma^{ln}\ell^bD_{b}h_{mn}\hat{W}_{kilj}+(n-3)\frac{D^{c}r}{r}\ell^b(\Omega_{cibj}+\Omega_{cjbi})\nonumber\\
&&-2\frac{\ell^bD_{b}r}{r^3}\gamma^{kl}\Omega_{ikjl}-\frac{1}{n}\frac{\ell^bD_b r}{r}w\big(h_{ij}-\gamma^{kl}h_{kl}\gamma_{ij}\big)\, .
\end{eqnarray}
From (\ref{Bianchiweyl2}), we arrive at
\begin{eqnarray}
&&\ell^a\ell^bn^c(\hat{D}_i\Omega_{ajbc}+\hat{D}_j\Omega_{aibc}+D_{b}\Omega_{ajci}+D_{b}\Omega_{aicj}-2D_c\Omega_{ajbi})
=c_2 w\ell^bD_bh_{ij}-2c_2 w\frac{\ell^bD_br}{r}h_{ij}\nonumber\\
&&-c_1 w\ell^b (D_{b}h_{ij}-\hat{D}_ih_{bj}-\hat{D}_jh_{bi}-2r\gamma_{ij}D^arh_{ab})+2c_2 w g_{ij}\ell^a\ell^bn^c(D_ch_{ab}-D_bh_{ac})\nonumber\\
&&-2\frac{n^cD_{c}r}{r}\ell^a\ell^b\Omega_{aibj}+2r\ell^aD_{a}r \gamma_{ij}\ell^b\ell^cn^dn^e\Omega_{bdce}+\frac{\ell^aD_{a}r}{r}\ell^bn^c(\Omega_{aicj}+\Omega_{ajci})\, .
\end{eqnarray}
Combining above two equations, we obtain
\begin{eqnarray}
\label{dkomegplusdio}
&&-\frac{1}{r^2}\ell^a(\hat{D}^{k}\Omega_{kiaj}+\hat{D}^{k}\Omega_{kjai})-\ell^a\ell^b n^c(\hat{D}_i\Omega_{ajbc}+\hat{D}_j\Omega_{aibc})+4n^c\ell^a\ell^bD_c\Omega_{ajbi}\nonumber\\
&&=2c_2 r^2\gamma_{ij}\ell^bh_{b}{}^{a}D_a w + \frac{1}{2} w \ell^bD_{b}(h_{ij}-\frac{1}{n}  \gamma_{ij}\gamma^{kl}h_{kl})-\frac{n-1}{2n} w\ell^b (\hat{D}_{i}h_{bj}+\hat{D}_{j}h_{bi})
\nonumber\\
&&+ 2c_2 w \gamma_{ij}\ell^b\hat{D}^kh_{bk}-4c_2 w r^2\gamma_{ij}\ell^a\ell^bn^cD_ch_{ab}-2c_2 wrD^ar\ell^b h_{ab}\gamma_{ij}-\frac{1}{n} w \frac{\ell^bD_br}{r}h_{ij}\nonumber\\
&&+2c_3 w \gamma_{ij}\gamma^{kl}\frac{\ell^bD_br}{r}h_{kl}+4\frac{\ell^bD_{b}r}{r^3}\gamma^{km}\gamma^{ln}h_{mn}\hat{W}_{kilj}-\frac{1}{r^2}\gamma^{km}\gamma^{ln}\ell^bD_{b}h_{mn}\hat{W}_{kilj}
\nonumber\\
&&-2\frac{\ell^bD_{b}r}{r^3}\gamma^{kl}\Omega_{ikjl}-\frac{1}{n}\frac{\ell^bD_b r}{r}w\big(h_{ij}-\gamma^{kl}h_{kl}\gamma_{ij}\big)-2r\ell^aD_{a}r \gamma_{ij}\ell^b\ell^cn^dn^e\Omega_{bdce}\nonumber\\
&&-2(n-4)\frac{n^cD^{c}r}{r}\ell^a\ell^b\Omega_{aibj}-(n-2)\frac{\ell^cD_{c}r}{r}\ell^an^b(\Omega_{aibj}+\Omega_{ajbi})\, .
\end{eqnarray}
To simplify the equation (\ref{gijninj}), we further  need  to consider the perturbation of $g^{MN}W_{MLNP}=0$. The latter gives
\begin{equation}
\label{bianchiperturbab}
\ell^a\ell^bn^cn^d\Omega_{acbd}+g^{ij}\ell^an^b\Omega_{aibj}=c_1 w \ell^an^bh_{ab}+c_2 w g^{ij}h_{ij}\, ,
\end{equation}
and
\begin{equation}
\label{bianchiperturbij}
c_3 w( h_{ij}- g_{ij}g^{kl}h_{kl}) -2c_2 w(\ell^an^bh_{ab}) g_{ij}-r^2g^{km}g^{ln}\hat{W}_{ikjl}h_{mn}-\ell^an^b(\Omega_{iajb}+\Omega_{ibja})+g^{kl}\Omega_{ikjl}=0\, .
\end{equation}
 Now we substitute  these results (\ref{dkomegplusdio}),(\ref{reducedBianchi}), (\ref{bianchiperturbab}), and (\ref{bianchiperturbij}) into Eq.(\ref{gijninj}), we find a lot of terms are beautifully canceled out, and  finally the equation becomes
\begin{eqnarray}
\label{ninj}
&&\ell^a\ell^b g^{kl}\nabla_{k}\nabla_{l}\Omega_{aibj}+g_{ij} \ell^a\ell^bg^{kl}\nabla_{k}\nabla_{l}(c_2 wh_{ab})=\frac{1}{r^2}\ell^a\ell^b\hat{\Delta}\Omega_{aibj}+\frac{1}{r^2}\ell^a\ell^b\hat{\Delta}(c_2 w h_{ab}g_{ij})\nonumber\\
&&+n\frac{D^cr}{r}\ell^a\ell^bD_c\Omega_{aibj}+n\frac{D^cr}{r}\ell^a\ell^bD_c(c_2 w h_{ab}g_{ij})-8\frac{\ell^eD_{e}r}{r}n^c\ell^a\ell^bD_c\Omega_{aibj}\nonumber\\
&&-8\frac{\ell^eD_{e}r}{r}n^c\ell^a\ell^bD_c(c_2 w h_{ab}g_{ij})-(n+8)\frac{(Dr)^2}{r^2}\Omega_{ij} +2(n+1)\frac{(Dr)^2}{r^2}\ell^a\ell^b(c_2 w g_{ij}h_{ab})\nonumber\\
&&-n\frac{\ell^c\ell^dD_c rD_dr}{r^2}n^b\ell^a(\Omega_{biaj}+\Omega_{aibj}-\frac{2}{n}g^{kl}\Omega_{akbl}g_{ij})
+2nc_2 w \frac{\ell^c\ell^dD_{c}rD_{d}r}{r^2}\Big( h_{ij}-\frac{1}{n}g^{kl}h_{kl}g_{ij}\Big)\, .
\end{eqnarray}
Substituting (\ref{nanb}) and (\ref{ninj}) into the perturbation equation (\ref{omegaaiaj}),  we arrive at
\begin{eqnarray}
&&\ell^a\ell^b\Big[{}^{2}\!\Box\Omega_{aibj}+{}^{2}\!\Box (c_2 w g_{ij}h_{ab})+\frac{1}{r^2}\hat{\Delta}\Omega_{aibj}+\frac{1}{r^2}\hat{\Delta}(c_2 w h_{ab}g_{ij})\Big]+(n-4)\ell^a\ell^b\frac{D^cr}{r}D_c\Omega_{aibj}\nonumber\\
&&+(n-4)\ell^a\ell^b\frac{D^cr}{r}D_c(c_2 w g_{ij}h_{ab})-8\frac{\ell^eD_{e}r}{r}n^c\ell^a\ell^bD_c\Omega_{aibj}-8\frac{\ell^eD_{e}r}{r}n^c\ell^a\ell^bD_c(c_2 w h_{ab}g_{ij}) \nonumber\\
&&-2\frac{{}^{2}\!\Box r}{r}\Omega_{ij}-(n+2)\frac{(Dr)^2}{r^2}\Omega_{ij}-w\Omega_{ij}
+\frac{2}{r^2}\gamma^{km}\gamma^{ln}\Omega_{mn}\hat{W}_{ikjl}-\frac{4\Lambda}{n}\Omega_{ij}
\nonumber\\
&&-n\frac{\ell^c\ell^dD_c rD_dr}{r^2}n^b\ell^a\big(\Omega_{biaj}+\Omega_{aibj}-\frac{2}{n}g^{kl}\Omega_{akbl}g_{ij}\big)
+2nc_2 w \frac{\ell^c\ell^dD_{c}rD_{d}r}{r^2}\Big( h_{ij}-\frac{1}{n}g^{kl}h_{kl}g_{ij}\Big)\nonumber\\
&&=0\, .
\end{eqnarray}
We are now further to deal with the above equation by considering the projection $\ell^a \ell^b$.
 Note that $\ell_a$ does not depend on the coordinates $z^i$,  thus one has $\hat{D}_i\ell^a=\hat{\Delta}\ell^a=0$. In addition, in general, $D_a\ell^b\ne 0$. To get the equation for  the gauge-invariant variable $\Omega_{ij}$, let us further define two quantities $\kappa_{\ell}$ and $\kappa_{n}$ as
\begin{equation}
\label{eq6.60}
D_a\ell_b=(\ell_an^c+n_a\ell^c)(\ell_bn^d+n_b\ell^d)D_c\ell_d=-\kappa_n\ell_a\ell_b-\kappa_{\ell}n_a\ell_b\, ,
\end{equation}
\begin{equation}
D_an_b=(\ell_an^c+n_a\ell^c)(\ell_bn^d+n_b\ell^d)D_cn_d=\kappa_n\ell_an_b+\kappa_{\ell}n_an_b\, ,
\end{equation}where
\begin{equation}
\kappa_{\ell} \equiv -n^d\ell^cD_c\ell_d\, ,\qquad \kappa_n \equiv -n^dn^cD_c\ell_d\, .
\end{equation}
From (\ref{eq6.60}) we can obtain
\begin{eqnarray}
{}^{2}\!\Box\ell_c
=-(\mathcal{L}_{\ell}\kappa_n+\mathcal{L}_n\kappa_{\ell}+2\kappa_n\kappa_{\ell})\ell_c\, ,
\end{eqnarray}
where $\mathcal{L}_{\ell}$ and $\mathcal{L}_n$ are the Lie derivatives along the vectors $\ell^a$ and $n^a$. By using these relations we can obtain
\begin{eqnarray}
D_c(\Omega_{aibj}\ell^a\ell^b)
=\ell^a\ell^b(D_c\Omega_{aibj})-2\kappa_n\ell_c(\Omega_{aibj}\ell^a\ell^b)-2\kappa_{\ell}n_c(\Omega_{aibj}\ell^a\ell^b)\, ,
\end{eqnarray}
and
\begin{eqnarray}
&&{}^{2}\!\Box (\Omega_{aibj}\ell^a\ell^b)
={}^{2}\!\Box\Omega_{aibj}\ell^a\ell^b-4\kappa_n\ell^cD_c(\Omega_{aibj}\ell^a\ell^b)
-4\kappa_{\ell}n^cD_c(\Omega_{aibj}\ell^a\ell^b)\nonumber\\
&&+8\kappa_{\ell}\kappa_n(\Omega_{aibj}\ell^a\ell^b)
-2(\mathcal{L}_{\ell}\kappa_n+\mathcal{L}_n\kappa_{\ell})(\Omega_{aibj}\ell^a\ell^b)\, .
\end{eqnarray}
Next let us consider the extrinsic curvature of the $n$-dimensional submanifold $\mathscr{N}^n$ (with the ``radius" $r$). The meaning curvature vectors of this submanifold (in abstract indices formalism)~\cite{Carter:1997pb, Cao:2010vj}: $$K^M=-n\frac{\nabla^Mr}{r}\, .$$
This meaning curvature vector is normal to the submanifold $\mathscr{N}^n$.
In the coordinates $\{y^a, z^i\}$, it  can be expressed as
\begin{equation}
K_M=K_a(d y^a)_M=-n\frac{D_ar}{r}(d y^a)_M\, .
\end{equation}
Thus the expansions of $\mathscr{N}^n$ along the normal vectors $\ell^M$ and $n^M$ are given by
\begin{equation}
\theta^{(\ell)}=-K_M\ell^M=n\frac{\ell^aD_ar}{r}\, ,\qquad \theta^{(n)}=-K_Mn^M=n\frac{n^aD_ar}{r}\, .
\end{equation}
With these relations we finally obtain the perturbation equation for the gauge-invariant variable $\Omega_{ij}$
\begin{eqnarray}
\label{masterOmegaij}
&&{}^{2}\!\Box\Omega_{ij}+\frac{\hat{\Delta}_L-2nK}{r^2}\Omega_{ij}
+\Big[4\kappa_n\ell^c
+4\kappa_{\ell}n^c-\frac{n+4}{n}\theta^{(\ell)}n^c-\frac{n-4}{n}\theta^{(n)}\ell^c
\Big]D_c\Omega_{ij}\nonumber\\
&&+2\Big[\mathcal{L}_{\ell}\kappa_n+\mathcal{L}_n\kappa_{\ell}-4\kappa_{\ell}\kappa_n
+\frac{n+4}{n}\kappa_n\theta^{(\ell)}+\frac{n-4}{n}\kappa_{\ell}\theta^{(n)}\Big]\Omega_{ij}\nonumber\\
&&-\Big[2\frac{{}^{2}\!\Box r}{r}+(n+2)\frac{(Dr)^2}{r^2}+w+\frac{4\Lambda}{n}\Big]\Omega_{ij}
-\frac{1}{n}\theta^{(\ell)}\theta^{(\ell)}C_{ij}
=0\, .
\end{eqnarray}
where $C_{ij}$ is defined in Eq.(\ref{Cij}), and $\hat{\Delta}_{L}$ is the Lichnerowicz operator on the Einstein manifold $(\mathscr{N}^n,\gamma_{ij})$ defined in Eq.(\ref{Lichnerowicz}).
In the same way we can obtain the equations for the gauge-invariant variable $\bar{\Omega}_{ij}$
\begin{eqnarray}
\label{masterbarOmegaij}
&&{}^{2}\!\Box\bar{\Omega}_{ij}+\frac{\hat{\Delta}_L-2nK}{r^2}\bar{\Omega}_{ij}
-\Big[4\kappa_{\ell}n^c
+4\kappa_{n}\ell^c+\frac{n+4}{n}\theta^{(n)}\ell^c+\frac{n-4}{n}\theta^{(\ell)}n^c
\Big]D_c\bar{\Omega}_{ij}\nonumber\\
&&-2\Big[\mathcal{L}_{n}\kappa_{\ell}+\mathcal{L}_{\ell}\kappa_{n}+4\kappa_{n}\kappa_{\ell}
+\frac{n+4}{n}\kappa_{\ell}\theta^{(n)}-\frac{n-4}{n}\kappa_{n}\theta^{(\ell)}\Big]\bar{\Omega}_{ij}\nonumber\\
&&-\Big[2\frac{{}^{2}\!\Box r}{r}+(n+2)\frac{(Dr)^2}{r^2}+w+\frac{4\Lambda}{n}\Big]\bar{\Omega}_{ij}
-\frac{1}{n}\theta^{(n)}\theta^{(n)}C_{ij}
=0\, .
\end{eqnarray}

Now, we turn to the variable $C_{MN}$. To get the perturbation of the gauge-invariant variable, let us first introduce an auxiliary tensor
\begin{equation}
C_{MLNP}=\Omega_{MLNP}+\Omega_{NLMP}-\frac{2}{n}(q^{ST}\Omega_{MSNT})q_{LP}\, ,
\end{equation}
its $``aibj"$ components can be written as
\begin{equation}
C_{aibj}=\Omega_{aibj}+\Omega_{biaj}-\frac{2}{n}\gamma^{kl}\Omega_{akbl}\gamma_{ij}\, .
\end{equation}
Another auxiliary tensor $H_{MN}$ is defined as
\begin{equation}
H_{MN}=q_{M}{}^Lq_{N}{}^Ph_{LP}-\frac{1}{n}(q^{LP}h_{LP})q_{MN}\, ,
\end{equation}
and its nontrivial components are
\begin{equation}
H_{ij}=h_{ij}-\frac{1}{n}\gamma^{kl}h_{kl}\gamma_{ij}\, .
\end{equation}
Considering the components ``$aibj$" of the equations (\ref{perturPenrose}) and multiplying $\ell^a n^b$ on the equation,
 we have
\begin{eqnarray}
\label{penroseperturbcij}
&&\ell^an^b\Box C_{aibj}-\frac{4\Lambda}{n}(\ell^an^bC_{aibj})
+\Big[\frac{2}{r^2}c_2 w \gamma^{kl}C_{ikjl}-\frac{8}{r^2}c_2 w \gamma^{km}\gamma^{ln}\hat{W}_{ikjl}H_{mn}-2c_2 w \Box H_{ij}\nonumber\\
&&+4(c_2)^2 w^2 H_{ij}+8c_2c_3 w^2H_{ij}+2c_1 w n^a\ell^b C_{aibj}-2c_3 w \ell^an^bC_{aibj}-4c_2 w \ell^an^bC_{aibj}\nonumber\\
&&+\frac{2}{r^2}\ell^an^bC_{akbl}\gamma^{km}\gamma^{ln}\hat{W}_{imjn}\Big]
-2\frac{D^{b}r}{r^3}\gamma^{km}\gamma^{lm}\hat{W}_{imjn}n^bD_{b}H_{kl}
+6\frac{(Dr)^2}{r^4}\gamma^{km}\gamma^{lm}\hat{W}_{imjn}H_{kl}\nonumber\\
&&-3\frac{(Dr)^2}{r^2}wH_{ij}-\frac{n-2}{n}w\frac{D^cr}{r}D_cH_{ij}
+\frac{n-1}{n}w\frac{D^cr}{r}\big(\hat{D}_ih_{cj}+\hat{D}_jh_{ci}-\frac{2}{n}\hat{D}^kh_{kc}\gamma_{ij}\big)
\nonumber\\
&&-4c_2D_cwD^{c}H_{ij}
=0\, .
\end{eqnarray}
Using (\ref{bianchiperturbij}) and the definition of $C_{ij}$, this equation becomes
\begin{eqnarray}
\label{eq6.75}
&&\ell^an^bg^{cd}\nabla_c\nabla_dC_{aibj}+\ell^an^bg^{kl}\nabla_k\nabla_lC_{aibj}- g^{cd}\nabla_c\nabla_d (2c_2 w H_{ij})- g^{kl}\nabla_k\nabla_l (2c_2 w H_{ij})-\frac{4\Lambda}{n}C_{ij}\nonumber\\
&&+\frac{n-2}{n} w C_{ij}+\frac{2}{r^2}C_{kl}\gamma^{km}\gamma^{ln}\hat{W}_{imjn}
+2c_2H_{ij}\Big( D^cD_c w +\frac{n-1}{n} w^2-\frac{4\Lambda}{n} w + n\frac{D^cr}{r}D_cw\Big)\nonumber\\
&&-2\frac{D^{b}r}{r^3}\gamma^{km}\gamma^{lm}\hat{W}_{imjn}D_{b}H_{kl}
+6\frac{(Dr)^2}{r^4}\gamma^{km}\gamma^{lm}\hat{W}_{imjn}H_{kl}
-\frac{n-2}{n}w\frac{D^cr}{r}D_cH_{ij}\nonumber\\
&&+\frac{n-4}{n}w\frac{(Dr)^2}{r^2}H_{ij}
+\frac{n-1}{n}w\frac{D^cr}{r}\big(\hat{D}_ih_{cj}+\hat{D}_jh_{ci}-\frac{2}{n}\hat{D}^kh_{kc}\gamma_{ij}\big)=0\, .
\end{eqnarray}
To simplify this equation, let us first deal with the D'Alembertian $\Box=g^{ab}\nabla_a\nabla_b+g^{ij}\nabla_i\nabla_j $.  The term concerning with the second part of the  D'Alembertian is
\begin{eqnarray}
&&\ell^an^bg^{kl}\nabla_{k}\nabla_{l}C_{aibj}=\ell^an^b\frac{1}{r^2}\hat{\Delta}C_{aibj} -\frac{(Dr)^2}{r^4}\gamma^{kl}C_{ikjl}+2\frac{D^{b}r}{r^3}\hat{D}^{k}C_{kibj}\nonumber\\
&&-2\frac{D^{c}r}{r}g^{ab}(\hat{D}_i\Omega_{acbj}+\hat{D}_j\Omega_{acbi}-\frac{2}{n}\hat{D}^k\Omega_{acbk}\gamma_{ij})
+(n-2)\frac{D^b rD^cr}{r^2}C_{bicj}
\nonumber\\
&&+n\ell^an^b\frac{D^cr}{r}D_cC_{aibj}-2(n+1)\frac{(Dr)^2}{r^2}(\ell^an^bC_{aibj})\, .
\end{eqnarray}
Note that we have from (\ref{Bianchiweyl})
\begin{eqnarray}
\label{eq6.76}
&&-\frac{1}{r^2}\hat{D}^kC_{kiaj} - D^bC_{biaj}=-\frac{1}{r^2}\gamma^{km}\gamma^{ln}\hat{W}_{kilj}D_aH_{mn}+4\frac{D_ar}{r}\gamma^{km}\gamma^{ln}\hat{W}_{kilj}H_{mn}\nonumber\\
&&+(3c_2 +c_3)wD_aH_{ij}-\frac{D_ar}{r}2(3c_2+2c_3)wH_{ij}-c_2 w (\hat{D}_{i}h_{aj}+\hat{D}_jh_{ai}-\frac{2}{n}\hat{D}^kh_{ak}\gamma_{ij})\nonumber\\
&&-\frac{D_ar}{r^3}\gamma^{kl}C_{ikjl}+(n-3)\frac{D^br}{r}C_{biaj}\, ,
\end{eqnarray}
and from (\ref{Bianchiweyl2})
\begin{eqnarray}
&&-\big(\hat{D}_i\Omega_{ajbc}+\hat{D}_j\Omega_{aibc}-\frac{2}{n}\gamma_{ij}\hat{D}^k\Omega_{akbc}\big)
-D_bC_{aicj}+D_cC_{aibj}
=\frac{D_cr}{r}C_{ajbi}\nonumber\\
&&-\frac{D_br}{r}C_{ajci}+c_1 w (D_cH_{ij}g_{ab}-D_bH_{ij}g_{ac})+c_2 w(D_bH_{ij}g_{ac}-D_cH_{ij}g_{ab})\nonumber\\
&&-2c_2 wH_{ij}(\frac{D_br}{r}g_{ac}-\frac{D_cr}{r}g_{ab})+c_1 wg_{ac} \big(\hat{D}_{i}h_{bj}+\hat{D}_{j}h_{bi}-\frac{2}{n}\hat{D}^kh_{bk}\gamma_{ij}\big) \nonumber\\
&&-c_1 w g_{ab}\big(\hat{D}_{i}h_{cj}+\hat{D}_{j}h_{ci}-\frac{2}{n}\hat{D}^kh_{ck}\gamma_{ij}\big)\, .
\end{eqnarray}
Further using
\begin{eqnarray}
g^{kl}\nabla_k\nabla_l H_{ij}=\frac{1}{r^2}\hat{\Delta}H_{ij}+n\frac{D^ar}{r}D_{a}H_{ij}-2(n+1)\frac{(Dr)^2}{r^2}H_{ij}+ 2\frac{D^cr}{r}\big(\hat{D}_ih_{cj}+\hat{D}_jh_{ci}-\frac{2}{n}\hat{D}^kh_{ck}\gamma_{ij}\big)\, ,
\end{eqnarray}
and (\ref{reducedpenrose}), we find the equation (\ref{eq6.76}) can be rewritten as
\begin{eqnarray}
&&\ell^an^bg^{kl}\nabla_{k}\nabla_{l}C_{aibj}=\ell^an^b\frac{1}{r^2}\hat{\Delta}C_{aibj}
-2\frac{D^{c}r}{r}\Big[-\frac{1}{r^2}\gamma^{km}\gamma^{ln}\hat{W}_{kilj}D_cH_{mn}\nonumber\\
&&+3\frac{D_cr}{r^3}\gamma^{km}\gamma^{ln}\hat{W}_{kilj}H_{mn} +(4c_2 +c_3-c_1)wD_cH_{ij}+\frac{D_cr}{r}(4c_2 +c_3)wH_{ij}\nonumber\\
&&+(c_1-c_2) w \big(\hat{D}_{i}h_{cj}+\hat{D}_jh_{ci}-\frac{2}{n}\hat{D}^kh_{ck}\gamma_{ij}\big)
\Big]
-(n-2)\frac{D^arD^br}{r^2}C_{aibj}\nonumber\\
&&+4\frac{D^{c}r}{r}
\ell^an^bD_cC_{aibj}
+n\ell^an^b\frac{D^cr}{r}D_cC_{aibj}-2(n+2)\frac{(Dr)^2}{r^2}(\ell^an^bC_{aibj})\, .
\end{eqnarray}
Thus substituting the above equation into  (\ref{eq6.75}), we have
\begin{eqnarray}
&&\ell^an^bg^{cd}\nabla_c\nabla_dC_{aibj}- g^{cd}\nabla_c\nabla_d (2c_2 w H_{ij})-\frac{4\Lambda}{n}C_{ij}
+\frac{n-2}{n} w C_{ij}\nonumber\\
&&+(n+4)\frac{D^{c}r}{r}D_cC_{ij}-(n+6)\frac{(Dr)^2}{r^2}C_{ij}
+\frac{1}{r^2}\hat{\Delta}C_{ij}+\frac{2}{r^2}C_{kl}\gamma^{km}\gamma^{ln}\hat{W}_{imjn}
\nonumber\\
&&-(n-2)\frac{\ell^c\ell^dD_crD_dr}{r^2}(C_{aibj}n^an^b)-(n-2)\frac{n^cn^dD_crD_dr}{r^2}(C_{aibj}\ell^a\ell^b)\nonumber\\
&&=0
\, .
\end{eqnarray}
Further consider the following relation
\begin{eqnarray}
&&\ell^an^b\nabla^c\nabla_cC_{aibj}- \ell^an^b\nabla^c\nabla_c(2c_2 wH_{ij})=\ell^an^b({}^{2}\!\Box C_{aibj}) -4\ell^an^b\frac{D^cr}{r}D_cC_{aibj}\nonumber\\
&&-\Big[2\frac{{}^{2}\!\Box r}{r}-6\frac{(Dr)^2}{r^2}\Big](\ell^an^bC_{aibj})-{}^{2}\!\Box (2c_2 w H_{ij}) +4\frac{D^cr}{r}D_c(2c_2 w H_{ij})\nonumber\\
&&+\Big[2\frac{{}^{2}\!\Box r}{r}-6\frac{(Dr)^2}{r^2}\Big](2c_2 w H_{ij})\, ,
\end{eqnarray}
we obtain the perturbation equation of the gauge-invariant variable $C_{ij}$
\begin{eqnarray}
\label{masterbarcij}
&&{}^{2}\!\Box C_{ij}+n\frac{D^{c}r}{r}D_cC_{ij}-\frac{\hat{\Delta}_L-2nK}{r^2}C_{ij}
-\Big[2\frac{{}^{2}\!\Box r}{r}+n\frac{(Dr)^2}{r^2}
+\frac{4\Lambda}{n}
-\frac{n-2}{n} w \Big]C_{ij}\nonumber\\
&&-2\Big(\frac{n-2}{n^2}\Big)\big[\theta^{(\ell)}\theta^{(\ell)}\bar{\Omega}_{ij}+\theta^{(n)}\theta^{(n)}\Omega_{ij}\big]=0
\, ,
\end{eqnarray}
Equations (\ref{masterOmegaij}), (\ref{masterbarOmegaij}) and (\ref{masterbarcij}) are main results of this section.
From these equations, we find the gauge-invariant perturbation variables $\Omega_{ij}$, $\bar{\Omega}_{ij}$ and $C_{ij}$ are coupled to each other. But
they will be decoupled in some special cases, for instance, the cases where $n=2$ or $r$ is a constant. This will be shown shortly.

\subsection{The decomposition of $\Omega_{ij}$, $\bar{\Omega}_{ij}$ and $C_{ij}$}
\label{subsec:decompos}

These three gauge-invariant variables  are symmetric and traceless tensors. But
in general, they are not transverse free. To see this,
for example,  we can have from Eq.(\ref{Aomegaij}) that
\begin{eqnarray}
&&2\hat{D}^i\Omega_{ij}=
\ell^a\ell^bD_a\Big(\hat{\Delta}F^{(1)}_{bj}+\hat{D}^i\hat{D}_jF^{(1)}_{bi}\Big)
-\hat{D}^i\hat{L}_{ij}\big(\ell^a\ell^bF^{(0)}_{ab}\big)\nonumber\\
&&=\ell^a\ell^bD_a\Big[\big(\hat{\Delta}+(n-1)K\big)F^{(1)}_{bj}\Big]
-\frac{n-1}{n}\big(\hat{\Delta}+K\big)\big(\ell^a\ell^b\hat{D}_jF^{(0)}_{ab}\big)
\, .
\end{eqnarray}
Clearly it does not vanish in general. The similar holds for $\bar{\Omega}_{ij}$ and $C_{ij}$.

These gauge-invariant quantities $\Omega_{ij}$, $\bar{\Omega}_{ij}$ and $C_{ij}$ can be viewed as symmetric rank two tensors on the Einstein manifold $(\mathscr{N}^n,\gamma_{ij})$.  Thus according to the decomposition theorem for symmetric tensor in~\cite{Ishibashi:2004wx}, we can decompose the tensor
$\Omega_{ij}$ as
\begin{equation}
\label{decomomegaij}
\Omega_{ij}=\Omega^{(2)}_{ij}+ \hat{D}_{i}\Omega^{(1)}_{j} + \hat{D}_{j}\Omega^{(1)}_{i}+\Omega_L \gamma_{ij}+ \hat{L}_{ij}\Omega_T\, ,
\end{equation}
where $\Omega^{(2)}_{ij}$ is a transverse traceless symmetric tensor, and  the scalar
$\Omega_L$ is the trace part of $\Omega_{ij}$, i.e.,
\begin{equation}
\Omega_{L}=\gamma^{ij}\Omega_{ij}\, ,
\end{equation}
which  is identically vanishing here.  The  scalar $\Omega_T$ satisfies
\begin{equation}
\label{deltaf}
\hat{\Delta}\Omega_{T}=f\, ,
\end{equation}
where  $f$ is a solution of
\begin{equation}
\label{fLijOmegaij}
(\hat{\Delta}+nK)f=\frac{n}{n-1}\hat{L}^{ij}\Omega_{ij}\, ,
\end{equation}
and has a vanishing integral on $\mathscr{N}^n$, i.e., $\int_{\mathscr{N}^n}f=0$. Note that the function $f$ always exists as shown in~\cite{Ishibashi:2004wx}.
Finally the vector $\Omega^{(1)}_{~i}$ is divergence free, and satisfies the following equation
\begin{equation}
\label{omegavector}
[\hat{\Delta}+(n-1)K]\Omega^{(1)}_{~i}=\hat{D}^j\Big(\Omega_{ij}-\frac{1}{n}\Omega_L\gamma_{ij}\Big)-\frac{n-1}{n}(\hat{D}_if+nK \hat{D}_i\Omega_{T})\, .
\end{equation}
We can do similar decomposition for tenors $\bar{\Omega}_{ij}$ and $C_{ij}$.


Taking this decomposition and using Eqs.(\ref{Aomegaij}), (\ref{Aomegabarij}), and (\ref{Acij}),  we can find the relations between these three gauge-invariant variables and
the Kodama-Ishibashi  variables as follows.
\begin{eqnarray}
\label{geometricexplain}
&& 2\Omega_{T}=-\ell^a\ell^bF^{(0)}_{ab}\, ,\nonumber\\
&& 2\Omega^{(1)}_{~i}=\ell^a\ell^bD_aF^{(1)}_{bi}\, ,\nonumber\\
&&2\Omega^{(2)}_{ij}=-\ell^a\ell^bD_a\Big[r^2D_b\Big(\frac{h^{(2)}_{Tij}}{r^2}\Big)\Big]\, ,\nonumber\\
&&  2\bar{\Omega}_{T}=-n^an^bF^{(0)}_{ab}\, ,\nonumber\\
&& 2\bar{\Omega}^{(1)}_{~i}=n^an^bD_aF^{(1)}_{bi}\, ,\nonumber\\
&& 2\bar{\Omega}^{(2)}_{ij}=-n^an^bD_a\Big[r^2D_b\Big(\frac{h^{(2)}_{Tij}}{r^2}\Big)\Big]\, ,\nonumber\\
&& 2C_{T}=F^{(0)c}_{~~c}=-2\ell^an^bF^{(0)}_{ab}\, ,\nonumber\\
&& 2C^{(1)}_{~i}=-D^c F^{(1)}_{ci}\, ,\nonumber\\
&& 2C^{(2)}_{ij}=D^{c}\Big[r^2D_{c}\Big(\frac{h^{(2)}_{Tij}}{r^2}\Big)\Big]\, .
\end{eqnarray}
Some remarks are in order here. (1). All the Kodama-Ishibashi  variables appear in (\ref{geometricexplain}) except for $F^{(0)}$. Note that in the case without matter sources,
$F^{(0)}$ is completely determined by $F_c^{(0)c}$ through Eq.(\ref{S4}). Therefore in principle, one can determine these Kodama-Ishibashi  variables once those Teukolsky-like
variables are known, and vice versa. (2). Those relations in (\ref{geometricexplain}) tell us that all the Kodama-Ishibashi variables come from some perturbations of projections of Weyl tensor of the spacetime through the two null vectors $\ell^M$ and $n^M$ and the projection tensor $q_M{}^{N}$. This implies that all the Kodama-Ishibashi variables have a same origin. (3). Those relations (\ref{geometricexplain}) also implies that the perturbation equations of the Kodama-Ishibashi variables are identical to the perturbation equations of the  Teukolsky-like variables found in this paper in the case without matter fields. (4). When matter fields are present, at the moment we are not sure whether
those relations in (\ref{geometricexplain}) are still valid or not, although we expect those relations or their generalizations holds as well.

In the next subsubsections, we will explicitly give the perturbation equations for these scalar, vector, and tensor parts of these Teukolsky-like
variables.

\subsubsection{Scalar type equation }

For the simplicity of those perturbation equations, we rewrite  Eqs. (\ref{masterOmegaij}), (\ref{masterbarOmegaij}), and (\ref{masterbarcij}) as
\begin{equation}
\label{SmasterOmegaij}
{}^{2}\!\Box\Omega_{ij}+X^cD_c\Omega_{ij}-\frac{\hat{\Delta}_L-2nK}{r^2}\Omega_{ij}+V_{\Omega}\Omega_{ij}-\frac{1}{n}\theta^{(\ell)}\theta^{(\ell)}C_{ij}=0\, ,
\end{equation}
\begin{equation}
\label{SmasterOmegabarij}
{}^{2}\!\Box\bar{\Omega}_{ij}+\bar{X}^cD_c\Omega_{ij}-\frac{\hat{\Delta}_L-2nK}{r^2}\bar{\Omega}_{ij}
+\bar{V}_{\bar{\Omega}}\Omega_{ij}-\frac{1}{n}\theta^{(n)}\theta^{(n)}C_{ij}=0\, ,
\end{equation}
and
\begin{eqnarray}
\label{SmasterCij}
&&{}^{2}\!\Box C_{ij}+Z^cD_cC_{ij}-\frac{\hat{\Delta}_L-2nK}{r^2}C_{ij}
+V_{C}C_{ij}\nonumber\\
&&-2\Big(\frac{n-2}{n^2}\Big)\big[\theta^{(\ell)}\theta^{(\ell)}\bar{\Omega}_{ij}-\theta^{(n)}\theta^{(n)}\Omega_{ij}\big]=0\, ,
\end{eqnarray}
by introducing  three vectors, $Z^c=nD^cr/r$,   $X^c$ and $\bar{X}^c$,
\begin{equation}
X^c=4\kappa_n\ell^c
+4\kappa_{\ell}n^c-\frac{n+4}{n}\theta^{(\ell)}n^c-\frac{n-4}{n}\theta^{(n)}\ell^c\, ,
\end{equation}
\begin{equation}
\bar{X}^c=-4\kappa_{\ell}n^c
-4\kappa_{n}\ell^c-\frac{n+4}{n}\theta^{(n)}\ell^c-\frac{n-4}{n}\theta^{(\ell)}n^c\, .
\end{equation}
and three  scalars, $V_{\Omega}$, $\bar{V}_{\bar{\Omega}}$, and $V_{C}$ , respectively,
\begin{eqnarray}
V_{\Omega}&=&2\Big(\mathcal{L}_{\ell}\kappa_n+\mathcal{L}_n\kappa_{\ell}-4\kappa_{\ell}\kappa_n
+\frac{n+4}{n}\kappa_n\theta^{(\ell)}+\frac{n-4}{n}\kappa_{\ell}\theta^{(n)}\Big)
\nonumber\\
&&-\Big[2\frac{{}^{2}\!\Box r}{r}+(n+2)\frac{(Dr)^2}{r^2}+w+\frac{4\Lambda}{n}\Big]\, ,
\end{eqnarray}
\begin{eqnarray}
\bar{V}_{\bar{\Omega}}&=&-2\Big(\mathcal{L}_{n}\kappa_{\ell}+\mathcal{L}_{\ell}\kappa_{n}+4\kappa_{n}\kappa_{\ell}
+\frac{n+4}{n}\kappa_{\ell}\theta^{(n)}-\frac{n-4}{n}\kappa_{n}\theta^{(\ell)}\Big)
\nonumber\\
&&-\Big[2\frac{{}^{2}\!\Box r}{r}+(n+2)\frac{(Dr)^2}{r^2}+w+\frac{4\Lambda}{n}\Big]\, ,
\end{eqnarray}
\begin{equation}
V_{C}=-2\frac{{}^{2}\!\Box r}{r}-n\frac{(Dr)^2}{r^2}
-\frac{4\Lambda}{n}+\frac{n-2}{n} w \, .
\end{equation}
 To get the equation for $\Omega_T$, let us consider the action of $\hat{L}^{ij}$ on Eq.(\ref{SmasterOmegaij}). By using following relation
\begin{eqnarray}
&&-\hat{L}^{ij}\hat{\Delta}_{L}\Omega_{ij}=\hat{L}^{ij}[(\hat{\Delta}-2nK)\Omega_{ij}+2\hat{W}_{ikjl}\Omega^{kl}]
\nonumber\\
&&=\hat{L}^{ij}(\hat{\Delta}-2nK)\Omega_{ij}+2\hat{D}^i\hat{D}^j(\hat{W}_{ikjl}\Omega^{kl})=\hat{\Delta}\hat{L}^{ij}\Omega_{ij} \nonumber\\
&&-2K\hat{D}^{i}\hat{D}^j\Omega_{ij}-2\hat{R}_{ikjl}\hat{D}^i\hat{D}^j\Omega^{kl}+2\hat{W}_{ikjl}\hat{D}^i\hat{D}^j\Omega^{kl}\nonumber\\
&&=\hat{\Delta}\hat{L}^{ij}\Omega_{ij}\, ,
\end{eqnarray}
we have
\begin{equation}
{}^{2}\!\Box(\hat{L}^{ij}\Omega_{ij})+X^cD_c(\hat{L}^{ij}\Omega_{ij})+\frac{\hat{\Delta}+2nK}{r^2}(\hat{L}^{ij}\Omega_{ij})
+V_{\Omega}(\hat{L}^{ij}\Omega_{ij})-\frac{1}{n}\theta^{(\ell)}\theta^{(\ell)}(\hat{L}^{ij}C_{ij})=0\, .
\end{equation}
Together with  Eq.(\ref{deltaf}) and Eq.(\ref{fLijOmegaij}), we find the equation for $\Omega_T$:
\begin{equation}
\label{omegaijsclar1}
(\hat{\Delta}+nK)\hat{\Delta}\Big\{\mathscr{E}[\Omega_T]\Big\}=0\, ,
\end{equation}
where $\mathscr{E}[\Omega_T]$ is defined as
\begin{equation}
\mathscr{E}[\Omega_T]={}^{2}\!\Box\Omega_T+X^cD_c\Omega_T+\frac{\hat{\Delta}+2nK}{r^2}\Omega_T
+V_{\Omega}\Omega_T-\frac{1}{n}\theta^{(\ell)}\theta^{(\ell)}C_T\, .
\end{equation}
This implies that $\mathscr{E}[\Omega_T]$ belongs to the kernel of the elliptic operator $(\hat{\Delta}+nK)\hat{\Delta}$.  When the eigenvalue of the operator  $(\hat{\Delta}+nK)\hat{\Delta}$ does not vanish, we have
\begin{equation}
\label{omegaijsclar}
{}^{2}\!\Box\Omega_T+X^cD_c\Omega_T+\frac{\hat{\Delta}+2nK}{r^2}\Omega_T
+V_{\Omega}\Omega_T-\frac{1}{n}\theta^{(\ell)}\theta^{(\ell)}C_T=0\, .
\end{equation}
This is the perturbation equation of the variable $\Omega_T$. On the other hand, when the eigenvalue of the operator  $(\hat{\Delta}+nK)\hat{\Delta}$ vanishes, the equation
(\ref{omegaijsclar1}) is trivially satisfied.

Similarly, we can get the equations for $\bar{\Omega}_T$ and $C_T$ as
\begin{equation}
{}^{2}\!\Box\bar{\Omega}_T+\bar{X}^cD_c\bar{\Omega}_T+\frac{\hat{\Delta}+2nK}{r^2}\bar{\Omega}_T
+\bar{V}_{\bar{\Omega}}\bar{\Omega}_T-\frac{1}{n}\theta^{(n)}\theta^{(n)}C_T=0\, ,
\end{equation}
and
\begin{eqnarray}
&&{}^{2}\!\Box C_T+Z^cD_cC_T+\frac{\hat{\Delta}+2nK}{r^2}C_T
+V_{C}C_T
\nonumber\\
&&-2\Big(\frac{n-2}{n^2}\Big)\big[\theta^{(\ell)}\theta^{(\ell)}\bar{\Omega}_{T}+\theta^{(n)}\theta^{(n)}\Omega_{T}\big]=0\, .
\end{eqnarray}
These are the scalar type equations for the gauge-invariant perturbation variables $\Omega_{ij}$, $\bar{\Omega}_{ij}$ and $C_{ij}$.  Let us remind here that these equations exactly hold for generic modes because we have applied operator $\hat{L}_{ij}$ on the equations (\ref{SmasterOmegaij}), (\ref{SmasterOmegabarij}), and (\ref{SmasterCij}) as in subsec.\ref{subsec:scalarmaster}.

\subsubsection{Vector type equation}

In the same way as in the previous subsection, we obtain the equations for $\Omega^{(1)}_{~i}$, $\bar{\Omega}^{(1)}_{~i}$,  and $C^{(1)}_{~i}$ as
\begin{equation}
\label{omegaijvector}
{}^{2}\!\Box\Omega^{(1)}_{~i}+X^cD_c\Omega^{(1)}_{~i}+\frac{\hat{\Delta}+(n+1)K}{r^2}\Omega^{(1)}_{~i}
+V_{\Omega}\Omega^{(1)}_{~i}-\frac{1}{n}\theta^{(\ell)}\theta^{(\ell)}C^{(1)}_{~i}=0\, .
\end{equation}
\begin{equation}
{}^{2}\!\Box\bar{\Omega}^{(1)}_{~i}+\bar{X}^cD_c\bar{\Omega}^{(1)}_{~i}+\frac{\hat{\Delta}+(n+1)K}{r^2}\bar{\Omega}^{(1)}_{~i}
+\bar{V}_{\bar{\Omega}}\bar{\Omega}^{(1)}_{~i}-\frac{1}{n}\theta^{(n)}\theta^{(n)}C^{(1)}_{~i}=0\, ,
\end{equation}
and
\begin{eqnarray}
&&{}^{2}\!\Box C^{(1)}_{~i}+Z^cD_cC^{(1)}_{~i}+\frac{\hat{\Delta}+(n+1)K}{r^2}C^{(1)}_{~i}
+V_{C}C^{(1)}_{~i}
\nonumber\\
&&-2\Big(\frac{n-2}{n^2}\Big)\big[\theta^{(\ell)}\theta^{(\ell)}\bar{\Omega}^{(1)}_{~i}+\theta^{(n)}\theta^{(n)}\Omega^{(1)}_{~i}\big]=0\, ,
\end{eqnarray}
respectively.
These are the vector type equations for the gauge-invariant perturbation variables $\Omega_{ij}$, $\bar{\Omega}_{ij}$ and $C_{ij}$. Note that
the Lichnerowicz operator $\hat{\Delta}_L$ does not appear in the scalar and vector type equations, the same as in the situations of the scalar and vector equations for the corresponding Kodama-Ishibashi variables discussed in Sec.\ref{sec:GivMaster}.  In addition, as the scalar type equations, these vector type equations holds for generic modes if spectral expansion is done.

\subsubsection{Tensor type equation}

By subtracting Eqs.(\ref{omegaijsclar}) and (\ref{omegaijvector}) from Eq.(\ref{SmasterOmegaij}), we obtain the tensor type equation of $\Omega_{ij}^{(2)}$ as
\begin{equation}
{}^{2}\!\Box\Omega^{(2)}_{ij}+X^cD_c\Omega^{(2)}_{ij}-\frac{\hat{\Delta}_L-2nK}{r^2}\Omega^{(2)}_{ij}+V_{\Omega}\Omega^{(2)}_{ij}
-\frac{1}{n}\theta^{(\ell)}\theta^{(\ell)}C^{(2)}_{ij}=0\, ,
\end{equation}
Similarly we can arrive at
\begin{equation}
{}^{2}\!\Box\bar{\Omega}^{(2)}_{ij}+\bar{X}^cD_c\bar{\Omega}^{(2)}_{ij}-\frac{\hat{\Delta}_L-2nK}{r^2}\bar{\Omega}^{(2)}_{ij}+\bar{V}_{\bar{\Omega}}\bar{\Omega}^{(2)}_{ij}
-\frac{1}{n}\theta^{(n)}\theta^{(n)}C^{(2)}_{ij}=0\, ,
\end{equation}
and
\begin{eqnarray}
&&{}^{2}\!\Box C^{(2)}_{ij}+Z^cD_cC^{(2)}_{ij}-\frac{\hat{\Delta}_L-2nK}{r^2}C^{(2)}_{ij}
+V_{C}C^{(2)}_{ij}
\nonumber\\
&&-2\Big(\frac{n-2}{n^2}\Big)\big[\theta^{(\ell)}\theta^{(\ell)}\bar{\Omega}^{(2)}_{ij}+\theta^{(n)}\theta^{(n)}\Omega^{(2)}_{ij}\big]=0\, .
\end{eqnarray}
Note that the Lichnerowicz operator $\hat{\Delta}_L$  is present in the above equations, as in (\ref{T}).

\subsection{Some special cases}

We have obtained the equations for the gauge-invariant variables $\Omega_{ij}$, $\bar{\Omega}_{ij}$ and $C_{ij}$. These equations form a closed system and in general
they are coupled to each other. In this subsection, we will show they are decoupled in four dimensions.  When $r$ keeps as a constant, these equations are also decoupled.

\subsubsection{Four dimensional case}

From the relation (\ref{Acij}), we have
\begin{eqnarray}
\label{cijtoki}
&&2C_{ij}=-(n-2)rD^crD_c\Big(\frac{h^{(2)}_{Tij}}{r^2}\Big)+ [\hat{\Delta}_L-2(n-1)K]\Big(\frac{h^{(2)}_{Tij}}{r^2}\Big)\nonumber\\
&&+(n-2)\frac{D^cr}{r}
\Big(\hat{D}_iF^{(1)}_{cj}+\hat{D}_jF^{(1)}_{ci}\Big)
-(n-2)\hat{L}_{ij}\Big(\frac{F^{(0)}}{r^2}\Big)\, .
\end{eqnarray}
Here we have used equations (\ref{T}), (\ref{V2}) and (\ref{S4}). In the four dimensional case,  the above equation reduces to
\begin{eqnarray}
2C_{ij}=-[\hat{\Delta}-2K]\Big(\frac{h^{(2)}_{Tij}}{r^2}\Big)\, .
\end{eqnarray}
Note that in the four dimension case, $h^{(2)}_{Tij}$ must vanish.  This means $C_{ij}=0$ when $n=2$. Thus,
Eqs. (\ref{masterbarcij}), (\ref{masterOmegaij}) and (\ref{masterbarOmegaij}) decouple from each other, and only two independent equations remain:
\begin{equation}
{}^{2}\!\Box\Omega_{ij}+X^cD_c\Omega_{ij}-\frac{\hat{\Delta}_L-2nK}{r^2}\Omega_{ij}+V_{\Omega}\Omega_{ij}=0\, ,
\end{equation}
and
\begin{equation}
{}^{2}\!\Box\bar{\Omega}_{ij}+\bar{X}^cD_c\bar{\Omega}_{ij}-\frac{\hat{\Delta}_L-2nK}{r^2}\bar{\Omega}_{ij}+\bar{V}_{\bar{\Omega}}\bar{\Omega}_{ij}=0\, .
\end{equation}
In fact, $\Omega_{ij}$ and $\bar{\Omega}_{ij}$ have only vector and scalar components in four dimensions. One can easily get the equations of these scalar and vector perturbations from the discussions in the previous subsection.

\subsubsection{Constant radius case}

When $r$ is a constant, we have
$D_ar=\theta^{(\ell)}=\theta^{(n)}=0$. In this case, Eqs. (\ref{masterbarcij}), (\ref{masterOmegaij}) and (\ref{masterbarOmegaij}) are decoupled, and these equations reduce to
\begin{eqnarray}
&&{}^{2}\!\Box\Omega_{ij}-\frac{\hat{\Delta}_L+2(n-1)K}{r^2}\Omega_{ij}
+4(\kappa_n\ell^c
+\kappa_{\ell}n^c
)D_c\Omega_{ij}\nonumber\\
&&+2\big(\mathcal{L}_{\ell}\kappa_n+\mathcal{L}_n\kappa_{\ell}-4\kappa_{\ell}\kappa_n
\big)\Omega_{ij}=0\, ,
\end{eqnarray}
\begin{eqnarray}
&&{}^{2}\!\Box\bar{\Omega}_{ij}-\frac{\hat{\Delta}_L+2(n-1)K}{r^2}\bar{\Omega}_{ij}
-4(\kappa_n\ell^c
+\kappa_{\ell}n^c
)D_c\bar{\Omega}_{ij}\nonumber\\
&&-2\big(\mathcal{L}_{\ell}\kappa_n+\mathcal{L}_n\kappa_{\ell}+4\kappa_{\ell}\kappa_n
\big)\bar{\Omega}_{ij}=0\, ,
\end{eqnarray}
and
\begin{eqnarray}
\label{cijconstantr}
{}^{2}\!\Box C_{ij}-\frac{\hat{\Delta}_L-2(n+1)K}{r^2}C_{ij}
=0
\, ,
\end{eqnarray}
respectively.
Here we have used the relation $(n-1)K/r^2=2\Lambda/n$ from Eq.(\ref{EinsteinSpacetime}) and Eq.(\ref{Riccicomponets}).

 Note that the equation of $C_{ij}$ is independent of the null vectors. As a result,  it is quite useful to discuss the stability of this kind geometry in higher dimensions. In addition, let us note that from Eq.(\ref{cijtoki}), one can see that $C_{ij}$ does not include vector part. Thus the tensor and scalar parts of Eq.(\ref{cijconstantr}) have forms
\begin{eqnarray}
{}^{2}\!\Box C^{(2)}_{ij}-\frac{\hat{\Delta}_L-2(n+1)K}{r^2}C^{(2)}_{ij}
=0
\, ,
\end{eqnarray}
and
\begin{eqnarray}
{}^{2}\!\Box C_{T}+\frac{\hat{\Delta}+2(n+1)K}{r^2}C_{T}
=0
\, .
\end{eqnarray}
Thus once the spectral expansion on the Einstein manifold $(\mathscr{N}^n,\gamma_{ij})$ is made, one can easily discuss  the stability of such kind of spacetime ($n>2$).

 Our discussion can be extended to the Einstein-Maxwell theory, and similar equations should exist. Thus one can study  the stability of the near horizon geometry of extremal black holes. The discussions in this section  support the conclusion by Durkee and Reall in a recent paper~\cite{Durkee:2010qu}.

\section{Summary and discussion}
\label{sec:summary}

In this paper, we have studied the linear perturbations of an $(m+n)$-dimensional spacetime with warped product metric (\ref{metric1}). By use of the gauge-invariant variables proposed by Kodama and Ishibashi, we have obtained the most general  perturbation equations for these variables, i.e., Eqs. (\ref{T}), (\ref{V1}), (\ref{V2}), (\ref{S1}), (\ref{S2}), (\ref{S3}), and (\ref{S4}). Here we have used the decomposition theorems of tensors on the submanifold ($\mathscr{N}^n,\gamma_{ij}$)   and spectral expansion method has not been used.  These equations are related by the perturbation equations of the Bianchi identity, i.e., Eqs.(\ref{Bianchi1}), (\ref{Bianchi2}), and (\ref{Bianchi3}).

When $m=2$, by using these perturbation equations, we are able to obtain the master equations for the tenor and vector perturbations. The master equation (\ref{scalarsingle}) for the scalar perturbation can be obtained only in the case without matter sources. In obtaining the master equation (\ref{scalarsingle}) of the scalar $\Omega$, we have not used the Fourier transformation with respect to the time coordinate in the usual way, instead taken use of the property of the Kodama vector.

By introducing three Teukolsky-like gauge-invariant variables $\Omega_{ij}$, $\bar{\Omega}_{ij}$, and $C_{ij}$ and considering perturbation equations of Penrose
wave equations, we have obtained the perturbation equations of the three Teukolsky-like variables in the $(2+n)$-dimensional Einstein spacetime. The three equations form a closed set of equations. In general the three equations are coupled to each other, and decouple only in some special cases, for instance, in four dimensional case or in the case with a constant warped factor. In particular, we have found that the  three Teukolsky-like gauge-invariant variables can be expressed in terms of the Kodama-Ishibashi gauge-invariant variables. This implies that the perturbation equations of three Teukolsky-like gauge-invariant perturbation variables are equivalent to the perturbation equations of the  Kodama-Ishibashi variables [see (\ref{geometricexplain})] . On the other hand, the relations (\ref{geometricexplain}) between the Teukolsky-like variable and Kodama-Ishibashi variables give the origin of those Kodama-Ishibashi gauge-invariant variables: they all come from the perturbation of some projections of Weyl tensor.

 With our perturbation equations of the Kodama-Ishibashi gauge-invariant variables,  we have obtained a complete gauge-invariant theory.  We can further make spectral analysis in a natural way.  For example,  the scalar-type and vector-type harmonic tensors introducing in \cite{Ishibashi:2011ws}
\begin{eqnarray}
\label{eq7.1}
&&\mathbb{S}_i=-\frac{1}{k}\hat{D}_i\mathbb{S}\, ,\nonumber\\
&& \mathbb{S}_{ij}=\frac{1}{k^2}\hat{D}_i\hat{D}_j\mathbb{S}+\frac{1}{n}\gamma_{ij}\mathbb{S}\, ,\nonumber\\
&&\mathbb{V}_{ij}=-\frac{1}{k^2}(\hat{D}_i\mathbb{V}_j + \hat{D}_j\mathbb{V}_i)\, ,
\end{eqnarray}
 are not necessary in our spectral expansion.  Thus some problems concerning with some special modes due to introducing these scalar-type and vector-type harmonic tensors
 can be avoided. In our mode expansion, for every mode, these gauge-invariant variables always exist and keep gauge-invariant, and those special modes can be easily dealt with.
 In this sense, our perturbation equations in terms of the Kodama-Ishibashi gauge-invariant variables make the perturbation theory be complete.

Let us stress here that we have established the relations (\ref{geometricexplain}) between the Teukolsky-like gauge-invariant variables and the  Kodama-Ishibashi variables only in the case of $(2+n)$-dimensional Einstein spacetime without matter sources. At the moment we are not sure whether these relations still valid or not in a general case with matter sources. In addition,  It would be great interesting to investigate whether one can construct some gauge-invariant variables from the Teukolsky-like variables, $\Omega_{ij}$, $\bar{\Omega}_{ij}$, and $C_{ij}$, so that corresponding equations for those gauge-invariant variables become decoupled each other.

Finally let us notice that in establishing the perturbations theory of spacetime, the decomposition theorems of tensor plays a crucial role. If $(\mathscr{N}^n,\gamma_{ij})$ is a closed Einstein manifold, Ishibashi and Wald have given a rigorous proof on these decomposition theorems~\cite{Ishibashi:2004wx}. However, when $(\mathscr{N}^n,\gamma_{ij})$ is noncompact, a rigorous proof is still absent for these decomposition theorems, although it is widely believed that these theorems are also valid.


\section*{Acknowledgments}

This work was supported in part by the National Natural Science Foundation of China with grants
No.10821504 and No.11035008 (RGC), and No.11205148 and No.11235010 (LMC).

\appendix
\section{Perturbation of Weyl tensor }
\label{sec:pweyl}
In this appendix, we give a detailed calculation to get the relations between the Teukolsky-like variables $\Omega_{ij}$, $\bar{\Omega}_{ij}$, and $C_{ij}$ and the Kodama-Ishibashi variables.
\subsection{The perturbation of Weyl tensor}
With the perturbation of metric $g_{MN}\rightarrow g_{MN}+h_{MN}$, one has
\begin{eqnarray}
\label{A1}
\delta R_{MNL}{}^{P}=-\frac{1}{2}\Big[(\nabla_M\nabla_N-\nabla_{N}\nabla_M)h_{L}^{~P}+(\nabla_M\nabla_Lh_{N}^{~P}-\nabla_N\nabla_Lh_{M}^{~P})
-(\nabla_M\nabla^Ph_{NL}-\nabla_N\nabla^Ph_{ML})\Big]\, ,
\end{eqnarray}
and $\delta R_{MNLP}=g_{SP}\delta R_{MNL}{}^{S}+R_{MNL}{}^{S}h_{SP}$. Thus the perturbation of Ricci tensor has the form
\begin{eqnarray}
\delta R_{MN}=-\frac{1}{2}\Big[\nabla_L\nabla_Mh_{N}^{~L}+\nabla_L\nabla_Nh_{M}^{~L}-\Box h_{MN}-\nabla_M\nabla_N h_{L}^{~L}\Big]\, ,
\end{eqnarray}
and the perturbation of scalar curvature $\delta R=g^{MN}\delta R_{MN}-R_{MN}h^{MN}$. The perturbation of Weyl tensor
\begin{eqnarray}
W_{MNLP}&=&R_{MNLP}-\frac{1}{n}(g_{ML}R_{PN}-g_{MP}R_{LN}-g_{NL}R_{PM}+g_{NP}R_{LM})\nonumber\\
&&+\frac{1}{n(n+1)}R(g_{ML}g_{PN}-g_{MP}g_{LN})\, .
\end{eqnarray}
can be expressed as
\begin{eqnarray}
&&\delta W_{MNLP}=\delta R_{MNLP}-\frac{1}{n}(h_{ML}R_{PN}-h_{MP}R_{LN}-h_{NL}R_{PM}+h_{NP}R_{LM})\nonumber\\
&&-\frac{1}{n}(g_{ML}\delta R_{PN}-g_{MP}\delta R_{LN}-g_{NL}\delta R_{PM}+g_{NP}\delta R_{LM})\nonumber\\
&&+\frac{1}{n(n+1)} R(h_{ML}g_{PN}+g_{ML}h_{PN}-h_{MP}g_{LN}-g_{MP}h_{LN})\nonumber\\
&&+\frac{1}{n(n+1)}\delta R(g_{ML}g_{PN}-g_{MP}g_{LN})\, .
\end{eqnarray}
For the Einstein spacetime with $R_{MN}=(2/n)\Lambda g_{MN}$, we have $\delta R_{MN}=(2/n)\Lambda h_{MN}$, the above equation is reduced to
\begin{eqnarray}
&&\delta W_{MNLP}=g_{SP}\delta R_{MNL}{}^{S}
-\frac{2\Lambda}{n^2}(g_{ML}h_{PN}-g_{MP}h_{LN}-g_{NL}h_{PM}+g_{NP}h_{LM})\nonumber\\
&&+R_{MNL}{}^{S}h_{SP}+\frac{2\Lambda}{n^2(n+1)}(h_{ML}g_{PN}-h_{MP}g_{LN}-h_{NL}g_{PM}+h_{NP}g_{LM})\, .
\end{eqnarray}
Note that one has for Einstein spacetime,
\begin{equation}
\label{AweylRiemann}
W_{MNLP}=R_{MNLP}-\frac{2\Lambda}{n(n+1)}(g_{ML}g_{PN}-g_{MP}g_{LN})\, .
\end{equation}
Thus we can obtain
\begin{eqnarray}
\label{A7}
\delta W_{MNLP}=g_{SP}\delta R_{MNL}{}^{S}+W_{MNL}{}^{S}h_{SP}
-\frac{2\Lambda}{n(n+1)}(h_{ML}g_{PN}-h_{NL}g_{PM})\, .
\end{eqnarray}

\subsection{Expressed by Kodama-Ishibashi Variables}
Substituting (\ref{A1}) into  (\ref{A7}), we have
\begin{eqnarray}
&&\Omega_{MNLP} \equiv \delta W_{MNLP} =-\frac{1}{2}\Big[(\nabla_M\nabla_N-\nabla_{N}\nabla_M)h_{LP}+(\nabla_M\nabla_Lh_{NP}-\nabla_N\nabla_Lh_{MP})\nonumber\\
&&-(\nabla_M\nabla_Ph_{NL}-\nabla_N\nabla_Ph_{ML})\Big]
+W_{MNL}{}^{S}h_{SP}\nonumber\\
&&-\frac{2\Lambda}{n(n+1)}(h_{ML}g_{PN}-h_{NL}g_{PM})\, .
\end{eqnarray}
By use of Eq. $(\ref{weylwarped})$, we get the $``aibj"$ components of $\Omega_{MNLP}$
\begin{eqnarray}
&&\Omega_{aibj}=-\frac{1}{2}\Big[(\nabla_a\nabla_i-\nabla_{i}\nabla_a)h_{bj}+(\nabla_a\nabla_bh_{ij}-\nabla_i\nabla_bh_{aj})\nonumber\\
&&-(\nabla_a\nabla_jh_{bi}-\nabla_i\nabla_jh_{ab})\Big]
-c_2 w g_{ab}h_{ij}-\frac{2\Lambda}{n(n+1)}h_{ab}g_{ij}\, .
\end{eqnarray}
Expressing the covariant derivative $\nabla_M$  by $D_a$ and $\hat{D}_i$, the above equation can be rewritten as
\begin{eqnarray}
\label{Aomegaaibj}
&&\Omega_{aibj}=-\frac{1}{2}\Big[\frac{D_ar}{r}(\hat{D}_jh_{bi}-\hat{D}_ih_{bj})+\frac{D_br}{r}(\hat{D}_jh_{ai}-\hat{D}_ih_{aj})
-\frac{D_ar}{r}D_bh_{ij}\nonumber\\
&&-\frac{D_br}{r}D_ah_{ij}+\gamma_{ij}rD^cr(D_{c}h_{ab}-D_bh_{ac}-D_ah_{bc})-2\Big(\frac{D_{a}D_{b}r}{r}-\frac{D_arD_br}{r^2}\Big)h_{ij}\nonumber\\
&&-D_a\hat{D}_jh_{bi}-D_b\hat{D}_ih_{aj}+D_aD_bh_{ij}+\hat{D}_i\hat{D}_jh_{ab}\Big]
\nonumber\\
&&-\frac{2\Lambda}{n(n+1)}(h_{ab}r^2\gamma_{ij})-c_2 w g_{ab}h_{ij}
\, .
\end{eqnarray}
Multiplying $\ell^a\ell^b$ on both sides of the above equation leads to
\begin{eqnarray}
\label{eqA11}
&&\ell^a\ell^b\Omega_{aibj}=-\frac{1}{2}\ell^a\ell^b\Big[-2\frac{D_ar}{r}D_bh_{ij}
-2\Big(\frac{D_{a}D_{b}r}{r}-\frac{D_arD_br}{r^2}\Big)h_{ij}+\gamma_{ij}rD^cr(D_{c}h_{ab}-2D_ah_{bc})\nonumber\\
&&-D_a\hat{D}_jh_{bi}-D_b\hat{D}_ih_{aj}
+D_aD_bh_{ij}+\hat{D}_i\hat{D}_jh_{ab}\Big]
-\frac{2\Lambda}{n(n+1)}(\ell^a\ell^ah_{ab}r^2\gamma_{ij})\, ,
\end{eqnarray}
From which we can read out the gauge-invariant variable $\Omega_{ij}$ defined in (\ref{Omegaijweyl}) as
\begin{eqnarray}
&&\Omega_{ij}=-\frac{1}{2}\ell^a\ell^b\Big[-2\frac{D_ar}{r}D_bh_{ij}
-2\Big(\frac{D_{a}D_{b}r}{r}-\frac{D_arD_br}{r^2}\Big)h_{ij}\nonumber\\
&&-D_a\hat{D}_jh_{bi}-D_b\hat{D}_ih_{aj}
+D_aD_bh_{ij}+\hat{D}_i\hat{D}_jh_{ab}\Big]\, .
\end{eqnarray}
Substituting the decomposition of $h_{ai}$ and $h_{ij}$ in Eqs.(\ref{hai}) and (\ref{hij}), the above expression can be rewritten as
\begin{eqnarray}
&&-2\Omega_{ij}=\ell^a\ell^b\Bigg{\{}-2\frac{D_ar}{r}D_bh^{(2)}_{Tij}
-2\Big(\frac{D_{a}D_{b}r}{r}-\frac{D_arD_br}{r^2}\Big)h^{(2)}_{Tij} +D_aD_bh^{(2)}_{Tij}\nonumber\\
&&+D_aD_b\Big(\hat{D}_ih_{Tj}^{(1)}+\hat{D}_jh_{Ti}^{(1)} \Big)-2\Big(\frac{D_{a}D_{b}r}{r}-\frac{D_arD_br}{r^2}\Big)\Big(\hat{D}_ih_{Tj}^{(1)}+\hat{D}_jh_{Ti}^{(1)} \Big)\nonumber\\
&&-2\frac{D_ar}{r}D_b\Big(\hat{D}_ih_{Tj}^{(1)}+\hat{D}_jh_{Ti}^{(1)} \Big)-D_a\Big(\hat{D}_jh_{bi}^{(1)}+\hat{D}_ih_{bj}^{(1)}\Big)+\hat{L}_{ij}\Big[h_{ab}-2D_ah_b
\nonumber\\
&&+D_aD_bh_T-2\Big(\frac{D_{a}D_{b}r}{r}-\frac{D_arD_br}{r^2}\Big)h_T-2\frac{D_ar}{r}D_bh_T\Big]\Bigg{\}}
\, .
\end{eqnarray}
We find that this gauge-invariant variable can be expressed in terms of the Kodama-Ishibashi variables defined in Eq.(\ref{giv}) as
\begin{eqnarray}
\label{Aomegaij}
2\Omega_{ij}=-\ell^a\ell^bD_a\Big[r^2D_b\Big(\frac{h^{(2)}_{Tij}}{r^2}\Big)\Big]
+\ell^a\ell^bD_a\Big(\hat{D}_iF^{(1)}_{bj}+\hat{D}_jF^{(1)}_{bi}\Big)
-\hat{L}_{ij}\big(\ell^a\ell^bF^{(0)}_{ab}\big)
\, .
\end{eqnarray}
Similarly, we can obtain
\begin{eqnarray}
\label{Aomegabarij}
2\bar{\Omega}_{ij}=-n^an^bD_a\Big[r^2D_b\Big(\frac{h^{(2)}_{Tij}}{r^2}\Big)\Big]
+n^an^bD_a\Big(\hat{D}_iF^{(1)}_{bj}+\hat{D}_jF^{(1)}_{bi}\Big)
-\hat{L}_{ij}\big(n^an^bF^{(0)}_{ab}\big)
\, .
\end{eqnarray}
Next we consider $C_{ij}$. From equation (\ref{Aomegaaibj}), we have
\begin{eqnarray}
&&-2\ell^an^b\Big[\Omega_{aibj}+\Omega_{ajbi}-\frac{2}{n}\gamma_{ij}\gamma^{kl}\Omega_{akbl}-2c_2 w\Big (h_{ij}-\frac{1}{n}\gamma_{ij}\gamma^{kl}h_{kl}\Big)\Big]
\nonumber\\
&&=-\ell^an^bD_b\Big(\hat{D}_ih_{aj}+\hat{D}_jh_{ai}-\frac{2}{n}\hat{D}^kh_{ak}\gamma_{ij}\Big)
-\ell^bn^aD_b\Big(\hat{D}_ih_{aj}+\hat{D}_jh_{ai}-\frac{2}{n}\hat{D}^kh_{ak}\gamma_{ij}\Big)\nonumber\\
&&-2\ell^an^b\frac{D_ar}{r}D_{b}\Big(h_{ij}-\frac{1}{n}\gamma_{ij}\gamma^{kl}h_{kl}\Big)
-2\ell^bn^a\frac{D_ar}{r}D_{b}\Big(h_{ij}-\frac{1}{n}\gamma_{ij}\gamma^{kl}h_{kl}\Big)\nonumber\\
&&-4\ell^an^b\frac{D_{a}D_br}{r}\Big(h_{ij}-\frac{1}{n}\gamma_{ij}\gamma^{kl}h_{kl}\Big)
+4\ell^an^b\frac{D_{a}rD_br}{r^2}\Big(h_{ij}-\frac{1}{n}\gamma_{ij}\gamma^{kl}h_{kl}\Big)\nonumber\\
&&+2\ell^an^bD_{a}D_{b}\Big(h_{ij}-\frac{1}{n}\gamma_{ij}\gamma^{kl}h_{kl}\Big)+2\hat{L}_{ij}(\ell^an^bh_{ab})\, .
\end{eqnarray}
With  $g^{ab}=-\ell^an^b-n^a\ell^b$, this equation can be transformed into
\begin{eqnarray}
&&-2\ell^an^b\Big[\Omega_{aibj}+\Omega_{ajbi}-\frac{2}{n}\gamma_{ij}\gamma^{kl}\Omega_{akbl}-2c_2 w\Big (h_{ij}-\frac{1}{n}\gamma_{ij}\gamma^{kl}h_{kl}\Big)\Big]
\nonumber\\
&&=D^a\Big(\hat{D}_ih_{aj}+\hat{D}_jh_{ai}-\frac{2}{n}\hat{D}^kh_{ak}\gamma_{ij}\Big)
+2\frac{D^br}{r}D_{b}\Big(h_{ij}-\frac{1}{n}\gamma_{ij}\gamma^{kl}h_{kl}\Big)
\nonumber\\
&&+2\frac{D^{b}D_br}{r}\Big(h_{ij}-\frac{1}{n}\gamma_{ij}\gamma^{kl}h_{kl}\Big)
-2\frac{(Dr)^2}{r^2}\Big(h_{ij}-\frac{1}{n}\gamma_{ij}\gamma^{kl}h_{kl}\Big)\nonumber\\
&&-D^{b}D_{b}\Big(h_{ij}-\frac{1}{n}\gamma_{ij}\gamma^{kl}h_{kl}\Big)-\hat{L}_{ij}(h_{c}{}^c)\, .
\end{eqnarray}
Substituting the decomposition of $h_{ai}$ and $h_{ij}$ in Eqs.(\ref{hai}) and (\ref{hij}) into the above equation yields
\begin{eqnarray}
&&-2\ell^an^b\Big[\Omega_{aibj}+\Omega_{ajbi}-\frac{2}{n}\gamma_{ij}\gamma^{kl}\Omega_{akbl}-2c_2 w\Big (h_{ij}-\frac{1}{n}\gamma_{ij}\gamma^{kl}h_{kl}\Big)\Big]
\nonumber\\
&&=-D^{c}D_{c}h^{(2)}_{Tij}+2\frac{D^cr}{r}D_{c}h^{(2)}_{Tij}+2\frac{D^{c}D_cr}{r}h^{(2)}_{Tij}-2\frac{D^{c}rD_cr}{r^2}h^{(2)}_{Tij}\nonumber\\
&&-D^{c}D_{c}\big(\hat{D}_ih_{Tj}^{(1)}+\hat{D}_jh_{Ti}^{(1)}\big)
+2\frac{D^cr}{r}D_{c}\big(\hat{D}_ih_{Tj}^{(1)}+\hat{D}_jh_{Ti}^{(1)}\big)\nonumber\\
&&+D^c\big(\hat{D}_ih_{cj}^{(1)}+\hat{D}_jh_{ci}^{(1)}\big)+2\frac{D^{c}D_cr}{r}\big(\hat{D}_ih_{Tj}^{(1)}+\hat{D}_jh_{Ti}^{(1)}\big)
\nonumber\\
&&-2\frac{D^{c}rD_cr}{r^2}\big(\hat{D}_ih_{Tj}^{(1)}+\hat{D}_jh_{Ti}^{(1)} \big)-\hat{L}_{ij}\Big[h_{c}^{~c}+D^{c}D_{c}h_T \nonumber\\
&&-2\frac{D^cr}{r}D_{c}h_T-2\frac{D^{c}D_cr}{r^2}h_T+ 2\frac{D^{c}rD_cr}{r^2}h_T -2D^ch_c\Big]
\, .
\end{eqnarray}
By using the gauge-invariant variables in Eq.(\ref{giv}), we have
\begin{eqnarray}
-2\ell^an^b\Big[\Omega_{aibj}+\Omega_{ajbi}-\frac{2}{n}\gamma_{ij}\gamma^{kl}\Omega_{akbl}-2c_2 w\Big (h_{ij}-\frac{1}{n}\gamma_{ij}\gamma^{kl}h_{kl}\Big)\Big]
\nonumber\\
=-D^{c}\Big[r^2D_{c}\Big(\frac{h^{(2)}_{Tij}}{r^2}\Big)\Big]+
D^c\Big(\hat{D}_iF^{(1)}_{cj}+\hat{D}_jF^{(1)}_{ci}\Big)
-\hat{L}_{ij}F^{(0)c}_{~~c}\, .
\end{eqnarray}
Thus we obtain the expression for $C_{ij}$ in terms of the gauge-invariant variables by Kodama and Ishibashi as
\begin{eqnarray}
\label{Acij}
2C_{ij}=D^{c}\Big[r^2D_{c}\Big(\frac{h^{(2)}_{Tij}}{r^2}\Big)\Big]-
D^c\Big(\hat{D}_iF^{(1)}_{cj}+\hat{D}_jF^{(1)}_{ci}\Big)
+\hat{L}_{ij}F^{(0)c}_{~~c}\, .
\end{eqnarray}
Eqs.(\ref{Aomegaij}), (\ref{Aomegabarij}) and (\ref{Acij}) are main results of this section.

\section{Penrose Wave Equation}
\label{sec:curvature}

The Penrose wave equation can be obtained by differential of the Bianchi identity. Consider the covariant derivative of the Bianchi identity
\begin{eqnarray}
\label{B1}
&&\nabla_T\nabla_SR_{MNLP}+\nabla_T\nabla_LR_{MNPS}+\nabla_T\nabla_PR_{MNSL}=0\, ,\nonumber\\
&&\nabla_P\nabla_SR_{MNLT}+\nabla_P\nabla_LR_{MNTS}+\nabla_P\nabla_TR_{MNSL}=0\, ,\nonumber\\
&&\nabla_L\nabla_SR_{MNTP}+\nabla_L\nabla_TR_{MNPS}+\nabla_L\nabla_PR_{MNST}=0\, .
\end{eqnarray}
The first equation in (\ref{B1}) minus the second and the third leads to
\begin{eqnarray}
&&\nabla_T\nabla_SR_{MNLP}+(\nabla_T\nabla_LR_{MNPS}-\nabla_L\nabla_TR_{MNPS})+(\nabla_T\nabla_PR_{MNSL}-\nabla_P\nabla_TR_{MNSL})
\nonumber\\
&&+(\nabla_P\nabla_LR_{MNST}-\nabla_L\nabla_PR_{MNST})-\nabla_P\nabla_SR_{MNLT}-\nabla_L\nabla_SR_{MNTP}\nonumber\\
&&=\nabla_T\nabla_SR_{MNLP}+R_{TLM}{}^UR_{UNPS}+R_{TLN}{}^UR_{MUPS}+R_{TLP}{}^UR_{MNUS}+R_{TLS}{}^UR_{MNPU}\nonumber\\&&
+R_{TPM}{}^{U}R_{UNSL}+R_{TPN}{}^{U}R_{MUSL}+R_{TPS}{}^{U}R_{MNUL}+R_{TPL}{}^{U}R_{MNSU}
\nonumber\\
&&+R_{PLM}{}^UR_{UNST}+R_{PLN}{}^UR_{MUST}+R_{PLS}{}^UR_{MNUT}+R_{PLT}{}^UR_{MNSU}\nonumber\\
&&-\nabla_P\nabla_SR_{MNLT}-\nabla_L\nabla_SR_{MNTP}=0\, .
\end{eqnarray}
One can obtain from the above equation that
\begin{eqnarray}
&&\Box R_{MNLP}+2R_{SLM}{}^UR_{UNP}{}^{S}+2R_{SLN}{}^UR_{MUP}{}^{S}+(R_{SLPU}+R_{SPUL})R_{MN}{}^{US}
\nonumber\\
&& +R_{L}{}^UR_{MNPU}+R_{P}{}^{U}R_{MNUL}
-\nabla_P\nabla^SR_{MNLS}-\nabla_L\nabla^SR_{MNSP}=0\, .
\end{eqnarray}
By considering the symmetry of Riemann tensor, we can rewrite the above equation as
\begin{eqnarray}
&&\Box R_{MNLP}+2R_{SLM}{}^UR_{UNP}{}^{S}+2R_{SLN}{}^UR_{MUP}{}^{S}+R_{SULP}R_{MN}{}^{SU}
 +R_{L}{}^UR_{MNPU}+R_{P}{}^{U}R_{MNUL}
\nonumber\\
&&+\nabla_P\nabla_MR_{LN}-\nabla_P\nabla_NR_{LM}-\nabla_L\nabla_MR_{NP}+\nabla_L\nabla_NR_{MP}=0\, .
\end{eqnarray}
Substituting Eqs.(\ref{EinsteinSpacetime}) and (\ref{AweylRiemann}), we can arrive at  the so called Penrose wave equation (\ref{penrosewave})
for the Einstein spacetime.  In fact, one can also obtain the wave equation (\ref{penrosewave}) from the so called de Rham wave equation~\cite{Bini:2002jx}.


\end{document}